\RequirePackage{fix-cm}
\documentclass[smallextended, natbib]{svjour3}       
\smartqed  
\usepackage{graphicx}
\usepackage{booktabs}
\usepackage{textcmds}
\usepackage{amssymb}
\usepackage{natbib,aas_macros,color}
\usepackage{mathtools}
\usepackage{siunitx}
\usepackage{glossaries}
\usepackage{tikz}
\usepackage{color}
\usepackage[colorlinks=true,citecolor=blue,urlcolor=blue,breaklinks]{hyperref}

\makeatletter
\def\cl@chapter{\@elt {theorem}}
\makeatother

\usepackage{cleveref}
\usepackage{csquotes}
\usepackage{pgf}
\usepackage{layouts}

\crefname{figure}{fig.}{figs.}
\Crefname{figure}{Fig.}{Figs.}
\crefname{section}{sec.}{secs.}
\Crefname{section}{Sec.}{Secs.}
\crefname{equation}{eq.}{eqs.}
\Crefname{equation}{Eq.}{Eqs.}
\crefname{table}{tab.}{tabs.}
\Crefname{table}{Tab.}{Tabs.}

\usetikzlibrary{external, positioning, calc, through, shadows, fadings, patterns, mindmap, trees, arrows, shapes, backgrounds, decorations.markings, decorations.pathmorphing, decorations.pathreplacing, calc, 3d, intersections, shapes.gates.logic.US, external}
\definecolor{VegaBlue}{HTML}{1F77B4}
\definecolor{VegaOrange}{HTML}{FF7F0E}

\newglossarystyle{listofconsts}{%
{\begin{longtable}{lp{\glsdescwidth}cp{\glspagelistwidth}}}%
{\end{longtable}}%
\renewcommand*{\glsgroupheading}[1]{}%
}

\newglossary{abbrev}{abs}{abo}{List of Abbreviations}
\newglossaryentry{MC}{
  name={MC}, 
  description={Monte Carlo}, 
  first={Monte Carlo (MC)},
  type=abbrev
}
\newglossaryentry{MCRT}{
  name={MCRT}, 
  description={Monte Carlo Radiative Transfer}, 
  first={Monte Carlo Radiative Transfer (MCRT)},
  type=abbrev
}
\newglossaryentry{SN}{
  name={SN}, 
  description={Supernova}, 
  first={supernova (SN)},
  type=abbrev,
  plural=SNe,
}
\newglossaryentry{SNIa}{
  name={SN~Ia}, 
  description={Type Ia Supernova}, 
  first={Type Ia Supernova (SN~Ia)},
  type=abbrev,
  plural=SNe~Ia,
}
\newglossaryentry{CMF}{
  name={CMF}, 
  description={Comoving Frame}, 
  first={comoving frame (CMF)},
  type=abbrev
}
\newglossaryentry{LF}{
  name={LF}, 
  description={Laboratory Frame}, 
  first={laboratory frame (LF)},
  type=abbrev
}
\newglossaryentry{RE}{
  name={RE}, 
  description={radiative equilibrium}, 
  first={radiative equilibrium (RE)},
  type=abbrev
}
\newglossaryentry{LTE}{
  name={LTE}, 
  description={local thermodynamic equilibrium}, 
  first={local thermodynamic equilibrium (LTE)},
  type=abbrev
}
\newglossaryentry{NLTE}{
  name={NLTE}, 
  description={local thermodynamic non-equilibrium}, 
  first={local thermodynamic non-equilibrium (NLTE)},
  type=abbrev
}
\newglossaryentry{TE}{
  name={TE}, 
  description={thermal equilibrium}, 
  first={thermal equilibrium (TE)},
  type=abbrev
}
\newglossaryentry{TRT}{
  name={TRT}, 
  description={Thermal Radiative Transfer}, 
  first={Thermal Radiative Transfer (TRT)},
  type=abbrev
}
\newglossaryentry{RT}{
  name={RT}, 
  description={Radiative Transfer}, 
  first={Radiative Transfer (RT)},
  type=abbrev
}
\newglossaryentry{RH}{
  name={RH}, 
  description={Radiation Hydrodynamics}, 
  first={Radiation Hydrodynamics (RH)},
  type=abbrev
}
\newglossaryentry{IMC}{
  name={IMC}, 
  description={Implicit Monte Carlo}, 
  first={Implicit Monte Carlo (IMC)},
  type=abbrev
}
\newglossaryentry{RNG}{
  name={RNG},
  description={Random Number Generator},
  first={Random Number Generator (RNG)},
  type=abbrev,
  plural=RNGs
}
\newglossaryentry{SPH}{
  name={SPH},
  description={Smoothed Particle Hydrodynamics},
  first={Smoothed Particle Hydrodynamics (SPH)},
  type=abbrev
}
\newglossaryentry{RW}{
  name={RW},
  description={Random Walk},
  first={Random Walk (RW},
  type=abbrev
}
\newglossaryentry{MRW}{
  name={MRW},
  description={Modified Random Walk},
  first={Modified Random Walk (MRW)},
  type=abbrev
}
\newglossaryentry{DDMC}{
  name={DDMC},
  description={Discrete Diffusion Monte Carlo},
  first={Discrete Diffusion Monte Carlo (DDMC)},
  type=abbrev
}
\newglossaryentry{IMD}{
  name={IMD},
  description={Implicit Monte Carlo Diffusion},
  first={Implicit Monte Carlo Diffusion (IMD)},
  type=abbrev
}
\newglossaryentry{SIMC}{
  name={SIMC},
  description={Symbolic Implicit Monte Carlo},
  first={Symbolic Implicit Monte Carlo (SIMC)},
  type=abbrev
}
\newglossaryentry{PDF}{
  name={PDF},
  description={Probability Density Function},
  first={Probability Density Function (PDF)},
  type=abbrev,
  plural=PDFs,
}
\newglossaryentry{CDF}{
  name={CDF},
  description={Cumulative Distribution Function},
  first={Cumulative Distribution Function (CDF)},
  type=abbrev,
  plural=CDFs,
}
\newglossaryentry{LHS}{
  name={LHS},
  description={left hand side},
  first={left hand side (LHS)},
  type=abbrev,
}
\newglossaryentry{RHS}{
  name={RHS},
  description={right hand side},
  first={right hand side (RHS)},
  type=abbrev,
}

\makenoidxglossaries

\newcommand{\simlt}{\,\hbox{\lower0.6ex\hbox{$\sim$}\llap{\raise0.6ex\hbox{$<$}}}\,}
\newcommand{\simgt}{\,\hbox{\lower0.6ex\hbox{$\sim$}\llap{\raise0.6ex\hbox{$>$}}}\,}

\journalname{Living Reviews in Computational Astrophysics}

\begin{document}

\title{Monte Carlo Radiative Transfer}

\author{Ulrich M. Noebauer$^{1,2}$
\and
Stuart A. Sim$^{3}$
}

\institute{$^1$
Max Planck Institute for Astrophysics, Garching, \\
Karl-Schwarzschild-Str. 1, \\ 
85741 Garching, \\
Germany \\
\email{unoebauer@mpa-garching.mpg.de} \\
~\\
$^2$MunichRe,\\ 
IT 1.6.1.1, K\"{o}niginstra{\ss}e 107,\\
 80802 M\"{u}nchen, \\
Germany\\
~\\
$^3$School of Mathematics and Physics, \\
Queen's University Belfast, University Road \\
Belfast BT7 1NN, \\
UK \\
\email{s.sim@qub.ac.uk}
}

\date{Received: 18 Nov 2018 / Accepted: 10 May 2019}

\maketitle

\begin{abstract}
The theory and numerical modelling of radiation processes and radiative
transfer play a key role in astrophysics: they provide the link between the
physical properties of an object and the radiation it emits. In the modern era
of increasingly high-quality observational data and sophisticated physical
theories, development and exploitation of a variety of approaches to the
modelling of radiative transfer is needed. In this article, we focus on one
remarkably versatile approach: \gls{MCRT}. We describe the principles behind
this approach, and highlight the relative ease with which they can (and have)
been implemented for application to a range of astrophysical problems. All
\gls{MCRT} methods have in common a need to consider the adverse consequences
of \gls{MC} noise in simulation results. We overview a range of methods used to
suppress this noise and comment on their relative merits for a variety of
applications.  We conclude with a brief review of specific applications for
which \gls{MCRT} methods are currently popular and comment on the prospects for
future developments.

\end{abstract}

\setcounter{tocdepth}{3}
\tableofcontents

\section{Introduction}
\label{Sec:Intro}

\subsection{The role of radiative transfer in astrophysics}

Much of astrophysics is at a disadvantage compared to other fields of physics.
While normally theories can be tested and phenomena studied by performing
repeatable experiments in the controlled environment of a lab, astrophysics
generally lacks this luxury. Instead, researchers have to mainly rely on
observations of very distant objects and phenomena over which they have no
control.  The vast majority of information about astrophysical systems is
gathered by observing their emitted radiation over the electro-magnetic
spectrum. Other messengers, such as neutrinos, charged particles and recently
gravitational waves, are also used but typically restricted to specific
astrophysical phenomena.

Given that the observation and interpretation of electro-magnetic radiation is
therefore the cornerstone of astrophysical research, a firm understanding of
how the observed signal forms and propagates is crucial. The framework of
\gls{RT} builds the theoretical foundation for this problem. It
combines concepts from kinetic theory, atomic physics, special relativity and quantum mechanics, and
provides a formalism to describe how the radiation field is shaped by the
interactions with the ambient medium.

Finding an analytic solution for \gls{RT} problems is usually very
challenging, a process that typically requires approximations and quickly
reaches its limits as the complexity of the problem increases. Thus,
numerical methods are normally employed instead. In such cases, one considers a
discretized version of the transfer equation, e.g.\ by replacing differentials
with finite differences, and uses sophisticated solution schemes to minimize
the inevitably introduced numerical errors. While being an established
approach, this often leads to very complex numerical schemes and faces some
particular challenges when scattering interactions have to be included or when
problems without internal symmetries require a fully multidimensional
treatment.

\gls{MC} methods offer a completely different approach to \gls{RT}
problems. Instead of discretizing the \gls{RT} equations, the
underlying \gls{RT} process is ``simulated'' by introducing a large
number of ``test particles'' (later referred to as ``packets'' in this
article). These test particles behave in a manner similar to their physical
counterparts, namely real photons. In particular, particles move, can scatter
or be absorbed during a \gls{MC} calculation. In the simulations, decisions
about the propagation behaviour of a particular test particle, e.g.\ when,
where and how it interacts, are taken stochastically. Seemingly, this leads to
a random propagation behaviour of the individual particles. However, as an
ensemble, the particle population can provide an accurate representation of the
transfer process and the evolution of the radiation field, provided that the
sample size is chosen sufficiently large. 

Given its design, the \gls{MC} approach to \gls{RT} offers a number of
compelling benefits. Due to its inspiration from the microphysical
interpretation of the \gls{RT} process, devising a \gls{MC} \gls{RT} scheme is
very intuitive and conceptually simple. This often leads to comparably simple
computer programs and involves moderate coding efforts: basic \gls{MCRT} routines to solve simple \gls{RT}
problems can be coded in only a few lines that combine a random number
generator with a number of basic loops (we provide a number of simple examples
of how this can be done as part of our discussions later in this article).
From a physical standpoint, a significant advantage of \gls{MC} methods is the
ease with which scattering processes are incorporated, a task which proves much
more challenging for traditional, deterministic solution approaches to \gls{RT}
problems. In addition, \gls{MCRT} calculations can be generalised with little
effort from problems with internal symmetries to problems with arbitrarily
complex geometrical characteristics. This feature makes \gls{MCRT} techniques
often the preferred choice for multidimensional \gls{RT} calculations.
Finally, the \gls{MCRT} treatment is often referred to as ``embarrassingly
parallel'' to describe its ideal suitability for modern high performance
computing in which the workload is distributed on a multitude of processing
units. Just as the photons they represent, the individual \gls{MC} particles
are completely decoupled and propagate independently of each other. Thus, each
processing unit can simply treat a subset of the entire particle population
without the need for much communication.\footnote{Note, however, that this
  situation changes when the data structures holding for example atomic data or
  the computational grid become too large to fit into the memory of a single
  computing node. Then these data structures have to be split and communicating
  \gls{MC} particles between threads becomes the performance bottleneck
  \citep[see, e.g.][for more details on possible parallelization schemes for
such situations]{Harries2015}.}

Of course, the \gls{MC} approach is not without its downsides. The most severe
disadvantage is a direct consequence of the probabilistic nature of \gls{MC}
techniques: inevitably, any physical quantity extracted from \gls{MC}
calculations will be subject to stochastic fluctuations. This \gls{MC} noise
can be decreased by increasing the number of particles, which naturally
requires more computational resources. Consequently, \gls{MC} calculations are
often computationally expensive. These costs further increase if
the optical thickness of the simulated environment is high. Since the
propagation of each particle has to be followed explicitly, the efficiency of
conventional \gls{MCRT} schemes suffers greatly if the number of interactions
the particles experience increases. Consequently, \gls{MCRT} approaches are
typically ill-suited for \gls{RT} problems in the diffusion
regime. Furthermore, as pointed
out by \citet{Camps2018}, care has to be taken when interpreting results of
\gls{MCRT} simulations applied to environments with intermediate to high
optical depth.  The need to explicitly follow the propagation of the individual
\gls{MC} particles is the cause for yet another drawback of \gls{MCRT}
approaches. In deterministic solution strategies to \gls{RT}, implicit
time-stepping is often used to improve numerical stability in situations with
short characteristic time scales. By design, conventional \gls{MCRT} schemes
require following the propagation of the individual particles in a
time-explicit fashion. It is thus very challenging to devise truly implicit
\gls{MCRT} approaches to overcome numerical stability problems.  In the course
of this review, we will highlight a variety of different techniques which have
been devised to address and alleviate each of these drawbacks.

\subsection{Scope of this review}

\gls{MC} techniques have become a popular and widely-used approach to address
\gls{RT} problems in many disciplines of physical and engineering research.
Covering all the different aspects and applications of \gls{MCRT} is beyond the
scope of this article and we refer readers to existing surveys of the
respective fields. Among these, we highlight the recent overview of \gls{MCRT}
in atmospheric physics by \citet{Mayer2009}, the seminal report by
\citet{Carter1975} and the book by \citet{Dupree2002}, which both discuss
\gls{MCRT} techniques to solving neutron transport problems, and to the
article by \citet{Rogers2006}, who reviews \gls{MCRT} methods in the field of
medical physics. In this article, we aim to provide an introduction to
\gls{MC} techniques used in astrophysics to mainly address photon transport
problems. While attempting to provide a general and comprehensive overview, we
take the liberty to put some emphasis on specific techniques used in our own
field of research, namely \gls{RT} in fast outflows, i.e.\ \gls{SN} ejecta,
accretion disc and stellar winds. We feel that this approach is appropriate
given that dedicated overviews of \gls{MCRT} methods for specific fields of
astrophysical research already exist. In particular, we refer the
reader to the reviews by \citet{Whitney2011} and \citet{Steinacker2013} on
\gls{MCRT} for astrophysical dust \gls{RT} problems.

\subsection{Structure of this review}

We have structured this review as follows: in \Cref{sec:RT} we briefly review
some fundamentals of radiative transfer theory that are relevant for our
presentation.  We begin the actual discussion of \gls{MCRT} methods with a
brief look at their history and review of their astrophysical applications in
\Cref{Sec:History}, and by introducing the basic concepts of a random number
generator and random sampling in \Cref{Sec:MCbasics}. In the following part,
\Cref{Sec:Quanta}, the basic discretization into \gls{MC} quanta or packets
will be introduced before their propagation procedure is explained in
\Cref{Sec:Propagation}.  In \Cref{Sec:macroatom}, we discuss how emissivity
by thermal and/or fluorescent processes can be incorporated in \gls{MCRT}
simulations.

Having introduced the basic \gls{MCRT} principles, the complications arising in
moving media, in particular the need to distinguish reference frames, are
discussed in \Cref{Sec:Flows}.  In \Cref{Sec:Estimators} we detail various
techniques to reconstruct important radiation field quantities from the
ensemble of \gls{MC} packet trajectories and interaction histories. Here,
particular emphasis is put on methods that reduce the inherent stochastic
fluctuations in the reconstructed quantities, such as biasing and volume-based
estimators.  In \Cref{Sec:imc_and_ddmc} advanced \gls{MC} techniques, such as
\gls{IMC} and \gls{DDMC}, are described which can be used to improve the
numerical stability of \gls{MCRT} calculations and their efficiency in optically
thick environments.  We conclude this review by touching upon the challenge of
coupling \gls{MCRT} to hydrodynamical calculations in \Cref{Sec:Dynamics} and
by presenting a hands-on example of applying \gls{MCRT} to \gls{SN} ejecta in
\Cref{Sec:example}.

\section{Radiative transfer background}
\label{sec:RT}

Before turning to the main focus of this review, a brief overview of the
fundamentals of \gls{RT} is in order to introduce the necessary nomenclature
and to define the basic physical concepts underlying \gls{MCRT}
calculations.  We assume the reader is already familiar with the principles of
\gls{RT} and so will not present a complete derivation. More rigorous
presentation of these principles are available in the literature, for example
in the books by \citet{Chandrasekhar1960}, \citet{Mihalas1978}, \citet{Rybicki1979},
\citet{Mihalas1984} and \citet{Hubeny2014}.

From a macroscopic perspective, \gls{RT} calculations rest on the
transfer equation\footnote{We
neglect general relativistic effects in this article.}
\begin{equation}
  \left( \frac{1}{c}\frac{\partial}{\partial t} + \nabla \cdot \mathbf{n} \right) I(\mathbf{x}, t; \mathbf{n}, \nu) = \eta(\mathbf{x}, t; \mathbf{n}, \nu) - \chi(\mathbf{x}, t; \mathbf{n}, \nu) I(\mathbf{x}, t; \mathbf{n}, \nu),
  \label{eq:transfer_eq}
\end{equation}
which encodes how the radiation field, expressed in terms of the specific
intensity $I$, varies with time, $t$ and in space, $\mathbf{x}$. The specific intensity
is defined in terms of the monochromatic energy $\mathrm{d}\mathcal{E}$ in the
frequency range $[\nu, \nu+\mathrm{d}\nu]$ streaming through a surface element
$\mathrm{d}\mathbf{A}$ during the time $\mathrm{d}t$ into the solid angle
$\mathrm{d}\Omega$ about the direction $\mathbf{n}$:
\begin{equation}
  \mathrm{d}\mathcal{E}(\mathbf{x}, t; \mathbf{n}, \nu) =
  I(\mathbf{x}, t; \mathbf{n}, \nu) \; \mathrm{d}\nu \; \mathrm{d} t
  \; \mathrm{d}\Omega \; \mathrm{d}\mathbf{A} \cdot \mathbf{n}.
  \label{eq:specific_intensity}
\end{equation}

The transfer equation can be interpreted as capturing the changes in the
radiation field induced by an imbalance of in- and outflows (left hand side)
and by interactions with the ambient material (source and sink terms on the
right hand side). This coupling to the surrounding material is described by two
material functions. The emissivity $\eta$ encodes how much energy is added to
the radiation field due to emission processes. The second term on the right
hand side of \Cref{eq:transfer_eq}, which involves the opacity
$\chi$, captures the opposite effect, namely how much radiation energy is
removed by absorptions. Emissivity and opacity are often combined into the 
source function
\begin{equation}
  S = \frac{\eta}{\chi}.
  \label{eq:source_function}
\end{equation}
The opaqueness of a medium along a given ray is usually quantified in terms of
the optical depth\begin{equation}
  \tau(l) = \int_0^l \mathrm{d}s \chi(\mathbf{x}(s), t; \mathbf{n}, \nu) \; ,
  \label{optical depth}
\end{equation}
which essentially measures the mean number of interactions a photon would
undergo along a trajectory $s$ from $\mathbf{x}(s=0)$ to $\mathbf{x}(s=l)$.

Scatterings can be incorporated into this description by formally splitting the
scattering process into an absorption which is immediately followed by an
emission. It should be noted, however, that the \gls{RT} problem is
often significantly complicated by the presence of scattering interactions
since these processes redistribute radiation in both frequency and direction
and introduce a non-local coupling to the ambient material.

It is often insightful to describe the radiation field not only in terms of the
specific intensity but also consider its moments. These involve a
solid angle average over the specific intensity and different powers of the
propagation direction. From the zeroth to the second moment, these quantities
have a clear physical interpretation. In particular, the zeroth
moment is identical to the mean intensity
\begin{equation}
  J(\mathbf{x}, t; \mathbf{n}, \nu) = \frac{1}{4 \pi} \int \mathrm{d} \Omega I(\mathbf{x}, t; \mathbf{n}, \nu),
  \label{eq:mean_intensity}
\end{equation}
which in turn is closely related to the energy density of the radiation field
\begin{equation}
  E(\mathbf{x}, t; \mathbf{n}, \nu) = \frac{4 \pi}{c} J(\mathbf{x}, t; \mathbf{n}, \nu).
  \label{eq:radiation_energy_density}
\end{equation}
The next higher moment is the vector quantity 
\begin{equation}
  \mathbf{H}(\mathbf{x}, t; \mathbf{n}, \nu) = \frac{1}{4 \pi} \int \mathrm{d} \Omega I(\mathbf{x}, t; \mathbf{n}, \nu) \mathbf{n}
  \label{eq:radiation_first_moment}
\end{equation}
and captures the radiation field energy flux 
\begin{equation}
  \mathbf{F}(\mathbf{x}, t; \mathbf{n}, \nu) = 4 \pi \mathbf{H}(\mathbf{x}, t; \mathbf{n}, \nu).
  \label{eq:radiation_energy_flux}
\end{equation}
Analogously, the second moment becomes a tensor 
\begin{equation}
  K^{ij}(\mathbf{x}, t; \mathbf{n}, \nu) = \frac{1}{4 \pi} \int \mathrm{d} \Omega I(\mathbf{x}, t; \mathbf{n}, \nu) n^i n^j
  \label{eq:radiation_second_moment}
\end{equation}
and relates to the radiation pressure
\begin{equation}
  P^{ij}(\mathbf{x}, t; \mathbf{n}, \nu) = \frac{4 \pi}{c} K^{ij}(\mathbf{x}, t; \mathbf{n}, \nu) 
  \label{eq:radiation_pressure}
\end{equation}
with each entry describing the flux of the radiation field momentum component
$i$ into the direction $j$.

We conclude this brief overview of basic \gls{RT} concepts by
introducing two important reference frames. As the name suggests, the \gls{LF}
is defined such that the laboratory is at rest. Consequently, it lends itself
naturally for convenient measurements of space and time. However, from the
perspective of interaction physics, a more natural frame is one in which the
interaction partner, i.e.\ the ambient material, is at rest. This frame is
typically referred to as the \gls{CMF}.  In general, it cannot be defined globally
whenever gradients in the fluid velocity are encountered.  However, a local
definition of the \gls{CMF}, which is advected by the fluid flow, can be
performed\footnote{Due to the local definition of the \gls{CMF}, it is not an
inertial frame \citep[see e.g. detailed discussion of this in][]{Mihalas1984}}.
Throughout this work, we adopt the nomenclature that quantities defined or
measured in the \gls{CMF} are designated with a subscribed zero. When changing
between these reference frames, certain transformation rules have to be obeyed.
Most importantly, these transformations lead to the Doppler effect 
\begin{equation}
  \nu_0 = \gamma \nu (1 - \mathbf{\boldsymbol{\beta} \cdot n})
  \label{eq:doppler_effect}
\end{equation}
and induce aberration
\begin{equation}
  \mathbf{n_0} = \left( \frac{\nu}{\nu_0} \right) \left[ \mathbf{n} - \left( 1 - \frac{\gamma \mathbf{\boldsymbol{\beta} \cdot n}}{\gamma + 1} \right) \gamma \boldsymbol{\beta} \right],
  \label{eq:angle_aberration}
\end{equation}
where $\boldsymbol{\beta} = \mathbf{v} /c$ is the ratio of the local
velocity to the speed of light and $\gamma = (1-\beta^2)^{-1/2}$ (with
$\beta = |\boldsymbol{\beta}|$).
Transformation rules for the other quantities introduced in this section, such
as the specific intensity, the opacity and emissivity 
\begin{align}
  I_0(\mathbf{x}_0, t_0; \mathbf{n}_0, \nu_0) &= I(\mathbf{x}, t; \mathbf{n}, \nu) \left( \frac{\nu_0}{\nu} \right)^3, \\
  \chi_0(\mathbf{x}_0, t_0; \mathbf{n}_0, \nu_0)  &= \chi(\mathbf{x}, t; \mathbf{n}, \nu)\frac{\nu}{\nu_0},\\
  \eta_0(\mathbf{x}_0, t_0; \mathbf{n}_0, \nu_0)  &= \eta(\mathbf{x}, t; \mathbf{n}, \nu) \left(\frac{\nu_0}{\nu}  \right)^2,
  \label{eq:transformation_laws_I_chi_eta}
\end{align}
have been first derived by \citet{Thomas1930} and are also discussed by
\citet{Mihalas1984}, for example.

\section{Historical sketch of the Monte Carlo method}
\label{Sec:History}

When Nicholas Metropolis suggested a name for the statistical method just
invented to study neutron transport through fissionable material
\citep{Metropolis1987}, he clearly drew inspiration from the game of chance
which is always played at the heart of \gls{MC} calculations. From
a historical perspective, Georges-Louis Leclerc, Comte de Buffon, is commonly
credited as having devised the first \gls{MC} experiment
\citep[cf.][]{House1968, Dupree2002, Kalos2008}. He considered a plane with a
superimposed grid of parallel lines and was interested in the probability that
a needle which is tossed onto the plane intersects one of the lines
\citep{Buffon1777}. It was later suggested, by Laplace, that such a
scenario may be used to experimentally determine the value of $\pi$
\citep{Laplace1812}. In 1873, the astronomer Asaph Hall reports in a short note
to the Messenger of Mathematics the successful execution of this experiment,
carried out in 1864 by his friend Captain O.\ C.\ Fox \citep{Hall1873}. A
detailed description of what is known today as ``Buffon's needle problem'' is
for example provided by \citet{Dupree2002} or \citet{Kalos2008}.

Notwithstanding these early rudimentary applications, the \gls{MC} method in
its modern form to solve transport problems has been established and shaped in
the 1940s, mostly by Stanis{\l}aw Ulam and John von Neumann \citep[see
e.g.][]{Metropolis1987}. Recognising the immense potential and utility of the
first large-scale electronic computers, which became operational at the time,
they harnessed the mathematical concept of ``statistical sampling'' to solve
the neutron transport problems in fissionable material, thus launching the
\gls{MC} method\footnote{According to Emilio Segr{\`e}, Enrico Fermi already
used statistical sampling to address neutron diffusion problems in the 1930s in
Rome. Doing the calculations by hand, he thus independently developed the
modern \gls{MC} method \citep[cf.][]{Anderson1986, Metropolis1987}.}.

With the growing availability of computational resources, which accompanied the
rapid development of computers, \gls{MC} methods became increasingly popular
and their application spread across many different scientific disciplines. In
the late 1960s, \gls{MC} calculations finally entered the astrophysics stage,
for example with the works by \citet{Auer1968}, \citet{Avery1968} and
\citet{Magnan1968, Magnan1970}.  \citet{House1968} review the status of these
early \gls{MC}-based \gls{RT} investigations.  In the time since, \gls{MC}
methods have become established, successful and reliable tools for the study of
a variety of astrophysical \gls{RT} phenomena.  These range all the way from
planetary atmospheres \citep[e.g.][]{Lee2017} to cosmological simulations of
reionization \citep[e.g.][]{Ciardi2001, Baek2009, Maselli2009, Graziani2013}.
The wide range of applications indicates the broad utility of \gls{MC} methods
for astrophysical applications. Many of these fields have in common needs that
involve a sophisticated treatment of scattering, complex (i.e.\ non-spherical)
geometries and/or complicated opacities.  For example, many astrophysical
\gls{MCRT} studies involve stellar winds \citep[e.g.][]{Lucy1983, Abbott1985,
  Lucy1987, Hillier1991, Lucy1993, Schmutz1997, Vink1999, Harries2000,
Vink2000, Sim2004, Watanabe2006, Lucy2007, Mueller2008, Lucy2010, Vink2011, Lucy2012, Lucy2012a, Muijres2012,
Muijres2012a, Surlan2012, Surlan2013, Mueller2014,Lucy2015, Noebauer2015a, Vink2018}, mass outflows from
disks \citep[e.g.][]{Knigge1995, Knigge1997, Long2002, Sim2005, Sim2005a,
Sim2008, Noebauer2010, Sim2010, Odaka2011, Sim2012, Higginbottom2013,
Kusterer2014, Hagino2015, Matthews2015, Matthews2016, Matthews2017,
Tomaru2018}, or supernovae \citep[e.g.][]{Lucy1987a, Janka1989, Mazzali1993,
  Lucy1999a, Mazzali2000, Lucy2005, Stehle2005, Kasen2006, Sim2007, Kromer2009,
  Jerkstrand2011, Abdikamalov2012, Jerkstrand2012, Wollaeger2013,
  Kerzendorf2014, Wollaeger2014, Bulla2015, Fransson2015, Jerkstrand2015,
  Roth2015, Botyanszki2018, Ergon2018, Sand2018}. In these environments a
  treatment of multiply overlapping spectral lines in high-velocity gradient
  flows are crucial. Others depend on accurate simulations of scattering, be it
  for high-energy processes \citep[e.g.][]{Pozdnyakov1983, Stern1995,
    Molnar1999, Cullen2001, Yao2005, Dolence2009, Ghosh2009, Ghosh2010,
  Schnittman2010, Tamborra2018} or from dust-rich structures
  \citep[e.g.][]{Witt1977, Yusef-Zadeh1984, Dullemond2000, Bjorkman2001,
    Gordon2001, Misselt2001, Juvela2003, Niccolini2003, Jonsson2006, Pinte2006,
    Bianchi2008, Pinte2009, Jonsson2010, Baes2011, Robitaille2011, Whitney2011,
  Lunttila2012, Camps2013, Camps2015, Gordon2017a, Verstocken2017}. Many of the
  applications primarily aim to calculate synthetic observables but \gls{MCRT}
  methods are also used to determine physical and/or dynamical conditions in
  complex multidimensional geometries, such as star forming environments,
  disc-like structures, nebulae or circumstellar material configurations
  \citep[e.g.][]{Wood1996, Och1998, Bjorkman2001, Ercolano2003, Kurosawa2004,
  Ercolano2005, Carciofi2006,Altay2008, Carciofi2008, Ercolano2008, Pinte2009,
Harries2011, Haworth2012, Harries2015, Hubber2016, Lomax2016, Harries2017}.
\gls{MCRT} schemes have also found use in astrophysical problems that require a
general relativistic treatment of radiative transfer processes
\citep[e.g.][]{Zink2008, Dolence2009, Ryan2015}.

\section{Monte Carlo basics}
\label{Sec:MCbasics}

At the heart of \gls{MCRT} techniques lies a large sequence of decisions about
the fate of the simulated quanta. These decisions reflect the underlying
physical processes and, as an ensemble, provide a representative description of the
transport process. On an individual level, this is realised by selecting from
the pool of possible outcomes based on a set of probabilities that encode the
underlying physics. This procedure is typically referred to as ``random
sampling'' and will be discussed below.

\subsection{Random Number Generation}
\label{Sec:RNG}

\begin{figure}[htb]
  \centering
  \includegraphics[width=\textwidth]{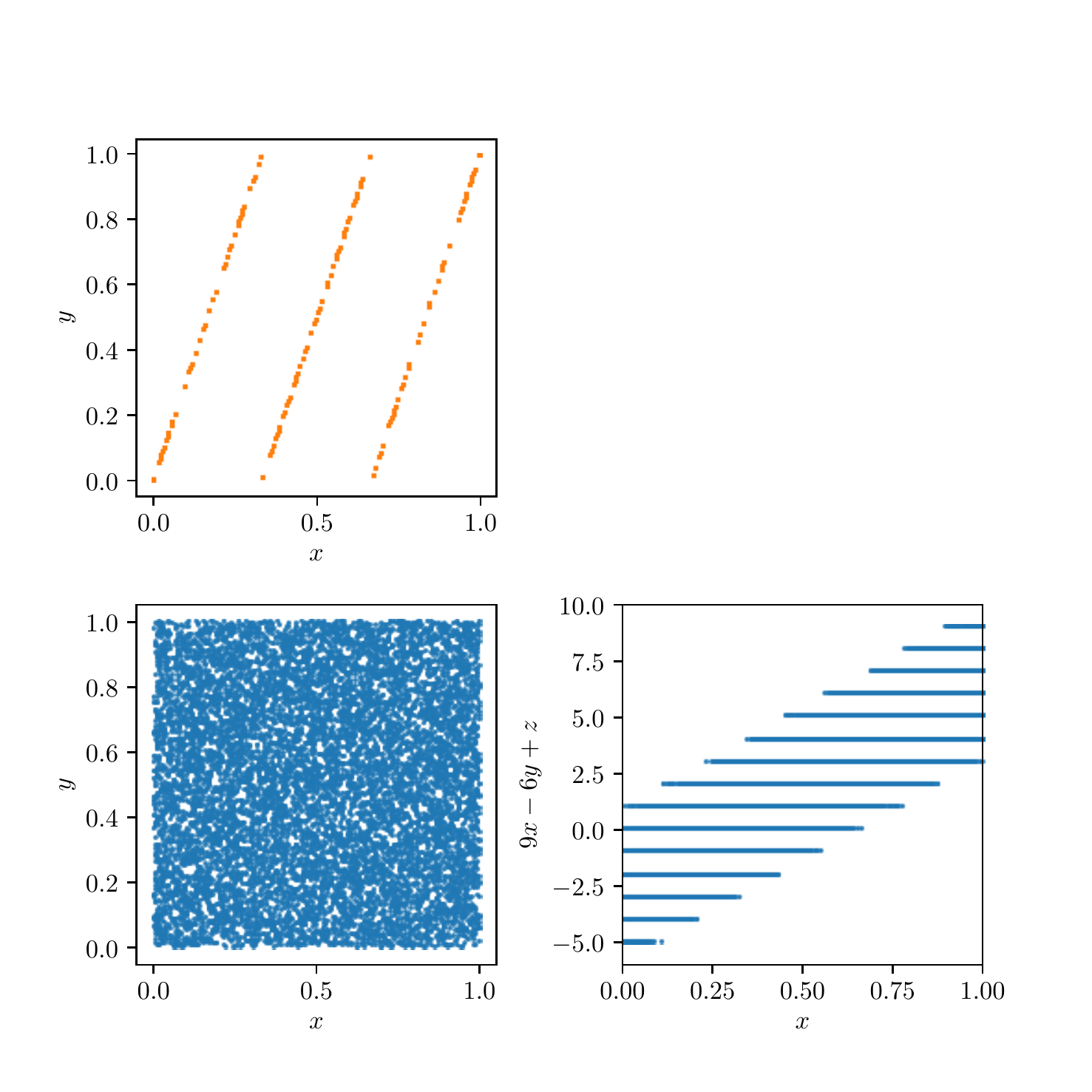}
  \caption{Illustration of two ``bad'' \glspl{RNG} that are based on the linear
    congruential scheme of \Cref{eq:mcbasics:lin_con_rng}. The upper left panel
    shows pairs, $(x,y) = (\xi_i, \xi_{i+1})$, of (normalised) sequential draws
    from an \gls{RNG} with a deliberately short period ($a = 3$, $c = 1$, $M = 257$
    and $\xi_0 = 11$). The lower panels contain draws from an \gls{RNG} with a
    much longer period of $M = 2^{31}$ based on sequential pairs
    (left) and triples ($(x,y,z) = (\xi_i, \xi_{i+1}, \xi_{i+2})$, right) . Due to the short period, the first \gls{RNG}
    performs poorly, exhibits strong correlations between successive draws and
    does not fill the unit square uniformly as seen in the upper left panel.
    The \gls{RNG} with the significantly longer period seems to perform much better:
    the unit square is filled evenly and no obvious correlations stand out when
    considering two successive draws (lower left panel). However, if three
    successive draws from this \gls{RNG} are examined, strong correlations become
    apparent as seen in the lower right panel. The infamous \texttt{RANDU} ($a
    = 65539$, $c = 0$, $M = 2^{31}$) was used to generate the data for this
    demonstration and illustrates an \gls{RNG} that fails lattice tests
  \citep[e.g.][]{Fishman1982}.}
  \label{fig:mcbasics:rng_props}
\end{figure}

Critical to the outline above is that some form of randomness is required to
perform the sampling, and thus the \gls{MCRT} calculation, on a computer.  True
randomness is difficult to achieve on a machine which is inherently
deterministic, but for many purposes ``pseudo-randomness'' is sufficient which
can be obtained via a so-called (pseudo) \gls{RNG}. Based on a starting value
(referred to as the seed), these algorithms provide sequences of numbers,
$\xi$, typically uniformly distributed over the interval $[0, 1[$.  Such
sequences are referred to as ``pseudo'' random since they share statistical
properties with true randomness but are still generated by relying on
deterministic prescriptions.  A well-known example of such algorithms is the
family of linear congruential methods. Based on a previous draw, $\xi_i$, and a
set of large numbers, $a$, $c$ and $M$, a new random number is generated
by\footnote{The resulting numbers can be mapped onto the unit interval $[0, 1[$
by dividing by $M$.}
\begin{equation}
  \xi_{i+1} = (a \xi_i + c) \; \mathrm{mod} \; M
  \label{eq:mcbasics:lin_con_rng}
\end{equation}
For the purpose of \gls{MCRT} applications, the ``pseudo''-randomness is not
problematic as long as the \gls{RNG} period, i.e.\ the lengths after which
repetitions occur\footnote{For the linear congruential methods as defined by
\Cref{eq:mcbasics:lin_con_rng}, the period can at most reach $M$.}, is large
and as long as the \gls{RNG} exhibits a good lattice structure. The latter
implies that $s$-tuples of successive \gls{RNG} draws, $(\xi_n,
\xi_{n+1},\ldots, \xi_{n+s-1})$, are evenly distributed within the
$s$-dimensional hypercube \citep[see e.g.][]{Kalos2008}, a property which a
number of early multiplicative congruential methods -- algorithms of the family
\Cref{eq:mcbasics:lin_con_rng} but with $c = 0$ -- lacked
\citep[first pointed out by][]{Marsaglia1968}.  \Cref{fig:mcbasics:rng_props}
illustrates some possible shortcomings of poorly performing \glspl{RNG}.
Popular examples for \glspl{RNG}, which fulfil the above requirements and are
well-suited for \gls{MCRT} applications include for example the Mersenne
Twister \citep{Matsumoto1998} or members of the \texttt{xorshift} family
\citep{Marsaglia2003}. 

\subsection{Random sampling}
\label{Sec:Sampling}

With the help of \glspl{RNG}, random numbers\footnote{For the sake of brevity
we will omit the attribute ``pseudo''.} can be rapidly produced on the computer
and used for sampling physical processes by mapping them onto the target
probability distribution. To illustrate the different sampling schemes, we
first introduce some basic concepts from statistics. For the sake of brevity,
we again refrain from a rigorous mathematical presentation and instead refer
the interested reader to the standard literature on the topic, e.g.\
\citet{Kalos2008}. 

\subsubsection{Sampling from an inverse transformation}

Consider a physical process with outcomes described by the random variable $X$.
We further assume, that $X$ is continuous and can take values from $[0,
  \infty[$. In this case, the probability that a certain event occurs, i.e.\
    that $X$ takes a value within $[x, x + \mathrm{d}x]$, is given by the
    so-called \gls{PDF} $\rho_{X}(x) \mathrm{d}x$, which fulfils the
    normalisation
\begin{equation}
  1 = \int_0^{\infty} \mathrm{d}x\,\rho_{X}(x).
  \label{eq:mcbasics:normalisation}
\end{equation}
With the density, the probability that $X$ takes any value less or equal to $x$
can be calculated, resulting in the \gls{CDF}
\begin{equation}
  f_{X}(x) = \int_0^{x} \mathrm{d}x' \rho_{X}(x').
  \label{eq:mcbasics:cumdist}
\end{equation}
Unlike the probability density, which is always positive but not necessarily
monotonous, the cumulative distributed function (by definition) is always monotonous.
Consequently, it can be used to establish a mapping between two probability
distributions via
\begin{equation}
  f_{Y}(y) = f_{X}(x).
  \label{eq:mcbasics:cumdist_equate}
\end{equation}
This is the fundamental concept of sampling one probability distribution
$\rho_{X}(x)$ using draws from another, $\rho_{Y}(y)$. Using the random
numbers $\xi$ provided by the \gls{RNG}, which are uniformly distributed between 0
and 1 and thus have $f_{\xi}(\xi) = \xi$, this simplifies to
\begin{equation}
  \xi = f_{X}(x) \; ,
  \label{eq:mcbasics:cumdist_rng}
\end{equation} 
which, after inversion, results in the sampling rule
\begin{equation}
  x = f_{X}^{-1}(\xi).
  \label{eq:mcbasics:inv_trans}
\end{equation}

Below, we illustrate this sampling process via examples of relevance to a number of physical
process in \gls{MCRT} applications. 

\paragraph{Example 1: Selecting directions}

Consider the situation of isotropic scattering of a photon (using a spherical
polar coordinate system). In this case, no propagation direction after the
interaction is preferred and the probability that the photon escapes into a
specific solid angle element $\mathrm{d} \Omega = \mathrm{d}\phi
\mathrm{d}\theta \sin \theta$ (with $\phi \in [0, 2\pi]$ and $\theta \in [0,
\pi]$) is constant
\begin{equation}
  \rho(\phi, \theta) \mathrm{d}\phi \mathrm{d}\theta \sin \theta = \mathrm{const}.
  \label{eq:mcbasics:const_solidangle_prop}
\end{equation}
Since the two angles are independent, and after the introduction of the
directional parameter
\begin{equation}
  \mu = \cos \theta,
  \label{eq:mcbasics:mu}
\end{equation}
this reduces to
\begin{align}
  \rho(\phi)\mathrm{d}\phi & = \mathrm{const}, \label{eq:mcbasics:phi_dens}\\
  \rho(\mu) \mathrm{d}\mu &= \mathrm{const}. \label{eq:mcbasics:mu_dens}
\end{align}
and finally results in the sampling rules
\begin{align}
  \phi &= 2 \pi \xi_1, \label{eq:mcbasics:phi_sample}\\
  \mu &= 2 \xi_2 - 1. \label{eq:mcbasics:mu_sample}
\end{align}
Thus to randomly select a direction of propagation, we draw two independent
random numbers ($\xi_1$, $\xi_2$) from the \gls{RNG} sequence and use them to
determine the direction via \Cref{eq:mcbasics:phi_sample} and
\Cref{eq:mcbasics:mu_sample}.

\paragraph{Example 2: Selecting interaction points}

A critically important application of random sampling is the decision when
photons interact. The probability that a photon interacts within $\mathrm{d}l$
after having covered a distance $l$ along its trajectory is given by the
probability density
\begin{equation}
  \rho(l) = \chi(l) \exp\left(-\int_0^l \mathrm{d}l' \chi(l')\right).
  \label{eq:mcbasics:normalized_int_prob}
\end{equation}
We omit a rigorous deviation of this result and refer to the literature
instead, in particular to \citet[][Sec.~6.3]{Kalos2008}. However, the
physics of this result can be quickly appreciated by
recognising it as the product of the probability of having travelled distance $l$ with no interaction (given by the exponential term) and the 
probability per unit length ($\chi$) of undergoing an interaction at the position reached after travelling $l$.
The inverse
transformation technique can be combined with \Cref{eq:mcbasics:normalized_int_prob} to determine the distance a 
photon will travel to the next interaction event leading to
\begin{equation}
  - \ln \xi = \int_0^l \mathrm{d}l' \chi(l').
  \label{eq:mcbasics:l_sampling_inhom}
\end{equation}
Here, we used the fact that $1 - \xi$ is equally distributed as $\xi$. In the case
of a homogeneous medium, $\chi$ is constant and the sampling rule simplifies to
\begin{equation}
  l = - \frac{\ln \xi}{\chi}.
  \label{eq:mcbasics:l_sampling}
\end{equation}

\subsubsection{Alternative sampling techniques}

In the examples above, the inverse transformation technique was used to sample
the involved physical processes since the underlying cumulative distribution
function could easily and analytically be inverted. Naturally, this is not
always feasible and in such cases, one has to rely on other sampling methods.
However, even if determining the cumulative distribution function is
analytically challenging, it can be done by means of numerical integration and
values for $f_{X}(x)$ pre-calculated for a number of monotonically increasing
$x_i$. Once these values, $f_{X}(x_i)$, are available, the distribution can be
sampled by first selecting a grid interval $[x_i, x_{i+1}]$ according
to\footnote{This is a common procedure to sample discrete probabilities
\citep[see e.g.][]{Carter1975}.}
\begin{equation}
  i = \mathrm{max}\left\{j;  f_{X}(x_j) \le \xi \right\}.
  \label{eq:mcbasics:interval_index}
\end{equation}
Since $\xi$ now lies between $f_{X}(x_i)$ and $f_{X}(x_{i+1})$, the final
sampling is performed by linear interpolation \citep[see e.g.][]{Carter1975}
\begin{equation}
  x = x_{i+1} - \frac{f_{X}(x_{i+1}) - \xi}{f_{X}(x_{i+1}) - f_{X}(x_i)} (x_{i+1} - x_{i}).
  \label{eq:mcbasics:linear_interp_sampling}
\end{equation}
Naturally, this approach only approximates the underlying probability
distribution and the accuracy increases with the number of grid points at which
$f_{X}$ is evaluated.

Another popular sampling technique which is applicable also to complex
distributions is the so-called rejection sampling method \citep[see, e.g.][for
a detailed description]{Carter1975,Kalos2008}. This approach is closely related
to \gls{MC}-based integration. We briefly illustrate its basic principles for
the example of one-dimensional distributions. In this case, pairs of random
numbers $(x, y) = (\xi_1, \xi_2)$ are generated\footnote{We assume that $\min
\rho_{X}(x) = 0$ and $\max \rho_{X}(x) = 1$. Otherwise, the draws for $\xi_2$
have to be scaled and shifted appropriately.}. If
\begin{equation}
  \xi_2 \le \rho_{X}(\xi_1)
  \label{eq:mcbasics:rejection_sampling}
\end{equation}
the trial $\xi_1$ is accepted as a valid sample of $\rho_{X}(x)$, otherwise it
is rejected and the procedure repeated until the desired number of samples is
obtained. This technique involves a certain level of unpredictability since not
every trial draw produces an accepted sample.  

In addition to the general sampling techniques outlined above, a
number of
specific schemes tailored to probability distributions of particular interest
are available. In the context of \gls{RT}, a prominent example is the
sampling of frequencies for a thermal radiation field from the normalized
Planck distribution,
\begin{align}
  b_{\nu} &= \frac{15}{\pi^4} \frac{x^3}{\exp(x) - 1}, 
  \label{eq:mcbasics:norm_planck}\\
  \nu &= \frac{k_{\mathrm{B}} T}{h}x.\label{eq:mcbasics:nondim_freq}
\end{align}
For this problem, Barnett and Canfield\footnote{Unpublished Lawrence Radiation
Laboratory internal report, cf.\ \citet{Fleck1971}} have proposed an
efficient sampling technique based on the series expansion of the Planck
function. This technique, which has been reviewed numerous times in the
literature (for example \citealt{Fleck1971}, \citealt{Carter1975} and \citealt{Bjorkman2001}),
relies on five uniform random numbers $\xi_0,\cdots, \xi_4$. The first one is
used in the minimization process
\begin{equation}
  L = \min \left\{l; \sum_1^l j^{-4} \geq \xi_0 \frac{\pi^4}{90} \right\},
  \label{eq:mcbasics:Lmin}
\end{equation}
providing $L$, which in turn is combined with the remaining random numbers to
give the final non-dimensional frequency
\begin{equation}
  x = -L^{-1} \ln(\xi_1 \xi_2 \xi_3 \xi_4).
  \label{eq:mcbasics:final_nondim_freq}
\end{equation}
According to \citet{Fleck1971}, $1.1$ trials (in terms of elements in the
summation in \Cref{eq:mcbasics:Lmin}) per calculated frequency are on average
required, resulting in an efficient and accurate algorithm for sampling a
thermal radiation field.

\section{Monte Carlo quanta}
\label{Sec:Quanta}

Unlike traditional approaches to \gls{RT} problems, \gls{MCRT} calculations do
not attempt to solve the \gls{RT} equation directly. Instead, a
\textit{simulation} of the \gls{RT} process is performed.  Specifically, the
radiation is discretized so that it may be represented by a large number of
\gls{MC} quanta.  During the initialization of such a simulation, each quantum
is assigned a position, an initial propagation direction, an energy and
frequency, if desired, a polarization vector, and some measure of {\em
importance} or {\em weight}. This last property essentially determines the
contribution of the individual quanta to the final results. After the
discretization and initialization, the quanta are propagated through the
computational domain to simulate the \gls{RT} process. In the following
sections, we highlight two discretization paradigms, namely the \textit{photon
packet} and the \textit{energy packet} scheme. These derive from different
interpretations of what the quanta represent and provide different
prescriptions for the choice and treatment of their weights. We then discuss
packet initialization. The process of propagating packets during the simulation
is described in \Cref{Sec:Propagation}.

\subsection{Discretization into photon packets}

Historically, \gls{MCRT} applications drew inspiration from nature's inherent
discretization of radiation and thus interpreted the fundamental \gls{MC}
quanta as photons. Indeed, in many early \gls{MCRT} studies performed in
astrophysics, such as \citet{Auer1968}, \citet{Avery1968} and \citet{Caroff1972},
the quanta are simply referred to as ``photons''. Although the number of
\gls{MC} photons that are introduced and considered is usually large in a
statistical sense, it is completely insignificant compared to the actual number
of real photons constituting the physical radiation field.  Thus, it
is inherent to this discretization scheme that the \gls{MC} photons, or
\textit{machine photons} as they are sometimes called
\citep[cf.][]{House1968}, actually represent a large number of physical
photons instead of individual ones. As a consequence, the \gls{MC} quanta are
typically referred to as \textit{photon packets} or simply \textit{packets}.

From this discretization perspective, packet weights can be interpreted as
encoding that the individual \gls{MC} packets represent many physical photons.
However, the weights are practically never assigned uniformly or held constant
during the simulation in \gls{MCRT} schemes that rely on the photon packet
discretization approach. These manipulations of packet weights lead to an often
dramatic reduction in variance (i.e.\ increase of statistics and reduction of
\gls{MC} noise) and belong to the more generic class of \textit{biasing}
techniques (see \Cref{Sec:biasing}). Considering \gls{MCRT}
applications in astrophysics, the majority relies on the photon packet
discretization scheme with non-uniform and variable packet weights. Prominent
examples certainly include the many \gls{MCRT} simulations performed in dust
\gls{RT} problems \citep[see, e.g., reviews by][]{Whitney2011, Steinacker2013}.

\subsection{Energy packets and indivisibility of packets}
\label{Sec:epackets}

The energy packet discretization approach has been mainly developed and shaped
by L.~Lucy. The basic interpretation was already given by \citet{Abbott1985},
but it was only after extending the approach and applying it to \gls{RE}
calculations \citep{Lucy1999}, that its full potential and benefits
were explored. The scheme was further generalized to include the possibility of
non-resonant interactions and of realising statistical equilibrium \citep[][see
\Cref{Sec:macroatom} for further details]{Lucy1999a, Lucy2002,
Lucy2003}.

Compared to the photon packet scheme introduced above, the energy packet
approach rests on a different interpretation of what \gls{MC} quanta
fundamentally represent: packets are now primarily thought of as parcels of
radiant energy and the packet energy also acts as its weight. Again, these
parcels of radiant energy represent many physical photons. At this point, the
difference between the photon and energy packet schemes seems very subtle,
almost semantic. However, the distinctiveness of this discretization scheme
becomes apparent once the treatment of packet weights is included into the
consideration.

The primary attraction of viewing the quanta as packets of radiation energy,
rather than bundles of photons, is the ease (and accuracy) which with energy
flows can be tracked during a simulation. For example, in \gls{RE} problems,
the combination of an energy packet discretization and an \textit{indivisible
packet} algorithm allow strict energy conservation to be imposed
\citep{Lucy1999a}.  While all other packet properties, in particular its
frequency, can change during the simulation, the packet energy, i.e.\ its
weight, is strictly held constant after the initial assignment (i.e.  the
  packets are \textit{indivisible}, and also \textit{indestructible}, excepting
  that they can exit through the boundaries of the computational domain). The
  \textit{indivisibility} property can readily be applied to all interactions,
  even those  that on first sight seem to require the splitting of packets or
  adjustment of weights. Instead of splitting, such events are handled by
  probabilistically sampling the different outcome channels (see the
    \textit{downbranching} scheme by \citealt{Lucy1999a} or the \textit{macro
    atom} approach by \citealt{Lucy2002, Lucy2003} which will both be described
    in detail in \Cref{Sec:rad-eqm}). In this process, a change in frequency
    assigned to the packets may occur, but the \gls{CMF} energy will always
    stay constant. A noteworthy property of indivisible energy packet schemes
    is that a \gls{MC} packet may represent a varying number of physical
    photons during its lifetime: the scheme does not enforce conservation of
    photon number (and nor should it: many physical radiation--matter processes
      e.g. recombination cascades or fluorescence do not conserve the number of
    photons).

Relying on this \textit{indivisible energy packet} formalism offers a number of
advantages as pointed out by \citet{Abbott1985} and \citet{Lucy1999}. Most importantly, it
enforces strict local energy conservation in \gls{RE} applications by
construction. However, we note that this energy conserving property does not
restrict the scheme to \gls{RE} problems: well posed sources and sinks of
radiative energy can be readily incorporated while maintaining strict energy
conservation (see \Cref{Sec:Dynamics}). In addition, the packet indivisibility
naturally controls the number of quanta tracked in an \gls{MCRT}
calculation and avoids the need to incorporate an elimination scheme for quanta
with small weights which may otherwise accumulate and slow-down the
calculation. The indivisible energy packet scheme has been widely used in
\gls{MCRT} calculations of \gls{RT} in mass outflows \citep[e.g.][]{Abbott1985,
Vink1999, Long2002, Sim2004, Sim2005, Carciofi2006, Carciofi2008, Noebauer2010}
and in \gls{SN} ejecta \citep[e.g.][]{Lucy2005, Kasen2006, Sim2007, Kromer2009,
Noebauer2012, Kerzendorf2014}.

We note that many of the advantages of \textit{indivisible energy packet}
schemes can still be retained when strict indivisibility is relaxed. In
particular, splitting of energy packets can be introduced in attempts to
improve statistics \cite[e.g.][]{Harries2015, Ergon2018} where strict energy
conservation is retained (i.e.  the algorithm is free to split an energy packet
at any point, provided that the sum of the energies of the newly created
packets matches that of the original unsplit packet). Similarly, there is no
requirement of the scheme that all packets have the same energy as each other:
the only rule is that the combined packet energies correctly sum to the total
energy / energy flow of the process under consideration.

\paragraph{Example: Packet scheme applied to Compton scattering}

To illustrate the manner in which physical processes are described in the
different packet approaches, we use the example of Compton scattering,
following \cite{Lucy2005}. Specifically, we consider Compton scattering of an
ensemble of high-energy photons by a population of low-temperature free
electrons (i.e.\ near-stationary in the \gls{CMF}). Each single Compton
scattering process can be roughly described (in the \gls{CMF}) as

\begin{equation}
e^-_{\mathrm{i}} + \gamma_{\mathrm{i}} \rightarrow e^-_{\mathrm{f}} + \gamma_{\mathrm{f}} \; ,
\end{equation}
where the electron initial state ($e^-_{\mathrm{i}}$) has close to zero kinetic energy but
the final state ($e^-_{\mathrm{f}}$) has non-zero kinetic energy (associated with the
recoil). Conversely, the post-scattering photon ($\gamma_{\mathrm{f}}$) will have less
energy ($E^{\gamma}_{\mathrm{f}} < E^{\gamma}_{\mathrm{i}}$) and lower frequency ($\nu^{\gamma}_{\mathrm{f}} <
\nu^{\gamma}_{\mathrm{i}}$) than the initial photon state ($\gamma_{\mathrm{i}}$). 

In a photon packet scheme, the manner in which this process can be simulated is
obvious: whenever one of the \gls{MC} photon packets undergoes such a Compton
scattering event, the number of photons it represents remains fixed but the
frequency of the photons represented by the packet is reduced (accordingly, the
packet then represents less energy).  

For an indivisibly energy packet scheme, the treatment is more subtle
\citep{Lucy2005}. Here we consider how energy flows through the problem: from
the initial energy of the incoming photon population ($\gamma_{\mathrm{i}}$) to the
combination of final photon population ($\gamma_{\mathrm{f}}$) and final electron kinetic
energy ($e^-_{\mathrm{f}}$).  In particular, a fraction $F_\gamma = E^\gamma_{\mathrm{f}} /
E^\gamma_{\mathrm{i}}$ of the incident photon energy is passed to the outgoing photon
while $F_e = E^e_{\mathrm{f}} / E^\gamma_{\mathrm{i}} = 1 - F_\gamma$ goes to the electron.
Thus, adopting the indivisible energy packet principle, an initial \gls{MC}
($\gamma_{\mathrm{i}}$) packet is converted to an outgoing $\gamma_{\mathrm{f}}$ packet with
probability $F_\gamma$ or to a packet representing the electron kinetic energy
with probability $F_e$. In either case, the energy represented by the packet
remains fixed (i.e. strict energy conservation), but the nature of the energy
has changed: in the first case the energy is still being carried by photons,
but now of lower photon frequency (in accordance with the $\gamma_{\mathrm{f}}$ state); in
the second case, the energy has been passed to the electron kinetic pool from
where its role in powering further emission of the material can be followed
using e.g. the $k$-packet formalism of \citet[][see also
\Cref{Sec:rad-eqm}]{Lucy2002}. 

This example primarily serves to illustrate the subtle difference between
photon-packet and energy-packet schemes but it is natural to wonder which
scheme is better. In general, there is no absolute statement to be made: both
approaches are valid and which is better suited will depend on the problem in
question. However, the relative merits are clear and can be stated fairly
simply for our example: the photon packet scheme will rigorously conserve
photon number (as does the physical Compton scattering process) and is well
suited if the aim of the simulation is to calculate the Comptonized photon
spectrum \citep[e.g.][]{Pozdnyakov1983, Laurent1999}, potentially following
many scattering events. On the other hand, multiple scattering in the photon
packet approach may lead to a proliferation of low-frequency photon packets
that carry very little energy, but still require the same computational effort
per scattering to simulate. This may not be ideal for e.g. applications in
which the primary interest in high-energy Compton scattering lies in its role
as a heating process (such as the modelling of \gls{SN} ejecta powered by
radioactive decay, as discussed by \citealt{Lucy2005}).  For such a
problem, the indivisible energy
packet scheme provides a simple means to determine the rate
at which energy flows into the electron pool with the computational effort
being invested in proportion to the energy carried by the photons,
rather than to the photon number.

\subsection{Initialisation of packets}
\label{sec:packet_init}

Closely related to the fundamental discretization of the radiation field into
discrete \gls{MC} packets is the initialization process. Here, the initial
packet properties are assigned by drawing from the spatial, directional and
spectral distribution of the radiation field by relying on the sampling
techniques presented in \Cref{Sec:Sampling}. The instantaneous values of these
properties\footnote{For the moment, we neglect polarization.}, i.e.\ the
position, direction, frequency\footnote{Throughout this review we generally
assume monochromatic packets for simplicity. Some of the techniques presented
here can also be generalized to polychromatic packets \citep[see, e.g.][for
more information on polychromatism]{Steinacker2013}.}, energy and
weight\footnote{We note that packet energies and weights are somewhat
interchangeable concepts. Thus, we will make use of both terminologies in this
review.}, fully describe the packet state during the entire \gls{MC}
simulation. If the effect of polarization is included in \gls{MCRT}
simulations, packets are additionally assigned appropriate values for the
Stokes vectors \citep[see, e.g.][]{Kasen2006, Whitney2011, Bulla2015}.

In the following, we briefly sketch the initialization process within the
indivisible energy packet scheme.  Note that the corresponding procedure is not
fundamentally different within the photon packet scheme. In the following
presentation, we distinguish between the initial assignment of properties for
packets that represent the radiation field in the domain at the onset of the
\gls{MC} simulation and for packets that represent the inflow of radiation into
the domain through the boundaries. 

We substantiate these concepts by highlighting the initialization of $N$
packets representing an initial thermal radiation field, at
temperature $T$, which is assumed to be
uniform within a cuboid volume $\Delta V$.
Despite its simplicity, this situation is often encountered in \gls{MCRT}
calculations. In the energy packet scheme, a commonly used practise involves
assigning a uniform packet energy. Thus, if $N$ packets are initialized, each
packet carries an energy of 
\begin{equation}
  \varepsilon = \frac{\Delta V a_{\mathrm{R}}T^4}{N} \; ,
  \label{eq:packet_energy}
\end{equation}
where $a_{\mathrm{R}}$ is the radiation constant.
The thermal radiation field is isotropic and as a consequence the initial
propagation direction of a packet can be assigned using previously presented
sampling rules, namely \Cref{eq:mcbasics:phi_sample} and
\Cref{eq:mcbasics:mu_sample}. Since we have assumed that the initial radiation
field is uniform over the volume $\Delta V$, locating the packets is trivially
done by 
\begin{align}
  x &= x_{0} + \xi_1 (x_{1} - x_0),\nonumber\\
  y &= y_{0} + \xi_2 (y_{1} - y_0),\nonumber\\
  z &= z_{0} + \xi_3 (z_{1} - z_0),\label{eq:uniform_initial_packet_location}
\end{align}
where $(x_{0}, y_{0}, z_{0})$ and $(x_{1}, y_{1}, z_{1})$ are
opposite corners of the cuboid volume.
Finally, the packet frequency is obtained from sampling the Planck function,
for example by relying on the technique given by \Cref{eq:mcbasics:Lmin} and
\Cref{eq:mcbasics:final_nondim_freq}. The initialization process has been
implicitly performed in the \gls{CMF}. Their \gls{LF} properties are obtained
by applying the appropriate frame transformations (see \Cref{sec:RT}). This
procedure will be revisited when discussing \gls{MCRT} in expanding media (see
\Cref{Sec:Flows}).

In applications for which radiation is streaming into the domain, \gls{MC}
packets are continuously created to represent the inflow of energy. A
frequently encountered example is that of a photosphere being located at the
inner boundary of a spherical domain which emits as a black body with
temperature $T_{\mathrm{phot}}$ (used for example in the \gls{MCRT} approaches
of \citealt{Mazzali1993} and \citealt{Kerzendorf2014} for studying \gls{SN}
ejecta). In this, case
\begin{equation}
  N = \frac{4\pi R_{\mathrm{phot}}^2 \sigma_{\mathrm{R}} T_{\mathrm{phot}}^4 \Delta t}{\varepsilon}
  \label{eq:photosphere_init_N}
\end{equation}
packets with energy $\varepsilon$ are initialized during the time interval
$\Delta t$ (here, $\sigma_{\mathrm{R}}$ is the Stefan-Boltzmann constant). Since these packets are launched from the photosphere, their
initial position is simply
\begin{equation}
  r = R_{\mathrm{phot}}.
  \label{eq:photosphere_init_r}
\end{equation}
If limb-darkening can be neglected, packets leave the photosphere along
directions drawn by\footnote{The difference between this case and the isotropic
initialization of the initial radiation field is that the procedure is based on
the flux in the former and on the energy density in the latter case.}
\begin{equation}
  \mu = \sqrt{\xi}.
  \label{eq;photosphere_init_mu}
\end{equation}
Finally, the initial packet frequency is again drawn from the Planck function.

We conclude this description by noting that in \gls{MCRT} applications packets
may also be initialized to represent the continuous radiative cooling of the
ambient material or to represent the emission from other sources within the
domain (e.g.\ in radiative non-equilibrium applications). The basic
initialization principles highlighted remain the same and can be simply
transferred to these applications.

\section{Propagation of quanta}
\label{Sec:Propagation}

The discretization paradigms and the initialization principles outlined above
(see \Cref{Sec:Quanta}) provide rules for the creation and launching of
\gls{MC} packets. The bulk of the computational effort involved in a \gls{MCRT}
calculation is spent in tracing the movement of these packets through the
ambient material to simulate the \gls{RT} process. During the propagation,
their trajectories are interrupted as the packets experience
radiation--matter interactions.  Depending on the nature of these interactions,
the packet properties may change or the propagation may even be terminated. The
\gls{MC} packets are thus followed until certain termination conditions are
met, e.g.\ the packet leaves the computational domain or has been active for a
pre-defined time (this aspect will be treated in detail in \Cref{Sec:Time}).
The propagation procedure of a \gls{MCRT} simulation is complete when all
packets representing the initial radiation field and the effects of sources of
radiative energy (e.g.\ inflows through boundaries, internal sources in
radiative non-equilibrium applications, etc.) have been processed this way.  In
the following, we first outline the fundamental propagation principles
(\Cref{Sec:Propagation_Basic}) and then
detail how basic absorption and scattering interactions are handled
(\Cref{sec:absorb1} -- \Cref{sec:propexample})
before
finally turning to the inclusion of time dependence (\Cref{Sec:Time}).

\subsection{Basic propagation principle}
\label{Sec:Propagation_Basic}

In the absence of general relativistic effects (which we neglect in this
review), a \gls{MC} packet propagates on a straight path in its propagation
direction $\textbf{n}$. In the simplest version of a \gls{MC} packet
propagation algorithm, the packet properties do not change along these
straight-flight elements of its path: interactions with the medium are treated
as discrete interaction events, and the primary aim of the \gls{MC} algorithm
is to determine where those interaction events occur.

Finding the interaction points depends on the opacity in the medium. Along its
trajectory, measured by $l$, a packet ($\mathrm{p}$), continuously accumulates optical depth
\begin{equation}
  \tau_{\mathrm{p}}(l) = \int_0^l \mathrm{d}l' \chi(l').
  \label{eq:packet_accumulated_optical_depth}
\end{equation}
Here the specific functional form of the opacity depends on the physical
interaction processes that are taken into consideration and can, in principle,
be very complicated. When the accumulated optical depth surpasses a threshold
value, $\tau$, the packet will undergo an interaction at the corresponding
location $l$. As anticipated in \Cref{Sec:Sampling}, this threshold value is
determined for each packet individually and probabilistically. In particular,
at the beginning of each packet trajectory segment, the packet is
assigned a randomly sampled
optical depth distance to the next interaction by 
\begin{equation}
  \tau = -\ln \xi.
  \label{eq:beer_lambert_sampling}
\end{equation}

Whenever a packet experiences an interaction, its properties may change
depending on the nature of the interaction process. In general, these
interactions can be broadly divided into scatterings and absorptions. In the
former, the packet is deflected and continues its propagation in a different
direction, possibly with a different energy and/or frequency. Depending on the
nature of the scattering process, the assignment of emergent packet properties
may become quite complex (e.g.\ in dust scattering applications).
Alternatively, if absorption occurs, the propagation is terminated and the
packet removed from the active pool\footnote{Note that in the indivisible
  energy packet scheme proposed by \citet{Lucy1999} for \gls{RE} applications,
packets are immediately re-emitted.}. 

For locating packet interactions using \Cref{eq:beer_lambert_sampling}, we
highlight an important property of the exponential distribution, namely its
\textit{infinite divisibility} \citep[see for example][]{Bose2002}. Probability
distributions with this property can be replaced by the ``distribution of the
sum of an arbitrary number of independent and identically distributed random
variables''\footnote{
\url{https://en.wikipedia.org/wiki/Infinite_divisibility_(probability)}}.  For
the purpose of \gls{MCRT}, this implies that, as long as the packet has not
interacted, one is at liberty to reset the comparison between accumulated and
interaction optical depth arbitrarily often. I.e.\ one can opt to redraw the
optical depth distance to the next interaction with
\Cref{eq:beer_lambert_sampling} and reset the tracking of accumulated optical
depth, \Cref{eq:packet_accumulated_optical_depth}, at the current packet
location. This property is often used when performing \gls{MCRT} simulations on
numerical grids (see \Cref{sec:matprop_grids}).

\begin{figure}[htbp]
  \centering
  \includegraphics[width=0.9\textwidth]{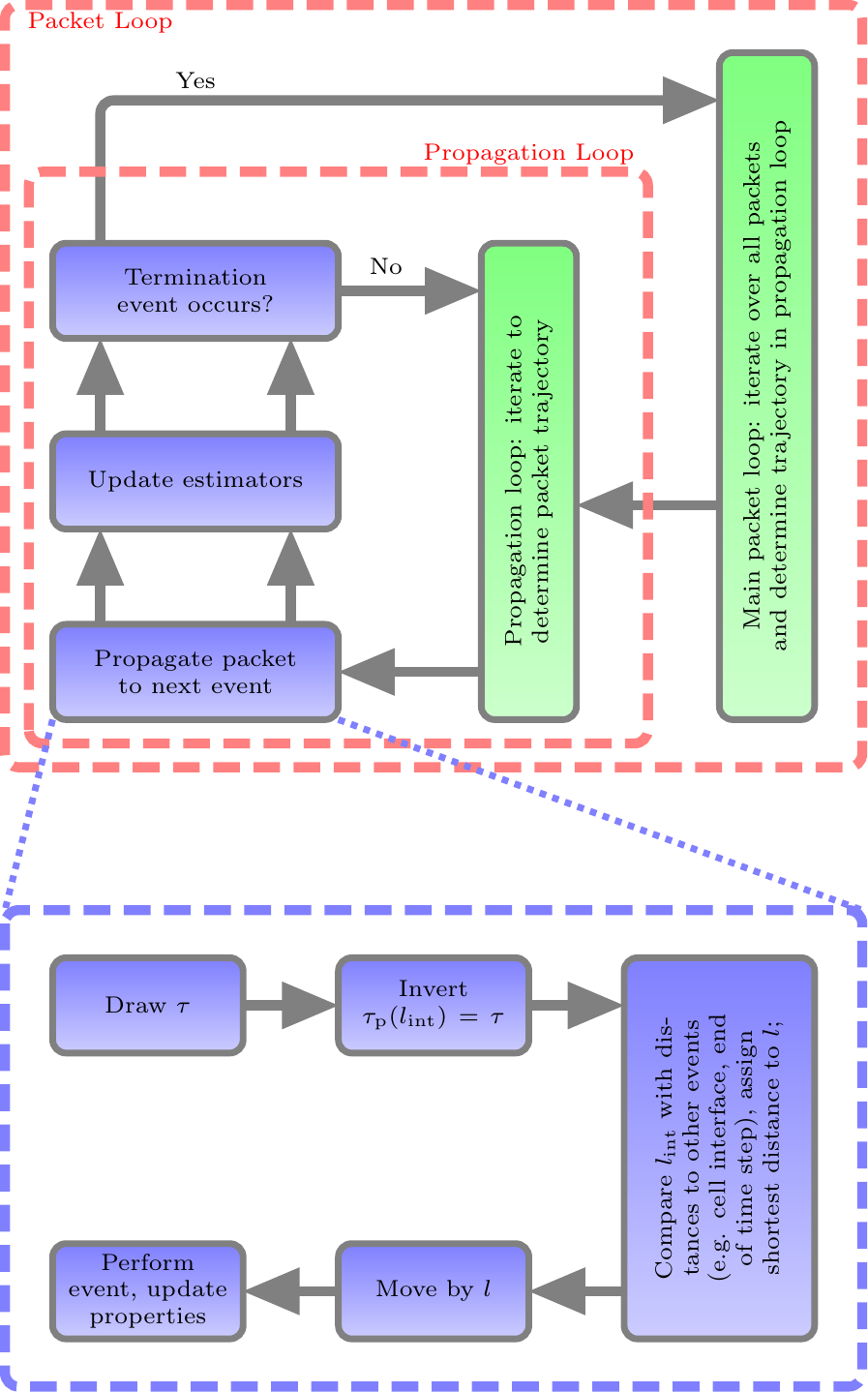}
  \caption{Illustration of the packet propagation process. At its core, the
    main processing loop works through the entire packet population,
    propagating each packet individually. In this loop, each packet is moved
    through the domain until a termination event occurs. Depending on the
    specific \gls{MCRT} application, this may for example be an absorption
    interaction, escaping through one of the domain boundaries or reaching the end
    of a pre-defined time interval. The instruction ``update estimators''
    refers to all activities related to recording packet properties that are
    used to reconstruct physical information about the radiation field from the
    packet ensemble. This procedure will be discussed in more detail in
    \Cref{Sec:Estimators}.}
  \label{fig:propagation_illustration}
\end{figure}

Following the generic propagation principles outlined above, each packet is
moved through the domain until a termination condition is reached. Depending on
the problem, this may be an absorption interaction, or the packet intercepting
a transparent domain surface through which it escapes to infinity, or simply
that a pre-defined amount of physical time has elapsed. The propagation
process, which is visually summarized in \Cref{fig:propagation_illustration},
is complete after all \gls{MC} packets have been processed in this manner.
During the propagation process, various events may be recorded or the change of
certain packet properties tracked.  These may then be used to reconstruct
physical properties of the radiation field (see \Cref{Sec:Estimators}). 

\subsection{Absorption as continuous weight degradation}
\label{sec:absorb1}

The general scheme outlined in the previous section can be applied to find
discrete interaction events for any physical contribution to the opacity. It is
particularly important (and widely used) for addressing scattering problems:
an accurate treatment of scattering is most easily formulated as an ensemble of
discrete interactions where the properties of the packets are changed at the
interaction point in accordance with the physics of the scattering process. The
scheme is also widely applied to true absorption processes, and this is
particularly attractive in applications that aim to exploit the
energy-conserving qualities of radiative equilibrium problems (see 
\Cref{Sec:macroatom}).

However, in some applications (see, e.g., the treatment of continuum absorption used by \citealt{Long2002} and references in
\citealt{Steinacker2013}) an alternative treatment of absorption is used.
Specifically, rather than treating absorption via discrete interaction events
it is simulated by continuous reduction of the energy carried by a \gls{MC}
packet as it propagates along its flight path. Specifically, whenever a packet
travels along a trajectory of length $l$, the energy it carries (weight, $w$)
is reduced according to
\begin{equation}
w(l) = w(l=0) e^{- \tau_{\mathrm{p,a}}},
\end{equation}
where 
\begin{equation}
  \tau_{\mathrm{p,a}}(l) = \int_0^l \mathrm{d}l' \chi_{\mathrm{a}}(l')
\end{equation}
is the optical depth associated with the {\it absorption} component of the
opacity ($\chi_{\mathrm{a}}$). Conceptually, the energy removed from the \gls{MC} packet
in this way is being transferred out of the radiation field by whichever
physical process(es) contribute to $\chi_{\mathrm{a}}$. For example, this might represent energy invested in heating or ionising the ambient medium.

There are several advantages for this approach to absorption compared to the
discrete scheme outlined above. First, it can reduce the \gls{MC} noise since
it replaces the stochastic identification of specific interaction points with a
smooth degradation of packet weight.  This seems especially attractive if
considering small contributions to the opacity (e.g.\ background continua):
using the discrete event algorithm to simulate such interactions would be
noisy since the number of events associated with a low opacity will be
small.\footnote{However, as we shall discuss later (see
Section~\ref{Sec:Estimators}), various techniques are available to alleviate the
issue of \gls{MC} noise in determining rates for rare physical processes.}
Second, it can greatly simplify the \gls{MC} algorithm for applications in which
pure absorption is dominant: in such cases, pure weight attenuation of packets
on straight trajectories may be sufficient to solve the problem.\footnote{In
such cases, the issue of how to handle the computational cost of packets with
weights attenuated to the point where they may become negligible can be handled
using strategies such as Russian Roulette (see
Sec~\ref{Sec:biasing}).}

However, there are some limitations associated with continuous weight
degradation. In particular, if the interaction
processes is associated (at the microphysical level) with some radiative
re-emission process, such as effective scattering/fluorescence in atomic or
molecular line transitions, this approach loses a direct connection between
the absorption and re-emission process. If important, the re-emission must be
simulated by injecting new \gls{MC} packets to represent it (see
Sec~\ref{Sec:emiss-simple}). For this reason, \gls{MCRT} applications that
depend on simulating e.g.\ atomic line interactions have found it more practical
to use a discrete interaction approach for this process \citep[similar to
e.g.][]{Abbott1985}. We note, however, that hybrid schemes have been
successfully employed where the continuous attenuation approach is used for
smooth continuum absorption opacity while a discrete interaction algorithm is
applied for atomic line absorption and electron scattering \citep[e.g.][]{Long2002}. Throughout most of this review we will focus on methods that adopt the
discrete interaction approach for treating both scattering and absorption but
note that many of the principles discussed in later chapters can be applied to
simulations that employ a weight-degradation approach to absorption, or a
combination of both.

\subsection{Material properties and numerical discretization}
\label{sec:matprop_grids}

To perform the packet propagation process, the local material state, such as
velocity, density and temperature, has to be accessible. It sets the local
opacity and thus determines the rate at which optical depth is accumulated
along the propagation path (cf.\ \Cref{eq:packet_accumulated_optical_depth}).
Moreover, the material state dictates the re-emission characteristics in
scattering interactions. Ideally, the material state is directly accessible in
closed analytic form such that the optical depth integration can be performed
analytically. In practise, however, the complex local dependence of the
material properties calls for a numerical integration. In addition, the
continuous material state is often not available but instead only a discrete
representation. This could, for example, be the snapshot of a hydrodynamical
simulation. Consequently, the packet propagation is typically realised by
introducing a computational grid, dividing the domain into cells on which the
matter state is discretely represented. Often, the material properties are
approximated as constant throughout the grid cells (although interpolation can
be used).

The packet propagation process can then proceed on the numerical grid along the
basic principles outlined above. If the material state is assumed to be
constant throughout the individual grid cells, determining the rate of
accumulation of $\tau$ is  generally simple\footnote{Additional complications
can arise from the frequency dependence of the opacity in fast flows (see
\Cref{Sec:Flows}).}. However, one does need to track when packets cross over
grid cell boundaries: at such points, quantities that depend on the material
state, such as opacities, have to be recalculated. Some codes, for algorithmic
convenience, also exploit the infinite indivisibility property of the
exponential distribution and re-draw the random optical depth from the usual
sampling law, \Cref{eq:beer_lambert_sampling}, when cell crossing occurs. 

\subsection{Absorption and scattering}
\label{Sec:Propagation_abs_scat}

Having outlined the principles of how packets can be propagated through the
simulation domain, we now discuss how interactions are handled.  In any real
application, the details of how interaction events are to be processed will
depend on the particulars of the radiation--matter physics being simulated. To
illustrate the general principles here, we adopt a number of simplifications,
namely that the medium is static and that all material functions are frequency
independent and isotropic. We also restrict the presentation to basic
absorption\footnote{In this illustration, we deviate from the indivisible
  energy packet scheme introduced by \citet{Lucy1999} and do not immediately
re-emit absorbed packets.} and coherent and isotropic scattering interactions.
Lifting these simplifications, in particular, including more complex
interaction processes, naturally complicates the individual steps of the
propagation process but the basic structure of the procedure remains the same.

We start by considering the packet propagation process in the presence of only
true absorptions, described by the opacity $\chi_{\mathrm{a}}$. As detailed
above, a packet interacts after accumulating the optical depth $\tau$, assigned
according to \Cref{eq:beer_lambert_sampling}. This decision in optical depth
space has to be translated into a physical location within the computational
domain by inverting \Cref{eq:packet_accumulated_optical_depth}. This is a
critical part of the \gls{MC} algorithm that can be very challenging when
the material functions are complicated functions of location, frequency and
direction. For our example, however, the translation is trivial since optical
depth and physical distance only differ by the opacity coefficient,
which is constant within a grid cell. Thus, a packet interacts after covering
the distance
\begin{equation}
  l = \frac{\tau}{\chi_{\mathrm{a}}}.
  \label{eq:absorption_location}
\end{equation}
For pure absorption, the propagation would end at this point with the packet
being discarded and the next packet of the active pool being treated. Note,
however, that many implementations, including those that enforce \gls{RE}
based on the indivisible energy packet formalism of
\citet{Lucy1999} do not terminate the packet at the absorption point: instead
absorption is immediately followed be re-emission, as will be elaborated in
\Cref{Sec:rad-eqm}.

Unlike in deterministic \gls{RT} approaches, where scattering generally
introduces complex non-local couplings \citep[see, e.g.][sec.\
11.1]{Hubeny2014}, including scattering does not pose any conceptual
difficulties for \gls{MCRT} approaches. In this case, the total opacity is a
sum of absorption and scattering terms:

\begin{equation}
\chi_{\rm tot} = \chi_{\mathrm{a}} + \chi_{\mathrm{s}} \; \; .
\end{equation}
The exercise of locating packet interaction points is as trivial as before: it
follows from \Cref{eq:absorption_location} adopting the total opacity
\begin{equation}
  l = \frac{\tau}{\chi_{\mathrm{\rm tot}}}.
  \label{eq:interaction_location}
\end{equation}
Once the interaction point is found, an additional random number experiment has
to be performed to decide the nature of the interaction. In particular, the
packet scatters if
\begin{equation}
  \xi \le \frac{\chi_{\mathrm{s}}}{\chi_{\mathrm{s}} +
    \chi_{\mathrm{a}}} \; .
  \label{eq:scattering_albedo}
\end{equation}
Otherwise it is absorbed and is treated as detailed above.  We
note that this procedure can easily be extended to situations in which more
than two outcomes are possible (e.g. if multiple mechanisms with different
scattering properties were relevant) by subdividing the interval $[0, 1[$ into
bins with sizes proportional to the relative strengths of the different
processes and sampling from this discrete set. If the packet scatters it
continues its propagation along a new direction with any relevant properties
(e.g. photon frequency) set by rules that describe the scattering process. For
example, in the simple case of isotropic coherent scattering, a new propagation
direction is assigned by the sampling rule 
\begin{align}
  \mu &= 2 \xi_1 - 1,\nonumber\\ 
  \phi &= 2\pi \xi_2,
  \label{eq:isotropic_scattering_dirs}
\end{align}
while the photon frequency would remain unchanged.  Of course, realistic
scattering processes may be neither isotropic nor coherent: but this merely
requires that the sampling procedures be altered to reflect the true
directional dependence and that the frequency be changed in the interaction. 

After the scattering event, the packet continues its propagation along the new
direction. A new distance to the next interaction event is drawn from
\Cref{eq:beer_lambert_sampling} and the tracking of the accumulated optical
depth of \Cref{eq:packet_accumulated_optical_depth} is re-initialized. In this
manner, the flow of packets can be followed including multiple scatterings in
arbitrary media.  We emphasise that the ease with which scattering interactions
can be treated is a major benefit of \gls{MCRT} approaches. 

\subsection{Propagation example}
\label{sec:propexample}

\begin{figure}[htb]
  \centering
  \includegraphics[width=\textwidth]{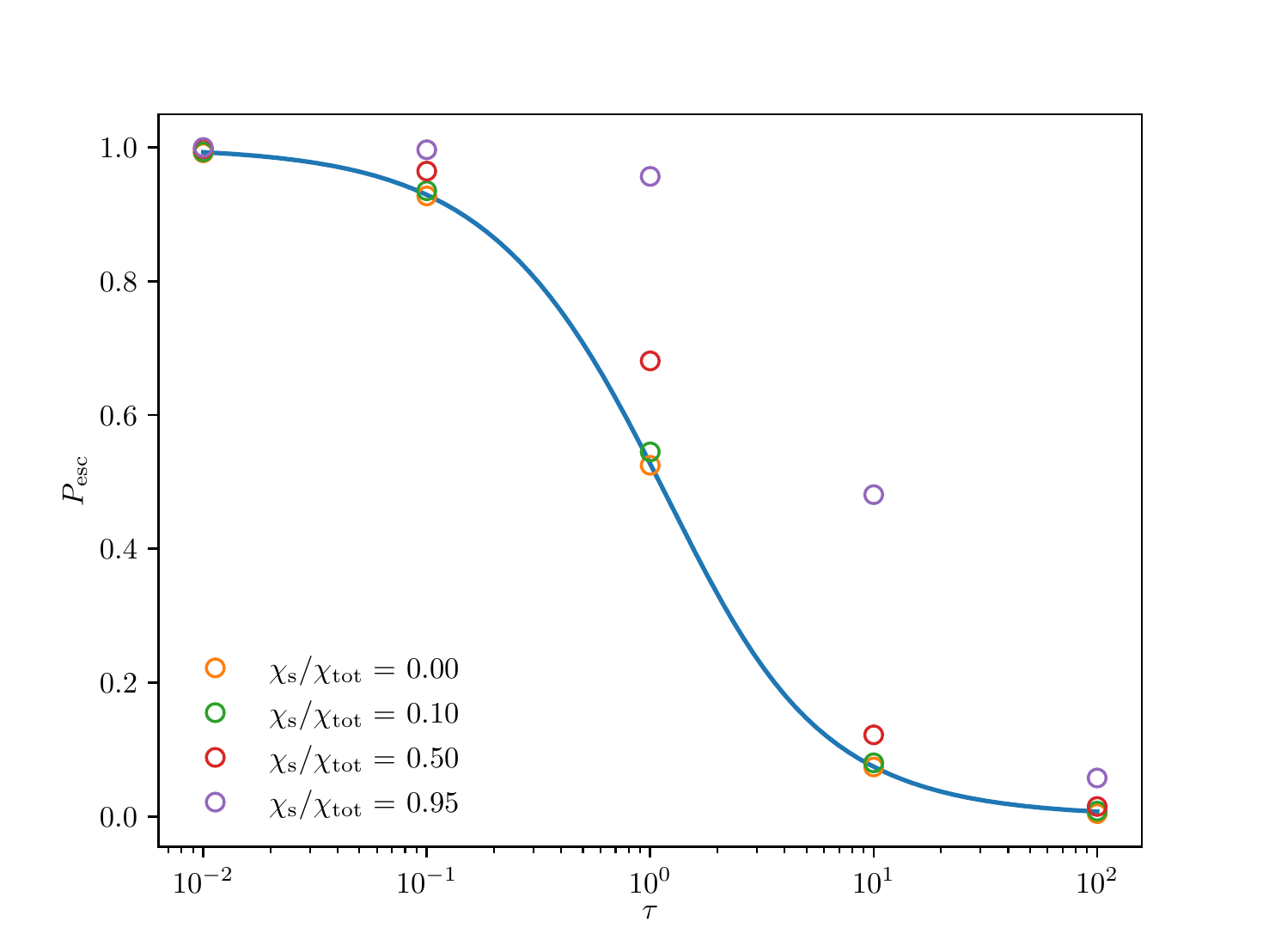}
  \caption{Results of a number of \gls{MCRT} simulations for the homogeneous
    sphere test problem. The escape probability is shown as a function of total
    optical depth of the sphere, cf.\ \Cref{eq:hom_sphere_tot_tau}, for
    different scattering to absorption strengths. For comparison, the analytic
  solution to the problem with pure absorption is shown as a solid line.}
  \label{fig:mcrt_hom_sphere_escape_cmp}
\end{figure}

\begin{figure}[htb]
  \centering
  \includegraphics[width=\textwidth]{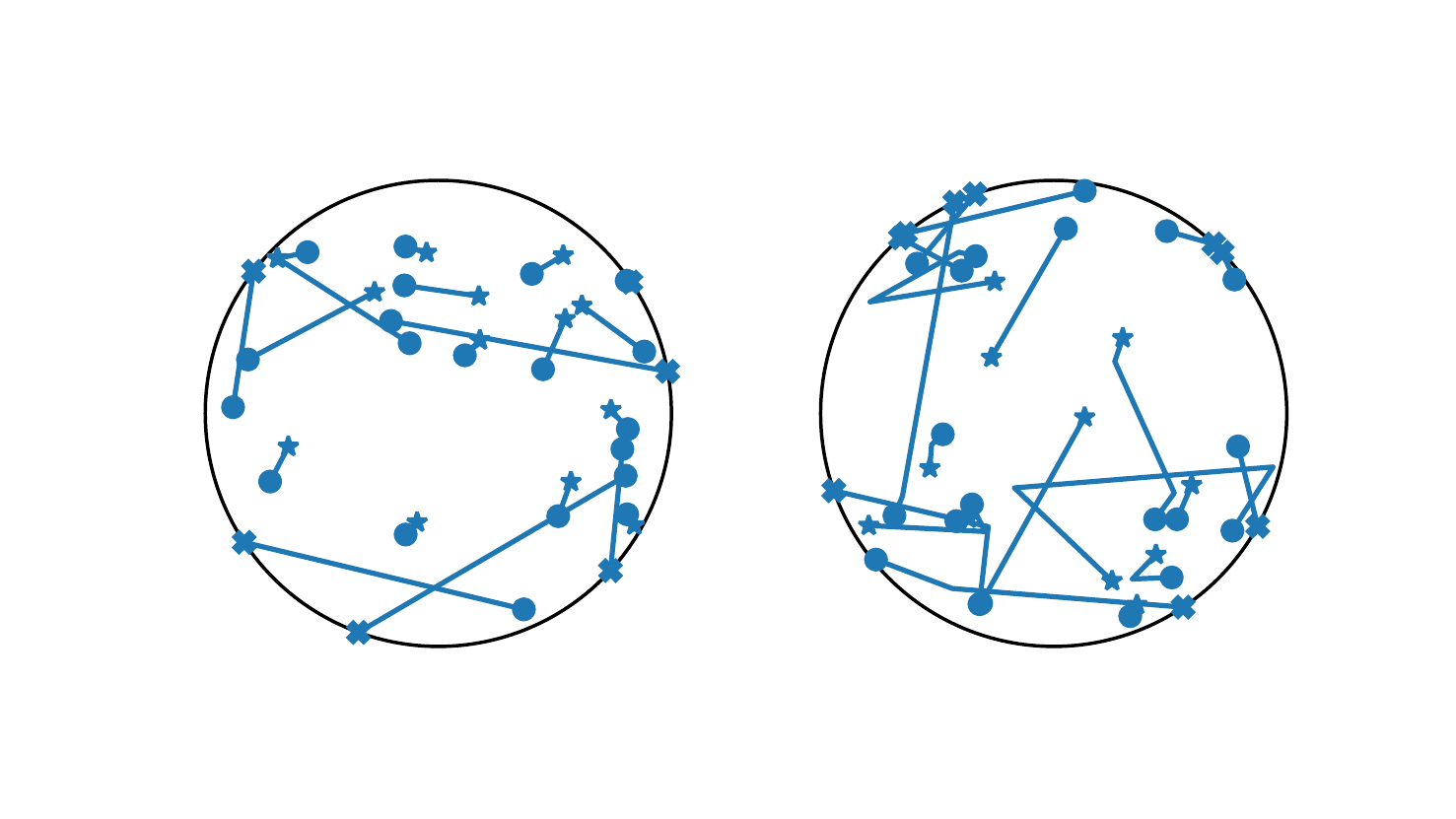}
  \caption{Illustration of the packet propagation process in the \gls{MCRT}
    simulations for the homogeneous sphere problem shown in
    \Cref{fig:mcrt_hom_sphere_escape_cmp}. A small subset of packet trajectories is
    visualized for the case $\chi_{\mathrm{s}} = 0$ (left panel) and
    $\chi_{\mathrm{s}} = \chi_{\mathrm{a}}$ (right panel). The total optical
    depth of the sphere is in both cases $\tau = 2$. Starting points for the
  trajectories are marked by small filled circles. The fate of the packet is
either marked by a star symbol (absorption) or by a cross (escape).}
  \label{fig:escape_prob_illustration}
\end{figure}

Using the principles described so far, simple \gls{MCRT} simulations can be
built using only a few lines of code. As one example, consider calculating the
escape probability of photons from a homogeneous sphere (uniform density and
  uniform emissivity) in which radiation may be absorbed or scattered (for
  simplicity we assume opacities and emissivities that are independent of
frequency and direction).  For the analytic solution to this test problem, see
\Cref{sec:test_hom_sphere}. A simple Python implementation of the \gls{MCRT}
simulation for this problem is part of our code repository distributed on
GitHub (see \Cref{sec:tool_collection}). In the outline of the \gls{MCRT}
procedure below, we specifically refer to the respective parts of the Python
code by providing the corresponding line numbers. The elements of this
calculation are:

\begin{itemize}
\item The \gls{MCRT} simulation begins by initialising $N$ packets and
  uniformly distributing them throughout the sphere. As this is a
  one-dimensional problem and since we are only interested in the escape
  probability, the packet state is essentially described by $r$ and $\mu$. The
  initial location in the uniform sphere is assigned by 
  \begin{equation}
    r = R \xi^{\frac{1}{3}}, 
    \label{eq:hom_sphere_init_r}
  \end{equation}
  where $R$ is the outer radius of the sphere (code line 90), and the initial
  direction is chosen isotropically (code line 92)
  \begin{equation}
    \mu = 2 \xi - 1.
    \label{eq:hom_sphere_init_mu}
  \end{equation}
\item After initialisation, the pool of packets is processed with each being
  propagated through the sphere following the principles outlined above. This
  includes drawing the random interaction $\tau$-values (code line 143) and
  calculating distances to boundary crossing (code line 145). Packet
  interaction occurs when the randomly drawn optical depth is reached before
  the boundary of the simulation (code line 149). Whenever a packet
  interacts, \Cref{eq:scattering_albedo} is used to decide whether the packet
  is absorbed (destroyed in this case) or scattered. Following scattering, a
  new direction (code line 172) is drawn with
  \Cref{eq:isotropic_scattering_dirs} and the propagation continues. 
\item Each packet is followed until it either is absorbed or escapes through
  the surface of the sphere, and the entire \gls{MCRT} simulation ends when all
  packets are processed in this manner. Finally, the escape probability is
  calculated by dividing the number of escaped packets by the total number of
  packets which have been initialised. 
\end{itemize}
\Cref{fig:mcrt_hom_sphere_escape_cmp} shows the result of a number of such
\gls{MCRT} simulations with varying total optical thickness of the sphere
\begin{equation}
  \tau = R (\chi_{\mathrm{s}} + \chi_{\mathrm{a}})
  \label{eq:hom_sphere_tot_tau}
\end{equation}
and different scattering to absorption strengths.
In the case of $\chi_{\mathrm{s}} = 0$, the \gls{MCRT} results agree very well
with the analytic predictions from
\Cref{eq:hom_sphere_escape_probability}.  As the scattering opacity increases,
the escape probability grows since the absorption probability is smaller for a
given optical thickness of the sphere. 
\Cref{fig:escape_prob_illustration} further illustrates the packet propagation
process by showing a small number of packet trajectories for two \gls{MCRT}
simulations at $\tau = 2$, with $\chi_{\mathrm{s}} = 0$ and $\chi_{\mathrm{s}}
= \chi_{\mathrm{a}}$.

\subsection{MCRT: time-dependent applications}
\label{Sec:Time}

The presentation so far has been centred on time-independent \gls{MCRT}
applications, i.e.\ on finding a steady-state solution. For many applications,
however, obtaining the detailed time-dependent evolution of the radiation field
is required. Conceptually, only few changes in the \gls{MCRT} techniques
outlined so far have to be made to include time dependence. Fundamentally, each
\gls{MC} quanta is assigned a time stamp, $t$, which tracks the current
simulation time.  During the initialisation process, the internal clock of each
packet is set depending on the physical process responsible for the packet
emission. If the packet represents the initial radiation field, $t$ is simply
set to the starting time of the current phase of the \gls{MCRT} simulation.  If
the packet models the effect of a particular source of radiative energy, the
time is sampled from the temporal emission profile of the source. In the
simplest case of a continuous emitter, $t$ is uniformly drawn from the duration
of the current \gls{MCRT} simulation step. After initialisation, as the packet
propagates, this time stamp is continuously updated. In particular, after
covering the distance $l$, the internal clock of the packet is advanced by 
\begin{equation}
  \Delta t = \frac{l}{c}.
  \label{eq:timedep_time_update}
\end{equation}
In time-dependent problems, a discretization of time is commonly introduced,
similar to the spatial grid covering the computational domain. On the one hand,
this defines time intervals over which packet properties will be tallied to
discretely reconstruct the time evolution of radiation field properties (see
\Cref{Sec:Estimators} for more details). On the other hand, the time
discretization is used to represent changes in the material state, which in
general will vary in time-dependent problems. In analogy to the spatial
discretization procedure, the material state in the different grid cell is
typically assumed to be constant during a time step. Before the start of the
next time step, the material state is then updated. This topic will be
revisited in more detail in \Cref{Sec:Dynamics}, when describing the coupling
of \gls{MCRT} schemes with full hydrodynamic calculations. Also in a later part
of this review, namely in \Cref{Sec:imc_and_ddmc}, we discuss problems arising
from very rapid changes in the material state and how to best address these
within \gls{MCRT} framework.

In addition to tracking the current time for each packet the basic propagation
scheme as outlined in
\Crefrange{Sec:Propagation_Basic}{Sec:Propagation_abs_scat} has to be further
extended to account for the subdivision of the \gls{MCRT} simulation into a
series of time steps.  Whenever the internal clock of a \gls{MC} packet has
progressed to the end of the current simulation time step, $t^{n+1}$, the
packet's propagation is interrupted. The instantaneous state of the packet,
i.e.\ its position, current frequency, energy, propagation direction and any
further properties, is stored and the next packet of the active population is
treated. At the end of the propagation process, all packets stored are
transferred to the next simulation cycle and the packets continue their
propagation at $t = t^{n+1}$ with the saved properties.

\section{Thermal and line emission in MCRT}
\label{Sec:macroatom}

The treatment of absorption and pure scattering processes as outlined above
are
relatively standard and the principles used are very well established.  In
contrast, the manner in which emission is handled in \gls{MCRT} schemes is
relatively varied and much of the sophistication and ongoing developments in
the \gls{MCRT} field relate to the manner in which this is done.

In this section we aim to review some of the approaches to treating emissions
within a computational domain. To be clear, this is distinct from questions of
how \gls{MC} quanta might be injected at some computational boundary: in
\Cref{sec:packet_init} we already reviewed how e.g.\ a population of packets
might be injected to represent a seed blackbody radiation field as might be
appropriate as an initial condition in a time-dependent simulation. Likewise,
we described how packets might be injected at some boundary surface to
represent a radiation source external to the simulation domain, for example a
photospheric boundary condition. Here, instead, our focus is on how emissivity
within the computational domain can be taken into account.

\subsection{Known emissivity}
\label{Sec:emiss-simple}

The most obvious case to handle is any for which the emissivity is externally
known (or can be easily estimated) without prior knowledge of the \gls{RT}
process within the domain. One such example might be a non-equilibrium
plasma that is predominantly heated and ionized by non-radiative processes.

In this case, the emissivity can simply be sampled using standard sampling
techniques (\Cref{Sec:Sampling}) to create a population of packets with
properties that represent the emission process (i.e. photon frequency,
weights/energies, propagation directions etc.).  This population is simply
injected alongside any packets due to external radiation field boundary
conditions, and their subsequent propagation followed in the same
manner.\footnote{In a time-dependent simulation, packets representing ongoing
emissivity can be gradually injected during the course of a numerical time step
(i.e. the time at which they are injected to the simulation is also a property
to be sampled).} 

\subsection{Radiative equilibrium (RE)}
\label{sec:radiative_equilibrium}

For several of the applications to which \gls{MCRT} has been applied, the
emissivity is not known a prior. Indeed, for many astronomical \gls{RT}
problems (e.g.\ stellar/disk atmospheres, winds, \glspl{SN} etc.), \gls{RE} is
a good approximation and the emissivity is effectively determined by the
radiation field itself (i.e.\ near-equilibrium is achieved between absorption
and emission of radiation).  In such cases, the emissivity usually cannot be
anticipated independently of a radiation transport simulation, which poses
a challenge for consistent modelling. In the following sections we
review methods that can be applied to problems with this requirement.

\subsection{Radiative equilibrium in MCRT by iteration}

One approach for \gls{RE} problems is to use an iteration scheme to determine
the conditions of the medium (temperature, ionization state, level populations
etc.) on which the emissivity depends. Here, an iterative sequence of \gls{RT} simulations
would be performed: in each iteration the current best estimate of the conditions in the
medium would be adopted to calculate the emissivity, and the outcome of the
\gls{RT} calculations\footnote{See \Cref{Sec:Estimators} for further details of
  how such information can be optimally extracted from a \gls{MCRT}
simulation.} used to make an improved estimate for those conditions in the
next cycle. This approach can be applied to schemes based on photon packets
and/or energy packets and it has been used for modelling at least some part of
the emissivity (e.g.\ the part associated with radiative cooling by
\citealt{Long2002}) and works well provided that the complexity of the problem is
not too severe.

However, as is well known from the history of modelling stellar atmospheres
\citep[cf.][]{Hubeny2014}, the non-local character of \gls{RT} problems can
lead to significant convergence problems for this type of iteration scheme,
especially when considering regions associated with high optical depth. In
particular, in its pure form, this scheme suffers from the issue that energy
conservation is only achieved asymptotically (i.e.\ once/if a converged
equilibrium solution is found). As a result, during the iteration process, over-
or under-estimated emissivities (due e.g.\ to unconverged temperatures) will
lead to spurious sources and sinks of radiation that might slow or inhibit
convergence.

\subsection{``On-the-fly'' radiative equilibrium in MCRT via indivisible
energy packets}
\label{Sec:rad-eqm}

As described/developed in the works of Lucy
\citep{Abbott1985,Lucy1999,Lucy2005}, it is actually very simple to rigorously
enforce the conservation of energy required by \gls{RE} via an
\textit{indivisible\footnote{One might argue that the key property here is that
the packets are \textit{indestructible} rather than \textit{indivisible}, but
we retain the more usual name for this approach for consistency.} energy
packet} \gls{MCRT} scheme. The principle is straight forward: \gls{RE} implies
that at all points there is (local) balance between the rates of absorption and
emission of energy from/to the radiation field. Since, in an \textit{energy
packet} discretizaion the \gls{MC} quanta already represent (local) bundles of
radiative energy, the condition of \gls{RE} is trivially enforced by insisting
that the \gls{MC} quanta are never destroyed, or otherwise degraded in weight,
in the course of the simulation. Thus, all interactions between \gls{MC}
packets and the medium -- even those representing pure absorption processes --
become effective scatterings controlled by rules devised for the \gls{MCRT}
simulation. The rules for how packets should be altered in these effective
scattering events depend on the physical process(es) being simulated
\citep{Lucy2002,Lucy2003,Lucy2005}: commonly, considerations of statistical
equilibrium and/or \gls{TE} will form the basis for formulating these rules. In
the following sections we will elaborate these principles more generally, but
for concreteness we first discuss one of the simplest specific examples. 

\subsubsection{Example: effective resonant scattering in a two-level
  atom}
\label{Sec:two-level}

\begin{figure}[htb]
  \centering
  \includegraphics[width=0.7\textwidth]{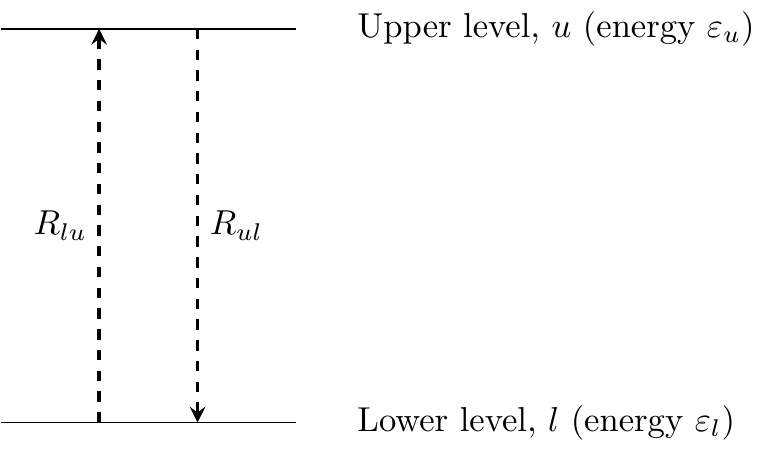}
  \caption{Energy level diagram: two-level atom with bound-bound transitions
    ($R_{lu}$ and $R_{ul}$).}
  \label{fig:two-level-atom}
\end{figure}

To illustrate the principle, we consider one of the simplest possible
radiation--matter interactions that might be of relevance to a radiative/statistical equilibrium problem, that of absorption by a spectral line in a
two-level atom (see \Cref{fig:two-level-atom}). For further simplicity we
restrict our discussion to the regime in which radiation dominates both the
rate of excitation and de-excitation and we assume that the emissivity is
isotropic.
If stimulated emission is treated as negative absorption, the emissivity of the
spectral line (photon frequency $\nu$) can simply be written \citep[cf.][page
118]{Hubeny2014}
\begin{equation}
\eta_{ul} = \frac{h \nu}{4 \pi} \psi_{ul}(\nu) R_{ul}
\label{Eqn:line-emiss}
\end{equation}
where the rate $R_{ul}$ 
\begin{equation}
R_{ul} = n_u A_{ul}
\label{eq:emiss}
\end{equation}
depends on the population of atoms in the upper level ($n_u$) and the
Einstein-A coefficient for the transition ($A_{ul}$). $\psi_{ul}(\nu)$ is the
emission profile function that determines the line shape (details of this
  function are not pertinent to our discussion here except to note that this
function is normalised over all frequencies). Thus to evaluate $\eta_{ul}$
directly we need to know the upper level population ($n_u$). If the
radiation field
itself controls excitation to $u$, however, we cannot reliably determine $n_u$
until {\it after} the radiation field has been simulated. Provided that
statistical equilibrium applies, however, we {\it do} know that the rate of
absorption of energy from the radiation field ($R_{lu}$) is equal to the rate
of emission:
\begin{equation}
R_{lu} = R_{ul}.
\label{eq:stat-eqm}
\end{equation}
This statement can effectively be used to replace a direct evaluation of
\Cref{eq:emiss} by imposing it as a rule of an indivisible energy packet
\gls{MCRT} simulation: whenever a \gls{MC} packet is absorbed by the line, it
is immediately (in situ) re-radiated by the line. In essence, this is
interpreting \Cref{eq:stat-eqm} as a traffic flow problem in which absorption
of \gls{MC} packets is the realisation of the $R_{lu}$ term and re-emission is
described by $R_{ul}$.

This particular example is almost trivial but, as will be elaborated below, the
basic idea of combining \gls{RE} (\textit{indivisible packets}) with
statistical equilibrium (\textit{traffic flow rules to process packet
interactions}) can be extended to much more sophisticated cases. Before
discussing more general cases, however, we pause to comment on some of the key
features that this simple example already highlights.

\subsubsection{Fluorescence and thermal emissivity via redistribution
parameters}

Early \gls{MCRT} implementations, such as \cite{Abbott1985}, applied the
two-level effective resonance line treatment of opacity, essentially as
outlined above.  The two-level approximation is relatively well-justified for
many of the strong metal lines in the ultraviolet (UV, such as C~{\sc iv}
\SI{1550}{\angstrom} or N~{\sc v} \SI{1242}{\angstrom}) that were relevant to
studies of stellar winds \citep{Abbott1985} and also later studies of
accretion disk winds \citep{Long2002, Kusterer2014}.  However, the two-level
atom approximation has limited utility and is not realistic for problems in
which flux redistribution via fluorescence and/or thermal reprocessing of
radiation is important. 

Various methods, with differing degrees of sophistication, can be employed to
simulate flux redistribution in indivisible packet \gls{MCRT}. One approach is
to assume that the radiation--matter interactions can be modelled as a
combination of resonance scattering and some form of complete flux
redistribution across the spectrum. In this approach, a redistribution
parameter, $\Lambda$, is introduced and used to determine the outcome of each
packet interaction by drawing a random number ($\xi$) and comparing: if $\xi >
\Lambda$ then the \gls{MC} packet is assumed to undergo coherent scattering
(i.e.\ a new direction is assigned but the \gls{CMF} frequency is conserved as
it would be in electron scattering or resonance line scattering); otherwise an
incoherent (effective) scattering is executed in which a new random direction
of propagation is assigned along with a new frequency. When the incoherent case
is selected, the new frequency must be drawn from some suitable normalised
emissivity distribution. One simple possibility is the \gls{LTE} thermal
emissivity ($\chi_{\nu} B_{\nu}$), but alternative choices could be made. The
redistribution parameter ($\Lambda$) can be set globally or made a function of
the interaction process (e.g. for line-scattering problems it might be
estimated by comparing the relative importance of collisional and radiative
de-excitation, similar to the considerations by \citealt{Long2002}). The
effectiveness of this approach naturally depends on the problem under
consideration. However, at least for some applications studied with \gls{MCRT}
it has been shown that this scheme is effective. In particular, as demonstrated
by \citet{Baron1996}, \citet{Pinto2000}, \citet{Pinto2000a}  and \citet{ Kasen2006}, flux redistribution in
\gls{SNIa} modelling can be quite effectively approximated via a simple
(thermal) redistribution parameter, achieving good agreement with more detailed
treatments without too much sensitivity to the particular value of $\Lambda$
adopted \citep[see also][]{Magee2018}. We note that, in the limit $\Lambda
\rightarrow 1$, it may appear that this type of approach seems very similar to
that outlined in \Cref{Sec:emiss-simple}: selecting post-interaction properties
of the \gls{MC} packets depends on knowing the material state sufficiently well
to estimate an emissivity distribution from which to draw e.g. photon
frequencies.  However, the notable difference is that here the absolute
\textit{normalisation} of the emissivity is not used: i.e. although the
emissivity distribution is used to select most properties, the packet energies
remain fixed by the \textit{indivisible packet} principle. As a consequence,
strict \gls{RE} is still enforced in the radiation/matter interactions.  

\subsubsection{Fluorescence and redistribution: Macro atom method}
\label{sec:fluorescence-ma}

Approaches similar to those outlined above (i.e.\ that treat interactions as
either coherent or fully redistributive) are easy to implement, fast to run and, with appropriate parameter choices, can capture many of the essential features
of scattering and redistribution. However, not all physical processes are
readily captured this way: for example fluorescence (and cascades) between
energy levels in an atom or ion certainly leads to a coupling of emission in
different parts of the spectrum, but it cannot be well described via a single
``redistribution emissivity'' that can be sampled for all interactions. In
general, we must acknowledge that every distinct radiation--matter interaction
can lead to its own distinct set of outcomes. For example, consider a
three-level atom in a problem for which statistical and \gls{RE} are assumed
(cf.\ \Cref{fig:three-level-atom}): if a \gls{MCRT} packet is absorbed in the
transition from the lowest level (1) to the highest (3), it is expected that
the range of outcomes following that event should at least include a
combination of re-emission in the $3 \rightarrow 1$ transition plus the cascade $3
\rightarrow 2$ and $2 \rightarrow 1$. It is therefore desirable to construct
sets of rules for processing packet interactions in \gls{MCRT} simulations that
can accurately describe this physics.

One way to handle the three-level atom example would simply be to split the
original interacting packet, whether a photon or energy packet, in proportion
to the number of emitted photons or corresponding energy flow for each of the
outgoing channels and continue the simulation. For the three-level case, this
is quite feasible but the drawback of such an approach in general is that, for
interactions with very many possible outcomes (e.g.\ atomic models with large
numbers of levels) this can lead to a proliferation of packets that is
computationally too expensive to follow. Moreover, it is non-trivial to
generalise that approach to handle e.g. the inverse: our three-level atom ought
to also be able to absorb $1 \rightarrow 2$ {\it and} $2 \rightarrow 3$, and
then emit $3 \rightarrow 1$ ({\it inverse} fluorescence). How can this process
be captured in such a redistribution scheme? 

Following \cite{Lucy1999a}, key steps towards finding a general solution stem
from the procedural approach outlined for the two-level atom above (see
\Cref{Sec:two-level}). In particular, a first generalisation of the
\textit{effective resonance scattering} treatment to multi-level atoms is to
extend the \textit{traffic flow} interpretation to include all possible
transitions out from an excited atomic state. Specifically, when an energy
packet is absorbed by a transition to some upper level $u$ of an atom, our
\gls{MC} rule is to re-emit that energy packet in \textit{any} of the transitions
out of that level to a lower level $l$ with probability $q_l$ proportional to
the rate of energy flow in the transition:
\begin{equation}
q_{u \rightarrow l} = \frac{R_{ul} \varepsilon_{ul}}{\sum_{k<u} R_{uk} \varepsilon_{uk}},
\end{equation}
where $\varepsilon_{uk}$ is the difference in the energies of the levels
($\varepsilon_{uk} = \varepsilon_u - \varepsilon_k$). Note that we use the energy flow
rates (rather than the pure transition rates) since our quanta are considered
parcels of fixed energy (not photons).  As shown by \cite{Lucy1999a}
and \cite{Mazzali2000}, this
\textit{downbranching} scheme already provides a huge improvement to the
modelling of flux redistribution in \gls{SN} spectra over a \textit{resonance
scattering approximation}. Indeed, as demonstrated by later studies
\citep[e.g. ][]{Kerzendorf2014}, this scheme performs extremely well even when compared
to more complete treatments such as the full macro atom approach
described below.

Nevertheless, the \cite{Lucy1999a} \textit{downbranching} scheme is still not
complete and does not address all the issues raised even by our simple
three-level atom example (e.g.\ while it will reproduce flux redistribution
  between the $1\rightarrow 3$ and $2 \rightarrow 3$ transitions -- because
  they share an upper level -- it does not connect the $1 \rightarrow 2$
transition to the cascade). A more complete solution that can incorporate all
transitions in multi-level atoms was later devised by \citet{Lucy2002,Lucy2003}
via what he called the \textit{macro atom} method. This approach provides a
generalised approach to formulating rules to process interactions of \gls{MC}
energy packets in accordance with the requirements of radiative and statistical
equilibrium. In essence, we can view all of the possible excited levels of the
matter as energy pools. Energy flows into/out of each pool via the set of
transitions into that energy level and the equilibrium condition (namely that
  the energy associated with each such pool is stationary) is satisfied by
  imposing a traffic flow set of rules to process interactions for each
  possible energy level. The extra sophistication compared to the
  \textit{downbranching scheme} is that we include the fact that physical
  processes represent not only energy flow to and from the radiation field, but
  also \textit{between} the various energy pools associated with the different
  available levels of the atoms/ions/molecules in the medium.  Expressions for
  the general \textit{macro atom} transition probabilities and their
  interpretation are derived by \citet{Lucy2002}. We will not repeat the
  general case here but, in the example below, illustrate its application to
  our example three-level atom in order to clarify the principles.

\subsubsection{Example and discussion: macro atom scheme for a three-level atom}

\begin{figure}[htb]
  \centering
  \includegraphics[width=0.7\textwidth]{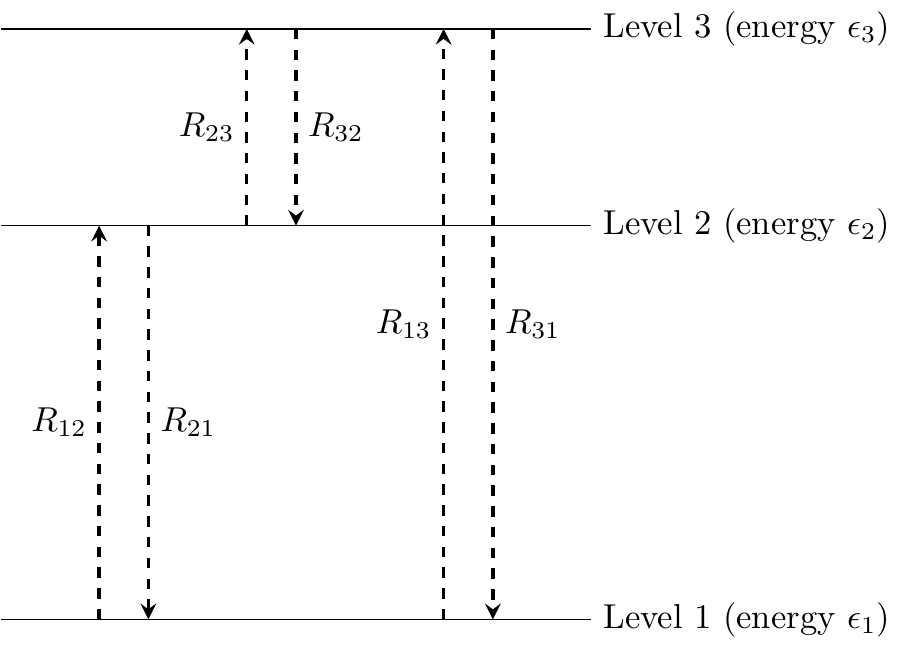}
  \caption{Energy level diagram for the simple three-level atom used
    as an example for
  the formulation of macro atom rules.}
  \label{fig:three-level-atom}
\end{figure}

Here we illustrate how the macro atom transition probabilities can be obtained
for a three-level atom assuming radiative and statistical equilibrium (see
\Cref{fig:three-level-atom}). For simplicity we consider only bound-bound
processes here and assume that all rates are dominated by radiation processes
(i.e.\ neglect collisions and the associated coupling to the thermal energy
pool; see \citealt{Lucy2003} and \Cref{Sec:thermal_pool} for the more general
case).
Our aim is to formulate a set of rules to use during an indivisible energy
packet \gls{MCRT} simulation whenever the random walk experiment determines
that a packet is absorbed by one of the three spectral lines (which have
frequencies $\nu_{12}$, $\nu_{13}$ and $\nu_{23}$, where $\nu_{ij} =
\varepsilon_{ij} / h$).

We start from the equations of statistical equilibrium applied to levels 1, 2
and 3 of the three-level atom. These can be expressed as:
\begin{align}
R_{21}+R_{31}-R_{12}-R_{13} & = 0,\label{Eqn:three-level-statem0}\\
R_{12}+R_{32}-R_{21}-R_{23} & = 0,\label{Eqn:three-level-statem1}\\
R_{13}+R_{23}-R_{31}-R_{32} & = 0 \;.
\label{Eqn:three-level-statem2}
\end{align}
For each of the transitions we can also specify the rate at which energy is
being transferred into or out of the population of atoms in each of the energy
states (i.e. the rates of energy flow into and out of the pool of excitation
energy associated with that state). Specifically, we identify
\begin{equation}
\dot{A}_{12} = \varepsilon_{12} R_{12} \; , \; \dot{A}_{13} = \varepsilon_{13} R_{13} , \; \dot{A}_{23} = \varepsilon_{23} R_{23} 
\label{Eqn:three-level-As}
\end{equation}
as the rates of energy flow from the radiation energy pool to the corresponding
excitation energy pool for each of the three transitions. Similarly, the rates
at which energy is converted back from excitation to radiation can be written
\begin{equation}
\dot{E}_{21} = \varepsilon_{12} R_{21} \; , \; \dot{E}_{31} = \varepsilon_{13} R_{31} , \; \dot{E}_{32} = \varepsilon_{23} R_{32} \;.
\label{Eqn:three-level-Es}
\end{equation}
Multiplying \Crefrange{Eqn:three-level-statem0}{Eqn:three-level-statem2} by
$\varepsilon_1$, $\varepsilon_2$ and $\varepsilon_3$ respectively, and using the
definitions from \Cref{Eqn:three-level-As} and
\Cref{Eqn:three-level-Es} leads, following some rearrangement\footnote{We make
the specific rearrangement such that all terms are positive: this is to
facilitate interpretation of the resulting equation in terms of energy flow
probabilities.}, to 
\begin{align}
\varepsilon_1 R_{21} + \varepsilon_1 R_{31} & =  \varepsilon_1 R_{12} + \varepsilon_1 R_{13},
\label{Eqn:three-level-MA0}\\
\dot{A}_{12} + \varepsilon_1 R_{12} + \varepsilon_2 R_{32} & = \dot{E}_{21} + \varepsilon_2 R_{23} + \varepsilon_1 R_{21},
\label{Eqn:three-level-MA1}\\
\dot{A}_{13} + \dot{A}_{23} + \varepsilon_1 R_{13} + \varepsilon_2 R_{23} & = \dot{E}_{31} + \dot{E}_{32} + \varepsilon_2 R_{32} + \varepsilon_1 R_{31}
\;.
\label{Eqn:three-level-MA2}
\end{align}
We now make a \textit{traffic flow} interpretation of these equations to
formulate rules for the interaction of packets in a \gls{MCRT} simulation.  In
particular, by definition, $\dot{A}_{12}$, $\dot{A}_{13}$ and $\dot{A}_{23}$
are the rates of energy flow from the radiation field to the atom levels via the
specific transitions: thus absorption of radiation \gls{MC} packets in our
simulation is interpreted as the realisation of these terms. Similarly, each of
the $\dot{E}$ terms represents the flow of energy to the radiation field in the
corresponding transition: i.e.\ in the \gls{MCRT} simulation, these terms
correspond to injecting radiation packets. The interpretation of the remaining
terms (all of form ``energy $\times$ rate'') can be made by recognising that
each such term appears twice: always once on the \gls{LHS} of one equation
and once on the \gls{RHS} of another. I.e.\ each of these terms can be regarded
as a source term for energy in one level of the macro atom (appearing as an
``incoming'' energy term on the \gls{LHS}, alongside the absorption from the
radiation field) but simultaneously a sink term for another energy level.
Accordingly, these terms are viewed as driving internal (radiationless)
transitions between levels of the macro atom -- they facilitate the energy flow
between states such that the underlying equations of statistical equilibrium are
conserved. These equations can thus be embedded in our \gls{MCRT} simulation
via the following algorithm:
\begin{itemize}
  \item[(A)] Whenever an active radiation packet is absorbed by any of the
    three transitions we view this as a discrete realisation of the
    corresponding $\dot{A}$ term in the macro atom equation. We say that this
    process has \textit{activated} a macro atom in the corresponding energy
    level.
  \item[(B)] We then inspect the sink terms (i.e.\ \gls{RHS} terms) for the
    activated level of the macro atom and use a random number to select an
    outcome with probabilities proportional to the energy flows implied by the
    system of macro atom equations. Thus, for example, if the macro atom is
    activated to level 2, with probability $\dot{E}_{21} / D_2$ we select
    emission in the $2\rightarrow 1$ transition, and with probabilities of
    $\varepsilon_2 R_{23} / D_2$ and $\varepsilon_1 R_{21}/ D_2$ we select internal
    macro atom transitions $2\rightarrow 3$ and $2\rightarrow 1$, respectively
    ($D_2 = \dot{E}_{21} + \varepsilon_2 R_{23} + \varepsilon_1 R_{21}$ is selected
    to normalise the probabilities correctly).
  \item[(C)]
  \begin{itemize}
    \item[(i)] If the selection corresponds to an emission $\dot{E}$ term, the
      macro atom deactivates and the radiation packet is returned to the main
      \gls{MC} simulation with new properties (photon frequency, direction
      etc.) set in accordance with the properties of the corresponding emission
      process.  The total energy carried by the packet (in the \gls{CMF})
      remains equal to that when the packet was absorbed (in accordance with
      the requirements of \gls{RE}).
    \item[(ii)] Alternatively, if an internal transition term is selected, the
      macro atom remains active but is switched from its current state to a new
      state in accordance with the selected term (e.g., selecting the
        $\varepsilon_2 R_{23} / D_2$ term results in a transition from macro atom
      state 2 to state 3, conceptually representing the ``sink'' on the
    \gls{RHS} of \Cref{Eqn:three-level-MA1} into the matching ``source'' term on
  the \gls{LHS} of \Cref{Eqn:three-level-MA2}). The algorithm then returns to
  step B and processes the activated macro atom again. This continues until
  deactivation occurs.
  \end{itemize}
\end{itemize}
By repeated application of these rule for packet interactions, the activation
and deactivation of macro atoms will encode both radiative and statistical
equilibrium on the effective emissivity in the simulation. Several specific
features of this macro atom algorithm are noteworthy of consideration for its
appropriateness to a range of applications. We comment on some of these in the
following.

First, we note that all rates $R_{ij}$ are directly proportional to the level
population $n_i$, which would imply that, like the normal line emissivity,
\Cref{eq:emiss}, determining the terms in the macro atom equations depends on
already knowing the level populations. However, because of the normalisation
process in step (B), this leading dependence cancels out from the transition
probabilities. Of course, additional effects (e.g.\ corrections for stimulated
emission to absorption rates or introduction of Sobolev escape probabilities;
see \Cref{Sec:line_ints_sobolev}) can still lead to dependencies on the
populations. Nevertheless, cancelling of the leading-order effect means that
the macro atom transition rates can be relatively well determined even in the
absence of a converged set of level populations.  This property can be rather
powerful when treating complex systems for which exact calculations of excited
state level populations (and therefore a direct evaluation of absolute
emissivities) is challenging: as shown by \citet[][fig. 5]{Lucy2002}, even for
a complicated ion such as Fe~{\sc ii} the \textit{macro atom} scheme produces
fairly accurate excited state effective emissivities without any iteration to
determine level populations.\footnote{Of course, the \textit{macro atom} scheme
can also be coupled to an iterative solution for the level populations to
provide accurate level populations upon convergence. Depending on the problem,
it may be anticipated that the use of the \textit{macro atom} scheme in such an
approach can aid convergence since it gives a relatively good estimate of the
true emissivity even before convergence of the level populations has been
achieved.}

Second, we note that the first of the set of macro atom \textit{traffic flow}
equations (\Cref{Eqn:three-level-MA0}) involves no activation ($\dot{A}$) terms
and no deactivation ($\dot{E}$) terms: it is a balance only between internal
transition rates. This makes sense because it follows from the equation of
statistical equilibrium for the lowest energy state: there are no channels for
absorption of energy directly to that state nor emission of energy directly
from it. Moreover, we note that the choice $\varepsilon_1 = 0$\footnote{This is,
  of course, the standard definition for the zero of (excitation) energy --
  namely that the energy of the lowest lying level (ground state) is defined to
be zero.} trivially satisfies \Cref{Eqn:three-level-MA0} and also eliminates
the corresponding internal transition terms from \Cref{Eqn:three-level-MA1} and
\Cref{Eqn:three-level-MA2}. Making use of this definition will therefore
(slightly) simplify the macro atom algorithm by effectively removing the need
to explicitly consider the ground state.

Third, it can be seen that both of the simpler treatments introduced earlier
for handling atomic line interactions are special cases of the full macro atom.
Specifically, the \textit{effective resonance scattering} approach used in
several early studies (example in \Cref{Sec:two-level}) is a two-level macro atom
with $\varepsilon_1 = 0$. The \textit{downbranching} scheme by \citet{Lucy1999a}
outlined in \Cref{sec:fluorescence-ma} is the macro atom scheme with all
internal transition terms suppressed (formally, this can be derived from the
general macro atom algorithm by assuming (i) downwards transition rate
coefficients dominate and (ii) for all transitions between upper level $u$ and
lower level $l$, $\varepsilon_u \gg \varepsilon_l$).

Repeated cycling through steps (B) and (C)(ii) in the algorithm above can make
it computationally inefficient, particularly when the scheme is extended to
also include coupling to the thermal pool. This can be addressed in several
ways, however. As noted by \cite{Lucy2002}, the macro atom algorithm can be
viewed as recursive application of the set of transition/deactivation
probabilities and recently, \citet{Ergon2018} have presented a Markov-chain
approach to the macro-atom machinery. This method effectively solves the
problem without the need to follow internal macro-atom state transitions which
can be a substantial advantage in terms of computational efficiency.

\subsubsection{The thermal energy pool}
\label{Sec:thermal_pool}

The \textit{macro atom} scheme is readily generalisable to include additional
energy pools relevant to the simulation. In particular, the thermal pool of
particle kinetic energies. In the nomenclature of \citet{Lucy2002},
interactions with the thermal pool are described as kinetic packet ($k$-packet)
events, and the processing rules are derived by considering energy flow into
and out of the $k$-packet pool. Here, the relevant ``transition'' processes are
all heating and cooling rates corresponding to a flow of energy into and out
of the thermal pool. These include, in particular, direct radiative
heating rate ($H_{\mathrm{R}}$), which includes processes such as free-free absorption,
heating by collisional de-excitation of e.g.\ atomic states ($H_{\mathrm{C}}$), and their
inverse cooling processes (rates $C_{\mathrm{R}}$ and $C_{\mathrm{C}}$). Energy flow through the
thermal pool is governed by an assumption of \gls{TE}
(analogous to the assumption of \textit{statistical} equilibrium for the atomic
energy levels descried in the macro atom scheme):
\begin{equation}
H_{\mathrm{R}} + H_{\mathrm{C}} = C_{\mathrm{R}} + C_{\mathrm{C}}.
\label{Eqn:HC-radeqm}
\end{equation}
Thus, whenever a physical process representing the flow of energy into the
thermal pool (e.g.\ absorption of a \gls{MC} radiation energy packet via a
heating process which is a realisation of the $H_{\mathrm{R}}$ term), the process
governing the fate of that energy is determined by randomly sampling all
available cooling processes with probabilities proportional to their respective
rates. This can lead either to direct remission of a radiation energy packet
with photon-frequency, direction etc. randomly assigned by one of the radiative
cooling processes (i.e.\ simulating part of the $C_{\mathrm{R}}$ term) or by activation of
a macro atom (associated with the collisional cooling process, $C_{\mathrm{C}}$). We note
that the scheme can also be applied to physical processes that involve
cross-talk between multiple energy pools: for example, bound-free processes,
which involve both changes in the populations of atomic states and
heating/cooling \citep[see][]{Lucy2003}.

\subsection{Indivisible energy packets beyond radiative equilibrium}
\label{Sec:radnoneq}

In the preceding sections we have discussed how equilibrium assumptions
(radiative, statistical, thermal) can be used to devise rules for \gls{MCRT}
algorithms that can handle complicated non-resonance
scattering/fluorescent processes without sacrificing rigorous conservation of
energy. Formulated in this way, however, such approaches will not be
immediately suitable for problems in which the corresponding equilibrium
assumption is \textit{not} satisfied. Nevertheless, the scheme can
be generalised to include appropriate source or sink terms.
Consider, for example, material in a stellar outflow that is irradiated (e.g.\
by the stellar photosphere) but is also heated by external non-radiative
($H_{\mathrm{E}}$) processes (examples might include hydrodynamical or
magnetohydrodynamical heating) and cooled by expansion ($C_{\mathrm{E}}$).  In such a
case, radiation processes may still be important but \gls{RE} is no longer well justified. If cooling
  time scales are sufficiently short that \gls{TE} can be sustained, the
  heating and cooling balance might be expressed at
  every point in the medium:
\begin{equation}
H_{\mathrm{R}}+H_{\mathrm{C}} +
H_{\mathrm{E}}=C_{\mathrm{R}}+C_{\mathrm{C}}+C_{\mathrm{E}}  \; .
\label{Eqn:heating-cooling-extra}
\end{equation}
In an indivisible energy packet \gls{MCRT} scheme, the $H_{\mathrm{R}}$ and $H_{\mathrm{C}}$ (local
radiative and collisional heating) terms are realised via the absorption of
radiation energy packets via radiation--matter processes that directly transfer
radiative energy to the kinetic particle pool ($H_{\mathrm{R}}$) and collisional
deexcitation of atomic states ($H_{\mathrm{C}}$). The $H_{\mathrm{E}}$ term can then be added to the
problem as an external source of packets injected in the course of the
simulation -- the rate of injection being determined by knowledge of the
external
process responsible for $H_{\mathrm{E}}$. In this particular case, because the external
source is a heating term, the energy of those new packets is initially injected
directly to the pool of kinetic energies (i.e.\ $k$-packet pool) and the
subsequent properties of those packets followed via the usual \textit{traffic
flow} interpretation of the equation.

As before, the three terms on the \gls{RHS} of \Cref{Eqn:heating-cooling-extra}
are the sink terms for the $k$-packet pool and so they are sampled to determine
the manner in which the energy flow out of the $k$-packet pool behaves. $C_{\mathrm{R}}$ and $C_{\mathrm{C}}$ can be simulated just as
before: packets are fed back to the radiation field ($C_{\mathrm{R}}$) or to the excitation
  energy of macro atom pools via collisional excitation ($C_{\mathrm{C}}$). The additional
  term, $C_{\mathrm{E}}$, can be treated as a true external sink term: i.e.\ with
  probability $C_{\mathrm{E}}/(C_{\mathrm{R}}+C_{\mathrm{C}}+C_{\mathrm{E}})$ energy packets that flow into the thermal
  pool are terminated. Alternatively, this term could be treated via a
  reduction in packet energies: i.e.\ one could opt to sample the \gls{RHS} of
\Cref{Eqn:heating-cooling-extra} only considering $C_{\mathrm{R}}$ and $C_{\mathrm{C}}$ but reduce
the energy of all packets processed through this channel by a factor of
$(C_{\mathrm{R}}+C_{\mathrm{E}})/(C_{\mathrm{R}}+C_{\mathrm{E}}+C_{\mathrm{C}})$. 
We note that, in the limit where the external sources and sink term ($H_{\mathrm{E}}$ and
$C_{\mathrm{E}}$  above) become dominant, an indivisible energy packet
simulation performed with this machinery will essentially reproduce the
elementary scheme explained in \Cref{Sec:emiss-simple}. 

The example above illustrates how the indivisible packet scheme can be altered
to take account of specific departures from \gls{RE}, and this is done in many
of the existing implementations of this method, particularly to account for
adiabatic cooling \cite[e.g.][]{Long2002, Kasen2006, Kromer2009a, Vogl2018}. In
principle a similar logic could be employed to deal with departures from
statistical equilibrium (affecting the macro atom transition rules) or \gls{TE}
(further affecting the $k$-packet transition rules). Specifically, if the
inflow and outflow rates are not in balance such that there is a net
rate-of-change of the energy reservoir, then terms representing the ongoing
accumulation (or loss) of energy from the pool could be built into the
formulation (i.e.\ retain terms including the derivatives of the level
populations and/or the kinetic temperature). Provided that values for those
derivatives are known they could also then be included in the packet flow. To
our knowledge, however, extensions of the macro atom/$k$-packet schemes that
consider such terms have not yet been implemented.

\section{MCRT: application in outflows and explosions}
\label{Sec:Flows}

A prominent field in astrophysics, in which \gls{MCRT} methods are very popular
and successful, is the study of fast mass outflows. For example, \gls{MCRT}
schemes can be used to calculate mass-loss rates and the structure of hot-star
winds \citep[see e.g.][]{Abbott1985, Lucy1993, Schaerer1994, Schmutz1997,
Vink1999, Vink2000, Sim2004, Mueller2008, Muijres2012a, Lucy2012a,
Noebauer2015a, Vink2018}, to determine synthetic light curves and spectra for
\glspl{SN} \citep[see e.g.][]{Mazzali1993, Lucy1999, Mazzali2000, Lucy2005,
  Kasen2006, Sim2007, Kromer2009, Wollaeger2013, Wollaeger2014, Kerzendorf2014,
Bulla2015, Magee2018} or to treat \gls{RT} in winds emanating from accretion
discs of cataclysmic variables \citep{Knigge1995, Long2002, Noebauer2010,
Kusterer2014, Matthews2015}  or active galactic nuclei \citep{Sim2005, Sim2010,
Sim2012, Higginbottom2013, Matthews2016, Matthews2017, Tomaru2018}. In
applications such as these, our current implicit assumption, namely that
\gls{RT} occurs in static media or in environments with material velocities low
enough to be safely ignored, can no longer be maintained.  Instead, special
relativistic effects play an important role and have to be taken into account.
In the following, we outline some important aspects of performing \gls{MCRT} in
moving media. While many of the described concepts are generic, the treatment
of line interactions using the Sobolev approximation (see
\Cref{Sec:line_ints_sobolev}) is specific to \gls{MCRT} in expanding media,
such as \gls{SN} ejecta or winds.

\subsection{The mixed-frame approach}
\label{Sec:mixed_frame}

As introduced in \Cref{sec:RT}, there are two fundamental frames of reference
for \gls{RT}, namely the \gls{LF} and the local rest frame (\gls{CMF}). Until
this point, we have largely ignored the distinction between these frames since
\gls{RT} was assumed to occur in static media or in low-velocity environments.
When the material velocities become large, however, this simplification is no
longer justified.  In these situations, \gls{MCRT} schemes often rely on a
so-called \textit{mixed-frame} approach \citep[see for example][]{Lucy2005}.
This exploits the fact that the handling of different tasks involved in
\gls{MCRT} simulations is easier in one or other of the two frames.
Specifically, the spatial and temporal mesh is usually defined in the lab
frame, making it most convenient for measuring distances and thus for tracking
packets and simulating their propagation.  Radiation--matter interactions, on
the other hand, are more easily described in the local rest frame of the
material. Here, the material functions take their simplest form.  Consequently,
\gls{MCRT} schemes adopting the mixed-frame approach propagate packets in the
\gls{LF} but treat all interactions in the \gls{CMF}. 

The inclusion of relativistic effects during the \gls{LF}--\gls{CMF}
transformation can be performed to varying degrees of accuracy. We first focus
on arguably the most important effect in the presence of matter flows, namely
the Doppler effect.  For this illustration, we again assume a simplified
situation of coherent and isotropic scattering. In moving media, these
assumptions will only ever hold in the \gls{CMF}. Whenever a \gls{MC} packet
interacts, its properties are transformed into the \gls{CMF}, where the
interaction is simulated and post-interaction properties are assigned.
Adopting the convention introduced in \Cref{sec:RT} and denoting quantities
measured in the \gls{CMF} by a $0$ subscript, the incident \gls{CMF} frequency
of an interacting packet is 
\begin{equation}
  \nu_{0,\mathrm{i}} = \gamma \nu_{\mathrm{i}} \left(1 -
    {\boldsymbol{\beta} {\mathbf{\cdot n_{\mathrm{i}}}}}\right).
  \label{eq:mixed_frame_cmf_nu}
\end{equation}
Likewise, it carries an energy  
\begin{equation}
  \varepsilon_{0,\mathrm{i}} = \gamma \varepsilon_{\mathrm{i}} \left(1 - {\boldsymbol{\beta} {\mathbf{\cdot n_{\mathrm{i}}}}}\right)
  \label{eq:mixed_frame_cmf_e}
\end{equation}
in the \gls{CMF}. Here, the additional subscripts (i and e) are used to denote
incident and emergent packet properties with respect to the interaction.
Performing interactions in the \gls{CMF} has the benefit that, for our example
of coherent scattering, energy is conserved in this frame and thus
\begin{equation}
  \varepsilon_{0, \mathrm{e}} = \varepsilon_{0, \mathrm{i}}.
  \label{eq:mixed_frame_cmf_e_conservation}
\end{equation}
Similarly, the packet frequency remains constant during the interaction in the
\gls{CMF}\footnote{For more complicated cases involving incoherent scattering
  and/or departures from equilibrium, the principles discussed in Chapter 7 can
  all be applied to the packet energy and frequency in the \gls{CMF}.}
\begin{equation}
  \nu_{0, \mathrm{e}} = \nu_{0, \mathrm{e}}.
  \label{eq:mixed_frame_cmf_nu_conservation}
\end{equation}
After drawing an emergent propagation direction, the packet is re-transformed
into the \gls{LF} and continues its propagation there with
\begin{align}
  \varepsilon_{\mathrm{e}} &= \gamma \varepsilon_{0,\mathrm{e}}
                             \left(1 + {\boldsymbol{\beta}
                             {\mathbf{\cdot n}_{0, \mathrm{e}}}} \right),\\
  \nu_{\mathrm{e}} &= \gamma \nu_{0,\mathrm{e}} \left(1 +
                     {\boldsymbol{\beta} {\mathbf{\cdot n}_{0, \mathrm{e}}}}\right).
\end{align}

In many applications, the frequency shifts due to the Doppler effect will be
the most significant consequence of the material motion. However, for a
detailed relativistic treatment, achieving at least $\mathcal{O}(v/c)$
accuracy, additional effects have to be taken into account. In particular,
aberration affects the propagation direction of packets and has to be taken
into account when transforming the incident direction into the \gls{CMF}
\begin{equation}
  \mathbf{n}_{0,\mathrm{i}} = \left( \frac{\nu}{\nu_0} \right) \left[\mathbf{n_{\mathrm{i}}} - \gamma\boldsymbol{\beta} \left( 1 - \frac{1}{c}\frac{\gamma}{\gamma + 1} \mathbf{n}_{\mathrm{i}} \cdot \mathbf{v} \right) \right],
  \label{eq:mixed_frame_cmf_prop_dir}
\end{equation}
and the emergent direction back into the \gls{LF} \citep[cf.][]{Mihalas2001}
\begin{equation}
  \mathbf{n_{\mathrm{e}}} = \left( \frac{\nu}{\nu_0} \right) \left[\mathbf{n}_{0,\mathrm{e}} + \gamma\boldsymbol{\beta} \left( 1 + \frac{1}{c}\frac{\gamma}{\gamma + 1} \mathbf{n}_{0, \mathrm{e}} \cdot \mathbf{v} \right) \right].
  \label{eq:mixed_frame_lf_prop_dir}
\end{equation}
Also, care has to be taken when calculating the accumulated optical depth.
Since the propagation is carried out in the \gls{LF}, the integration is best
performed in this frame as well. For this, opacities have to be properly
transformed into the \gls{LF} using \Cref{eq:transformation_laws_I_chi_eta}.
The optical depth integration is further complicated in the presence of matter
flows by the continuous change in \gls{CMF} frequency as packets propagate
through material with varying velocities.  Thus,
\Cref{eq:packet_accumulated_optical_depth} becomes 
\begin{equation}
  \tau(l) = \int_{\nu_0(0)}^{\nu_0(l)} \mathrm{d}\nu_0 \frac{\mathrm{d}l}{\mathrm{d}\nu_0} \frac{\nu_0}{\nu} \chi_0(\nu_0),
  \label{eq:opt_depth_integ_moving_media}
\end{equation}
which accounts for the \gls{CMF} frequency change along the trajectory by
$\mathrm{d}l/\mathrm{d}\nu_0$ and includes a Doppler factor $(\nu_0/\nu)$ due
to the transformation of the opacity into the \gls{CMF}. Performing this
integration for non-trivial opacity laws and general flow patterns is very
challenging and computationally demanding since it has to be carried out
frequently during the propagation process of each packet.  However, for an
important class of astrophysical applications of \gls{MCRT}, the
\textit{Sobolev approximation} can be adopted and drastically reduces the
complexity of the integration by turning it into a purely local problem. We
elaborate on this approach in \Cref{Sec:line_ints_sobolev}. 

Finally, we note that one also has to decide in which frame the \gls{MC}
packets are launched during the initialisation. Often, the \gls{CMF} is the
natural choice for this process, e.g.\ when representing a thermal radiation
field. In such cases, packet properties are drawn in the \gls{CMF} and then
transformed into the \gls{LF} using the rules given here and in \Cref{sec:RT}
before starting the propagation.

\subsection{Line interactions in outflows}
\label{Sec:line_ints_sobolev}

As described above, treating frequency-dependent opacities in the presence of
large material velocities is challenging. However, in the case of bound-bound
processes, the situation can be significantly simplified with the so-called
\textit{Sobolev approximation} \citep{Sobolev1960}. Indeed, \gls{RT} through
fast expanding mass outflows is the classical example for the use of the
Sobolev approximation. We refrain from a detailed description of Sobolev theory
since it is a widely used technique in astrophysical \gls{RT} problems (see
e.g.\ \citealt{Castor2007} for a general overview of the approximation and
\citealt{Rybicki1978, Rybicki1983, Hummer1985, Jeffery1993, Jeffery1995} for
various extensions of the original formulation) but highlight some key aspects
and describe how a Sobolev line interaction scheme can be easily incorporated
into \gls{MCRT} simulations for fast outflows.  An illustrative overview of
this approximation can be found in \citet{Lamers1999}.

Bound-bound processes are fundamentally resonant processes in the sense that only
photons, and thus \gls{MC} packets, with \gls{CMF} frequencies in a small
window can perform these interactions. The frequency dependence of the
bound-bound opacity is encoded in the line profile function, $\psi$,
which is narrowly peaked about the natural line frequency 
\begin{equation}
  \nu_{lu} = \frac{1}{h} (\varepsilon_u - \varepsilon_l),
  \label{eq:natural_line_frequency}
\end{equation}
corresponding to the energy separation of the two atomic levels,
$\varepsilon_{l}$ and $\varepsilon_{u}$, connected by the line transition. The
width of the line transition is mainly a result of the turbulent and thermal
motion of the atoms. Together with the local gradient of the velocity field it
can be translated into a characteristic length scale, the so-called
\textit{Sobolev length}, over which the photon shifts into and out of resonance
with the line in the \gls{CMF}. This scale is compared with the typical length
over which the plasma state, and thus the frequency-independent parts of the
line opacity change. If the Sobolev length is much smaller, the line profile
can be approximated by a delta-distribution around $\nu_{lu}$. As a
consequence, the optical depth integration in
\Cref{eq:opt_depth_integ_moving_media} collapses and turns into an expression
that only depends on the material state at the so-called \textit{Sobolev
point}. At this location, the \gls{CMF} frequency of the photon exactly equals
$\nu_{lu}$. We note that in addition to the condition concerning plasma state
changes, the traditional Sobolev approximation can only be applied to
environments with a monotonous velocity gradient\footnote{See
  \citet{Rybicki1978} for an extension to non-monotonous flows.}.

\begin{figure}[htb]
  \centering
  \includegraphics[width=\textwidth]{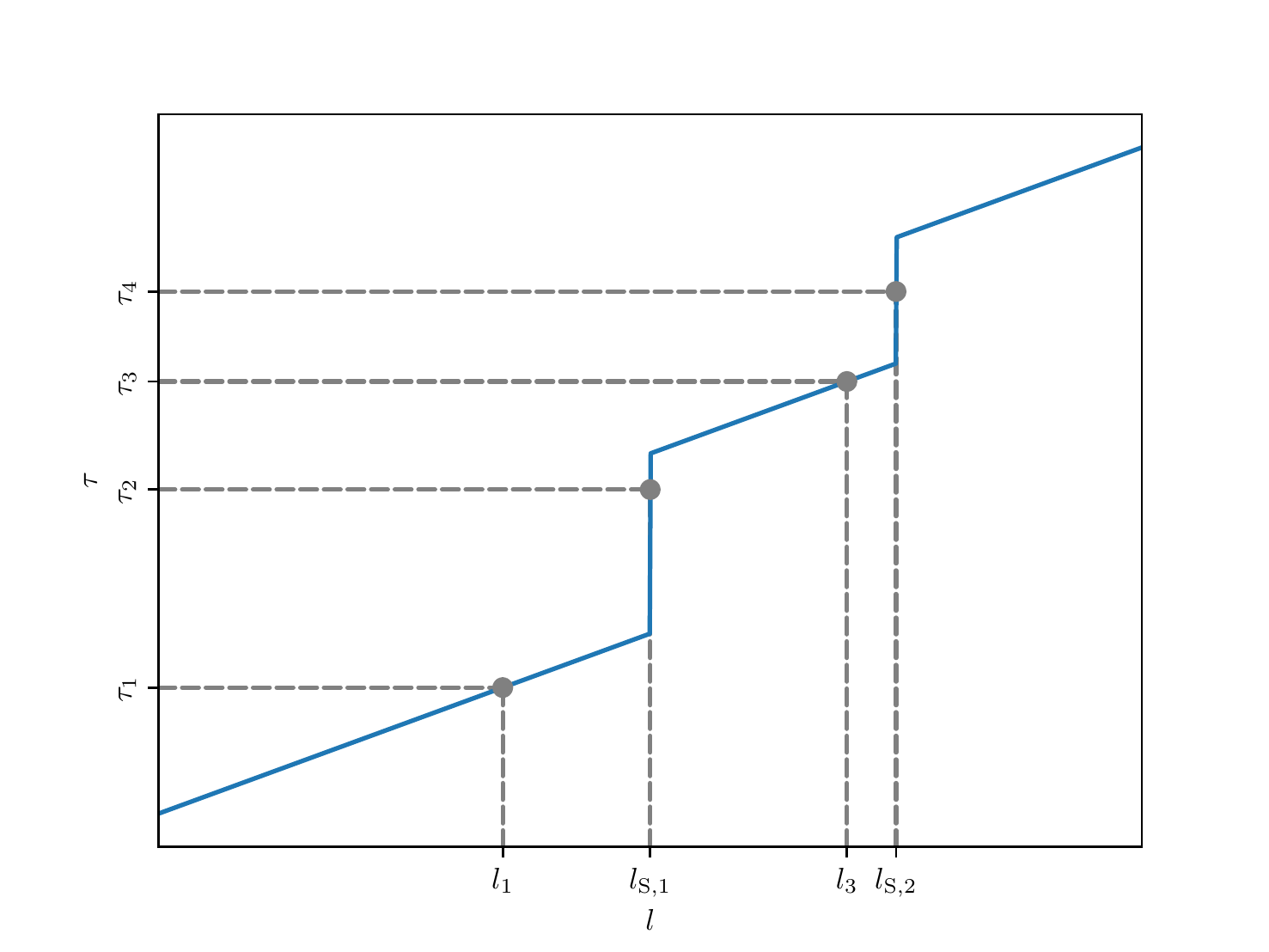}
  \caption{Illustration of the optical depth accumulation process devised by
    \citet{Abbott1985} and \citet{Mazzali1993} for \gls{MCRT} simulations relying on the
    Sobolev approximation. At the Sobolev points, $\mathrm{S}_1$ and
    $\mathrm{S}_2$, the accumulated optical depth is instantaneously
    incremented by the full line optical depth. In addition four possible
    outcomes of the decision process about the next interactions are shown. For
    $\tau_1$ and $\tau_3$, the packet would experience a continuum interaction
    at $l_1$ and $l_3$ respectively. In the remaining cases, i.e.\ for $\tau_2$
    and $\tau_4$, the packet undergoes a line interaction at the respective
    Sobolev points $l_{\mathrm{S},1}$ and $l_{\mathrm{S}, 2}$. This figure is
    loosely adapted from \citet[][Fig.\ 1]{Mazzali1993}.}
  \label{fig:sob_optical_depth_accumulated}
\end{figure}

Whenever the Sobolev approximation can be adopted, \gls{MCRT} simulations that
include line interactions are dramatically simplified. In addition to reducing
the calculation of the line optical depth to a local problem, line overlaps are
eliminated within the Sobolev approximation. Since photons continuously red
shift in monotonously expanding flows (or blue shift in compression flows),
they successively scan over the possible line transitions in the \gls{CMF}
one-by-one.  This reduces the book-keeping effort in \gls{MCRT} simulations and
suggests storing the line transitions in a frequency-ordered list
\citep{Lucy1999a}. In outlining the basic \gls{MCRT} propagation routine for
such cases, which has been developed by \citet{Abbott1985} and \citet{Mazzali1993}, we
assume that in addition to line interactions, only frequency-independent
continuum processes, such as electron scattering in the Thomson limit,
contribute to the total opacity. As the packet propagates, it continuously
accumulates optical depth due to continuum processes (see
\Cref{Sec:Propagation_Basic}). Whenever the packet reaches the Sobolev point of
the next line transition, i.e.\ when its local \gls{CMF} frequency equals the
natural frequency of the line, the accumulated optical depth is instantaneously
incremented by the full line opacity of the transition
\begin{equation}
  \tau_{\mathrm{s}} = \frac{\pi e^2}{m_{\mathrm{e}} c}f_{lu} n_l \left( 1 - \frac{n_u}{n_l} \frac{g_l}{g_u} \right)_{r_{\mathrm{s}}} \left( \frac{\mathrm{d}l}{\mathrm{d}\nu_0} \right)_{r_{\mathrm{s}}}.
  \label{eq:sobolev_tau}
\end{equation}
Here, $e$ and $m_{\mathrm{e}}$ denote the elementary charge and the mass of the
electron, $f_{lu}$ is the absorption oscillator strength of the transition from
the lower to the upper energy lever, $l \rightarrow u$, and $n_{l,u}$ and
$g_{l,u}$ are the population numbers and statistical weights of these levels.
The subscribed $r_{\mathrm{s}}$ denotes that the plasma state entering the
optical depth calculation is evaluated locally at the Sobolev point. In
one-dimensional geometries, the derivative in \Cref{eq:sobolev_tau} simplifies
to 
\begin{equation}
  \left(\frac{\mathrm{d}l}{\mathrm{d}\nu_0} \right)_{r_{\mathrm{s}}} = \frac{c}{\nu_{lu}}\left( \frac{1}{(1-\mu^2) v / r + \mu^2 \mathrm{d}v/\mathrm{d}r} \right)_{r_{\mathrm{s}}}.
  \label{eq:sobolev_1d_gradient}
\end{equation}
The optical depth accumulation procedure is further illustrated in
\Cref{fig:sob_optical_depth_accumulated}.
The decision about the nature of the next interaction a \gls{MC} packet
experiences is based on whether the assigned optical depth value is surpassed
between Sobolev points or during the instantaneous increments at one of these
resonance points. In the former, the packet is simply moved by the distance $l$
given by \Cref{eq:packet_accumulated_optical_depth}, which now only involves
continuum opacity. At the interaction site, the packet properties are changed
according to the specific continuum process. If the optical depth value is
reached at a Sobolev point, however, the packet is moved to this location and
performs the corresponding line interaction. It should be noted that the
re-emission direction for the packet should be assigned according to the
Sobolev escape probability
\begin{equation}
  \rho(\mathbf{n}) \propto \frac{1 - \exp(-\tau_{\mathrm{s}}(\mathbf{n}))}{\tau_{\mathrm{s}}(\mathbf{n})}.
  \label{eq:sobolev_escape_prob}
\end{equation}
For the particular case of an homologous flow,
\begin{equation}
  \frac{v}{r} = \frac{\mathrm{d}v}{\mathrm{d}r} = \mathrm{const.}
  \label{eq:homologous_flow_velocity} \; ,
\end{equation}
the Sobolev optical depth becomes independent of direction and the escape
probability is isotropic. However, in more general velocity fields the Sobolev
escape probability will vary with direction and must be sampled whenever
directions for re-emission of packets are needed.

Many \gls{MCRT} applications in outflows adopt the Sobolev approximation and
follow a line interaction scheme similar to the one just outlined. Examples
include the studies by \citet{Abbott1985, Lucy1993, Vink1999,
Sim2004,Noebauer2015a} dealing with hot star winds, or the works by
\citet{Long2002} performing \gls{MCRT} in disc winds and \citet{Mazzali1993,
Mazzali2000, Kasen2006, Sim2007, Kromer2009, Kerzendorf2014} who use \gls{MCRT}
in \gls{SN} ejecta. There are several studies that treat line interactions
without relying on the Sobolev approximation such as
\citet{Knigge1995} and
\citet{Kusterer2014}.  Here, the conceptual and computational effort is, however,
significantly higher.

\begin{figure}[htb]
  \centering
  \includegraphics[width=\textwidth]{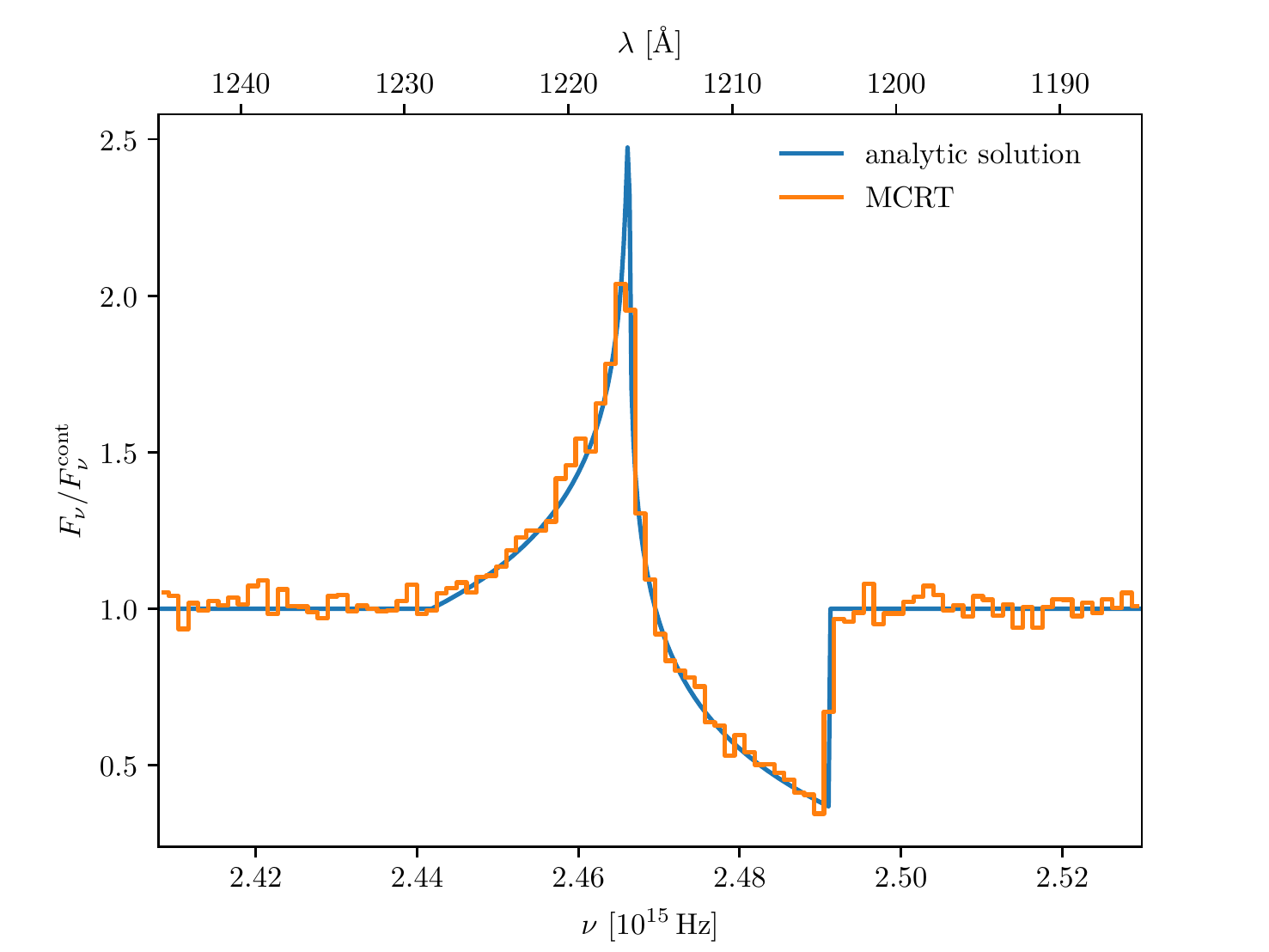}
  \caption{H Lyman $\alpha$-line profile determined by a simple mixed-frame \gls{MCRT}
    simulation for the test setup described in \Cref{Sec:mixed_frame} and
    \Cref{sec:line_test}. The analytic prediction (blue) is obtained by a
    formal integration of the \gls{RT} problem according to the
    principles outlined by \citet{Jeffery1990}. Note that the emergent spectra
    have been normalized by dividing by the incident thermal radiation field
    $F^{\mathrm{cont}}_{\nu}$.}
  \label{fig:sobolev_line_profile}
\end{figure}

To demonstrate the use of the Sobolev approximation in \gls{MCRT}, we describe
a simple test simulation to calculate the H Lyman $\alpha$ line profile formed in a
homologous flow composed of only neutral hydrogen in \Cref{sec:line_test}.
This leads to the line profile shown in \Cref{fig:sobolev_line_profile}.
For comparison, the analytic solution for this test problem is
included.\footnote{It was obtained by a formal integration of the \gls{RT}
problem according to the scheme outlined by \citet{Jeffery1990}.}
The collection of tools available as part of this review contains a simple
Python implementation of this \gls{MCRT} simulation (see
\Cref{sec:tool_collection}).

\subsection{MCRT and expansion work}

In \gls{RE}, packet energy is conserved in the \gls{CMF} during interactions, which
partly motivates the introduction of the mixed-frame approach for \gls{MCRT} in
moving media. We emphasize, however, that packet energy conservation does not
necessarily hold in the \gls{LF}. In fact, depending on the flow of radiation
relative to the moving ambient material, photons may either lose or gain energy
in interactions. This is a crucial process in astrophysical applications
involving strong mass outflows, for example hot-star winds. Here, photons
collectively lose energy in interactions by performing expansion work,
ultimately driving and maintaining the outflow \citep[cf.][]{Lamers1999,
Puls2008}. In the following, we briefly demonstrate that \gls{MCRT} schemes
adopting the mixed-frame approach readily capture this work term (indeed this
was one of the original motivating factors for the approach; \citealt{Abbott1985}).

When a packet interacts, the mixed-frame \gls{MCRT} approach switches to the
local \gls{CMF} to perform the interaction and update the packet properties
before returning into the \gls{LF} and continuing the propagation. Considering
again isotropic and resonant scatterings for illustrative purposes, the
\gls{LF} energy of a packet changes during the interaction from
$\varepsilon_\mathrm{i}$ to $\varepsilon_\mathrm{e}$ according to 
\begin{equation}
  \varepsilon_{\mathrm{e}} = \varepsilon_{\mathrm{i}} \frac{1 -
    \boldsymbol{\beta} \mathbf{\cdot n}_{\mathrm{i}}}{1 - \boldsymbol{\beta} \mathbf{\cdot n}_{\mathrm{e}}}.
  \label{eq:energy_after_scatter_LF}
\end{equation}
Depending on the orientation of the propagation direction prior and after the
scattering relative to the material flow, the packet thus gains or loses energy
in the \gls{LF}.

In astrophysical mass outflows, radiation is typically emitted from a source at
the base of the flow, for example by a central star or an accretion disc. As a
consequence, photons predominantly propagate in the direction of the flow
before they interact.  Electron scatterings in the Thomson limit or resonant
line interactions, two processes playing important roles in mass outflows, are
either approximately isotropic or exhibit at least a re-emission profile that
is forwards-backwards symmetric. Thus, a radiation field that is initially
predominantly aligned with the expanding flow will be partly diffused by the
interactions and packets, on average, lose energy in the \gls{LF}. This process
can be further illustrated by considering the mean packet energy, $\bar
\varepsilon_{\mathrm{e}}$, after the
first interaction, which is obtained by averaging over all re-emission
directions. Specifically, in terms of the incident and emergent
direction cosines ($\mu_{\mathrm{i}}$ and $\mu_{\mathrm{e}}$),
\begin{equation}
  \bar \varepsilon_{\mathrm{e}} = \frac{1}{2} \varepsilon_{\mathrm{i}}
  \int_{-1}^{1} \frac{1 - \mu_{\mathrm{i}} \beta}{1 - \mu_{\mathrm{e}}
    \beta}\mathrm{d}\mu_{\mathrm{e}} \; ,
  \label{eq:mean_energy_after_scatter_wind_integral}
\end{equation}
where, for simplicity, we have assumed spherical symmetry and have neglected
aberration.  Performing the integration gives
\begin{equation}
  \bar \varepsilon_{\mathrm{e}} = \frac{1}{2 \beta} \varepsilon_{\mathrm{i}} (1 - \mu_{\mathrm{i}} \beta) \left( \log (1 + \beta) - \log (1 - \beta) \right).
  \label{eq:mean_energy_after_scatter_wind}
\end{equation}
For small values of $\beta$, this reduces to 
\begin{equation}
  \frac{\bar \varepsilon_{\mathrm{e}}}{\varepsilon_{\mathrm{i}}} = 1 - \mu_{\mathrm{i}} \beta + \mathcal{O}(\beta^2),
  \label{eq:mean_energy_after_scatter_small_beta}
\end{equation}
but in general, 
\begin{equation}
  \frac{\bar \varepsilon_{\mathrm{e}}}{\varepsilon_{\mathrm{i}}} < 1 \quad \text{for} \quad \beta > 0
  \label{eq:mean_energy_after_scatter_wind_behaviour}
\end{equation}
holds for incident photons propagating in the direction of the flow, i.e.\ with
$\mu_{\mathrm{i}} = 1$.  

\begin{figure}[htb]
  \centering
  \includegraphics[width=\textwidth]{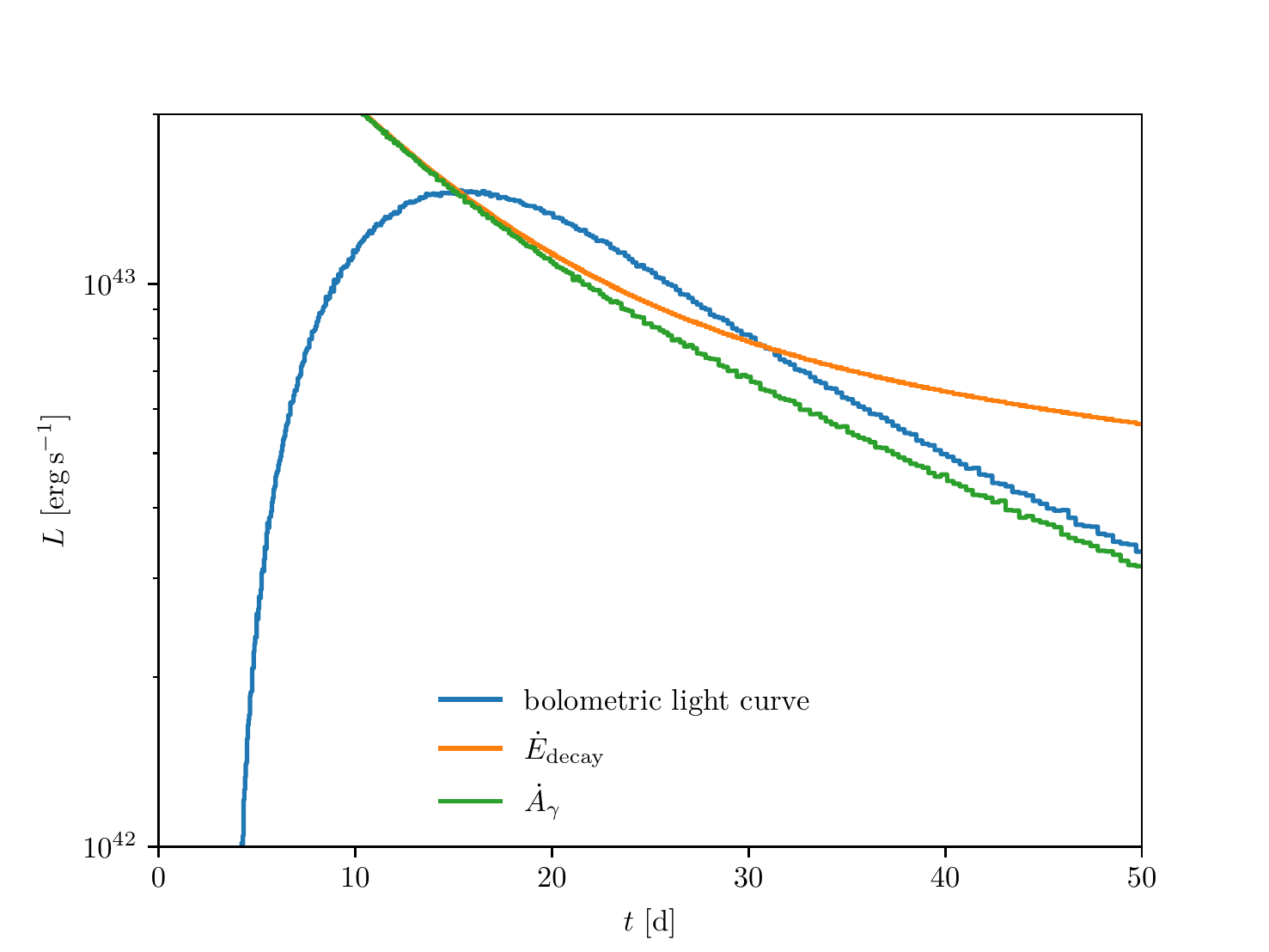}
  \caption{Bolometric light curve obtained with the code \textsc{Mcrh}
  \citep{Noebauer2012} for the test problem devised by \citet{Lucy2005}. In
  addition to the evolution of the radiation emerging from the model \gls{SN}
  (blue), the rate of energy released in the radioactive decay chain of
  $^{56}$Ni (orange) and the rate at which $\gamma$ packets deposit their
  energy in the ultraviolet-optical-infrared radiation field (green) are shown.
  See \Cref{sec:lucy2005_test} for more details on the test problem setup and
the \gls{MCRT} calculation.}
  \label{fig:lucy2005_test_lc}
\end{figure}

\begin{figure}[htb]
  \centering
  \includegraphics[width=\textwidth]{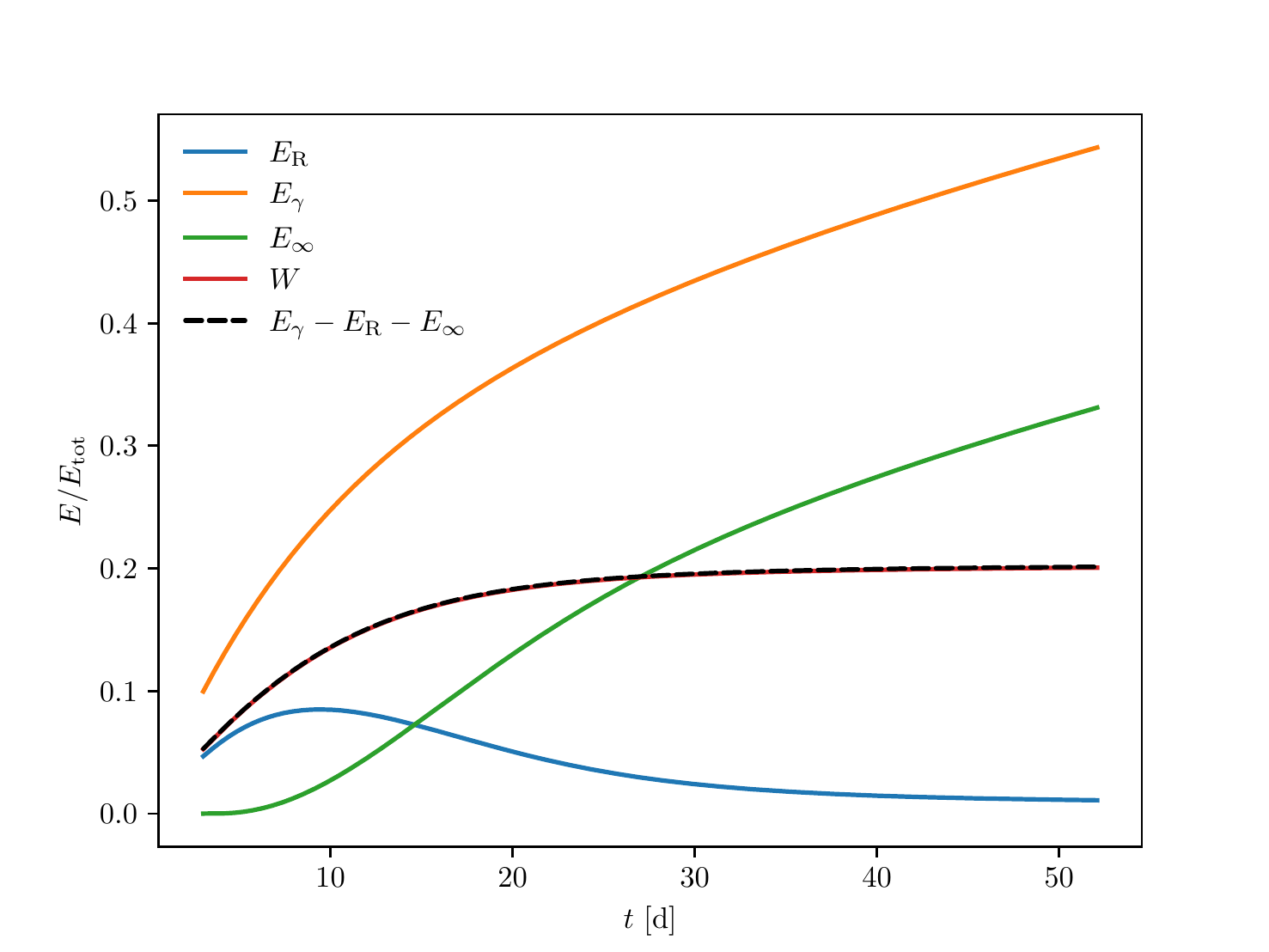}
  \caption{Energy flows in the test problem devised by
    \citet{Lucy2005}. $E_{\gamma}$ shows the energy that has been
    released by radioactive decays, $E_{\mathrm{R}}$ indicated the energy currently
    stored in the radiation field, and $E_{\infty}$ is the energy that
    has escaped to infinity.
In
    addition to the work term, $W(t)$ obtained directly from the \gls{MCRT} simulation
    by balancing the emergent and incident \gls{LF} packet energies in the
    interactions (red), the imbalance between $E_{\gamma}$ and $E_{\infty}$ and
  $E_{\mathrm{R}}$ is shown (dashed black). These two quantities agree very well,
demonstrating the mixed-frame \gls{MCRT} approach captures the expansion work
term and conserves energy.}
  \label{fig:lucy2005_test_eflows}
\end{figure}

As a final illustration for the energy loss experienced by \gls{MC} packets in
\gls{MCRT} calculations in mass outflows, we present the model \gls{SN}
calculation described by \citet{Lucy2005}. This test, which is described in
detail in \Cref{sec:lucy2005_test}, constitutes a simplified and idealised view
of \gls{RT} in \gls{SNIa} ejecta.  We use the code \textsc{Mcrh}
\citep[see][]{Noebauer2012} to perform the \gls{MCRT} simulation for this test
problem as described in \Cref{sec:lucy2005_test}. In
\Cref{fig:lucy2005_test_lc}, the synthetic bolometric light curve from this
test calculation is shown. Following \citet{Lucy2005}, we illustrate the
different energy flow terms in the simulation in
\Cref{fig:lucy2005_test_eflows}. In particular, the energy currently stored in
the radiation field, $E_{\mathrm{R}}(t)$, is shown. In addition, the total energy
that has escaped through the ejecta surface, $E_{\infty}(t)$, the total energy
generated in radioactive decays, $E_{\gamma}(t)$, and the total work performed
by the radiation field, $W(t)$, are given. These three quantities represent
cumulative measures over the time interval from 0 to $t$. All quantities are
simply reconstructed from the \gls{MCRT} simulation by counting appropriate
packet energies. In particular, the work term is obtained by summing up the
difference between incident and emergent packet energies in the \gls{LF} during
each scattering.  In addition, \Cref{fig:lucy2005_test_eflows} contains the
imbalance between the source and sink terms of radiation energy, i.e.\ between
$E_{\gamma}(t)$ on the one hand and $E_{\infty}(t)$ and $E_{\mathrm{R}}(t)$ on
the other hand. This quantity perfectly follows the reconstructed work term,
$W$.
For more technical details about this test, we refer to the original works by
\citet{Lucy2005} and \citet{Noebauer2012}.

\section{Extracting information from MCRT simulations}
\label{Sec:Estimators}

With the algorithms outlined above, the flight paths of \gls{MC} quanta can be
determined and tracked during \gls{MCRT} simulations. In general, the
individual trajectories are not of primary interest. Rather, meaningful
information that effectively represents the radiation field needs to be
extracted from them. In some cases, only radiation escaping from the simulation
box may be of interest to construct synthetic spectra, light curves or images.
For other applications, the most important outcome may be a characterisation of
the radiation field internal to the system. In this part of the review, we
present a number of common approaches that can be used to extract physical
information from \gls{MCRT} simulations. We preface this by a brief discussion
of \gls{MC} noise, which is a fundamental, often undesired property of
\gls{MCRT} simulations that motivates the design of the extraction techniques
described below.

\subsection{MC noise}

\gls{MCRT} simulations are probabilistic by nature.  Consequently, results
obtained with these approaches will generally be subject to stochastic
fluctuations. This fundamental and inherent  property of \gls{MC} calculations
is often referred to as \textit{Monte Carlo noise} or simply
\textit{noise}.
Here, we briefly present the basic behaviour of this noise component and
discuss the implications for devising techniques to extract or reconstruct
physical information from a \gls{MC} simulation. More details about this
subject may be found in the standard literature, e.g.\ in \citet{Kalos2008}.

\begin{figure}[htb]
  \centering
  \includegraphics[width=\textwidth]{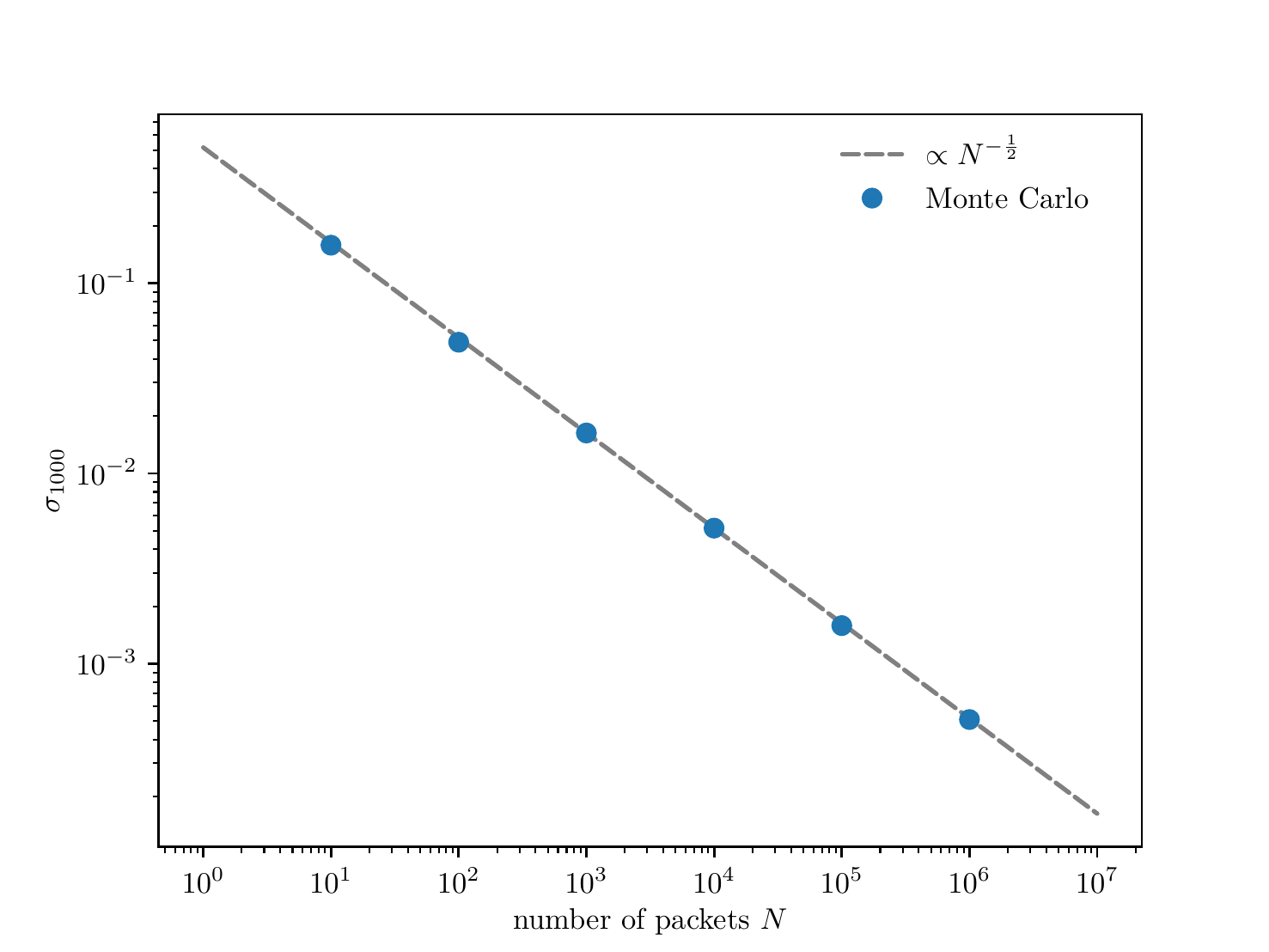}
  \caption{Illustration of the fundamental \gls{MC} noise behaviour,
    \Cref{eq:basic_mc_error}, in case of determining the escape probability
    from a homogeneous sphere (cf.\ \Cref{sec:test_hom_sphere}). In particular
    a sphere with optical depth $\tau_{\mathrm{sp}} = 1$ was considered and the
    escape probability determined for different numbers of \gls{MC} quanta.
    The standard error $\sigma_{1000}$ was determined after repeating each
    experiment 1000 times and follows the expected behaviour almost perfectly.}
  \label{fig:hom_sphere}
\end{figure}

In general, one exploits the law of large numbers when reconstructing physical
information from \gls{MCRT} calculations. To illustrate this, we consider
extracting a specific physical property from the simulation (e.g.\ the escape
  probability from a homogeneous sphere: see \Cref{sec:test_hom_sphere}). This
  quantity has to be related to a particular behaviour or property of the
  \gls{MC} quanta (e.g.\ a packet emerges from the sphere or not). We now
  assume that the process of the quanta performing this behaviour or its
  property taking a specific value is expressed by the random variable $X$ with
  a probability density of $\rho_X(x)$. Considering the fate or measuring a
  property of an individual packet will not result in conclusive statements
  about $X$ (i.e.\ we cannot make meaningful statements about the escape
  probability by considering whether one particular packet emerged from the
sphere or not). However, if this process is repeated and performed for the
entire packet population, the law of large numbers ensures that the resulting
average converges towards the expectation value of $X$\footnote{The law of
  large numbers states that this convergence proceeds almost surely
\citep[cf.][]{Kalos2008}.}.  Consequently, the extraction of physical
information from a \gls{MCRT} calculation can typically be described
mathematically by 
\begin{equation}
  G_N = \frac{1}{N} \sum_{i=1}^N X_i.
  \label{eq:generic_reconstruction}
\end{equation}
To estimate the uncertainty associated with such an \gls{MC} estimate, we rely
on the central limit theorem.\footnote{The applicability of this theorem is not
  a necessity. Qualitatively equivalent estimates can be derived when only
  weaker statements can be made about the random processes \citep{Kalos2008}.}
  According to this statement, the \gls{MC} estimator process $G_N$ will be
  governed by a normal distribution in the limit of infinite contributions ($N
  \rightarrow \infty$) and its standard deviation or standard error will follow
\begin{equation}
  \sigma_G = \frac{\sigma_X}{\sqrt{N}}.
  \label{eq:basic_mc_error}
\end{equation}
Here, $\sigma_X$ denotes the standard error of the individual process $X_i$.
The first natural implication of this fundamental \gls{MC} error behaviour is
that the accuracy improves when the number of contributions to the estimator
given by \Cref{eq:generic_reconstruction} increases. This is typically achieved
by increasing the number of \gls{MC} quanta in the simulation and is
illustrated in \Cref{fig:hom_sphere} by the example of determining the escape
probability for the homogeneous sphere problem (see
\Cref{sec:test_hom_sphere}). 
However, since this improvement scales only with $N^{-1/2}$, efficient and
effective reconstruction techniques strive towards maximising the number of
contributions to the estimator procedure for a given computational effort,
which is often equivalent to considering a certain number of \gls{MC} quanta in
a simulation. These methods are also often referred to as acceleration
techniques since they achieve a certain signal-to-noise level with fewer quanta
and thus by spending less computational effort. In the following, a variety of
such strategies for the extraction of physical information from \gls{MCRT}
calculations and reducing the stochastic fluctuations are presented.  We focus
first on simple counting techniques, then turn to the volume-based estimator
approaches introduced by \citet{Lucy1999} and finally introduce the widely-used
acceleration concept of \textit{biasing}.

\subsection{Direct counting of packets}
\label{Sec-directcounting}

Undoubtedly the most obvious and straight-forward approach to reconstruct a
physical property of the radiation field (or any associated \gls{RT} process)
from the ensemble of packet trajectories is to simply count the relevant packet
properties or packet interaction events. For example, a simple synthetic image
of an astrophysical system can be produced by recording all packets emerging
from the domain through the surface element $\Delta S_i$ during a time interval
$\Delta t$ then binning them according to their propagation directions
\begin{equation}
  L(\Delta S_i, \Delta \Omega_j) = \frac{1}{\Delta t} \sum_k \varepsilon_k.
  \label{eq:rec:direct_counting_image}
\end{equation}
Here, the summation only involves packets that escape through the surface
element $\Delta S_i$ into the solid angle element $\Delta \Omega_i$. 

Likewise, internal properties can be reconstructed by querying the positions of
all packets at a given instant during a time-dependent simulation and then
summing up the relevant properties of all packets that are currently located
within a certain control volume $\Delta V$. In particular, the radiation field
energy density in a grid cell $i$ with the volume $\Delta V_i$ can be
determined by
\begin{equation}
  E_i = \frac{1}{\Delta V_i} \sum_j \varepsilon_j,
  \label{eq:rec:direct_counting_E}
\end{equation}
following this strategy. Here, the summation over $j$ involves all packets
that at $t=t^n$ are within cell $i$. By analogy, any quantity that involves
radiation--matter interactions, such as the amount of absorbed radiant energy,
can be determined by counting all absorptions packets perform during a certain
time interval (e.g.\ the duration of a simulation time step).  While this
reconstruction approach is very intuitive, it is also the least sophisticated
and does not optimally use the information contained in \gls{MCRT} simulations.
In general, a large number of packets will be needed to achieve acceptable
results since the approach requires that a sufficient number of packets
propagate into the desired direction, are at a certain location or have
performed a particular interaction (or combination of all these).  Fulfilling
this requirement becomes even more challenging when fully three-dimensional
and/or frequency-dependent calculations are performed. As a consequence, the
utility of the direct packet counting technique is often limited due to strong
noise in the reconstructed quantities.  Still, this approach can be of use as a
reference when testing and verifying more complex reconstruction techniques.
Moreover, the quality of direct counting estimates can be vastly improved when
combined with biasing techniques, which will be introduced below. 

\subsection{Volume-based estimators}
\label{Sec:volume_estimators}

\citet{Lucy1999} introduced a technique to reconstruct properties of the
internal radiation field that is less vulnerable to noise than direct counting
approaches since information along the entire packet trajectory is used instead
of only a momentary snapshot of the packet distribution. These techniques have
been refined by \citet{Lucy1999a, Lucy2003, Lucy2005} and are often referred to
as {\it volume-based} estimators\footnote{\citet{Och1998} presented reconstruction
schemes which use control surfaces.}. The effective use of such estimators has
been a key consideration for many \gls{MCRT} studies relying on Lucy's approach
\citep[e.g.][]{Sim2004, Kasen2006, Sim2007, Kromer2009, Harries2011,
Noebauer2012, Kerzendorf2014, Harries2015}

The volume-based estimator approach rests on the idea that instead of
considering packets at certain discrete instances, time-averaged estimates of
radiation field properties can be constructed by incorporating information from
the full packet propagation path. The fundamental notion is that the packet
flight histories form an ensemble of trajectory elements that statistically
represent the radiation field.  To better illustrate this principle, we follow
\citet{Lucy1999} and repeat the formulation of a volume-based estimator for the
radiation field energy density. 

\subsubsection{Example: Formulation of volume-based estimator for the radiation
energy density.}

We consider the trajectory of a packet with energy $\varepsilon$ propagating
during a simulation time interval of $\Delta t$\footnote{For a time-dependent
  calculation, the appropriate $\Delta t$ will be the duration of the current
  time step. In a time-independent/steady-state calculation, it will be the
implied length of the time interval being simulated.}.  In general, this
trajectory will consist of multiple separate segments that correspond to the
flight of the packet between events in the simulation (i.e., neglecting general
relativistic light bending, the photon packet trajectory will be a sequence of
straight line segments between scattering/interaction points, cell boundary
crossing events etc.).  We denote the time the energy packet spends on any
particular segment $j$ by $\delta t_j$.  Each packet trajectory segment
contributes to the total radiation energy content with its packet energy,
weighted by the relative time spent on that trajectory: i.e.\ $\Delta E =
\frac{\delta t_j}{\Delta t} \varepsilon_j$.  Thus the implied total energy
density for a grid cell $i$ of volume $\Delta V_i$ in a simulation may be
constructed from a volume-based estimator obtained by summing over all
trajectory elements of all packets that were active in the cell:
\begin{equation}
  E_i = \frac{1}{\Delta V_i} \sum_{j \in \Delta V_i} \frac{\delta t_j}{\Delta t} \varepsilon_j.
  \label{eq:rec:vbase_est_E}
\end{equation}
As packets propagate at the speed of light, the estimator can be expressed in
terms of trajectory segment length $l = c \delta t$
\begin{equation}
  E_i = \frac{1}{\Delta t c \Delta V_i} \sum_{j \in \Delta V_i} l_j \varepsilon_j.
  \label{eq:rec:vbase_est_E_l}
\end{equation}
Here and in \Cref{eq:rec:vbase_est_E}, the summation includes the trajectory
segments $j$ of all packets that lie within the cell $i$.  We stress that the
ensemble of segments contributing to these sums includes all packet trajectories
between events, both physical interactions, like scatterings, and numerical
events, such as grid cell boundary crossings. We also note that, although our
presentation derives from a time-based formulation, the resulting estimator
depends only on the ratio $\varepsilon_j / \Delta t$ and so can be applied
without adjustment to steady-state \gls{RT} problems (i.e.\ where time steps
need not explicitly appear in the algorithm). In such cases the problem will
involve a fixed luminosity, and the packet energies will be normalised to
correspond to a pre-determined or arbitrary time interval. In
\Cref{eq:rec:vbase_est_E}, the choice of this normalisation time interval will
be inconsequential: the value of the estimator will depend only on the ratio of
the packet energies to the duration of the simulation time interval (i.e.\ sensitive to the luminosity
but neither to the absolute packet energies nor absolute time interval). 

The advantage of the volume-based estimator scheme compared to simple direct
counting measurements is two-fold. First, a single packet can contribute to the
estimators in multiple cells, provided that its trajectory intersects these
cells during the time step. Second, the same packet can in principle contribute
repeatedly to the estimator in a specific cell, if it is scattered in the cell
or backscattered from a different cell. Both features of the volume-based
estimator scheme drastically increase statistics and thus reduce the amount of
\gls{MC} noise in the reconstructed quantity. Also, this technique reduces the
risk of obtaining undetermined results. In the direct counting approach, at
least one packet must reside in the cell at the instant considered to obtain a
non-zero result. This condition is mitigated to the much less restrictive
requirement that at least one packet has resided in the cell at any point
during the time step.

\subsubsection{Constructing volume-based estimators: radiation field quantities}
\label{sec-volumeestimators}

Having established a volume-based estimator reconstruction scheme for $E$ and
having appreciated the benefits such an approach offers, other radiation field
properties can be determined in a similar manner. For this purpose, the
relationship
\begin{equation}
  E = \frac{4 \pi}{c} J
  \label{eq:rec:E_J_rel}
\end{equation}
can be used together with \Cref{eq:rec:vbase_est_E_l} to obtain an
estimator for the mean intensity \citep[cf.][]{Lucy1999}
\begin{equation}
  J_i = \frac{1}{4 \pi \Delta V_i \Delta t} \sum_{j \in \Delta V_i} l_j \varepsilon_j.
  \label{eq:rec:vbased_est_J}
\end{equation}
Once the summation is restricted to contributions of trajectory segments which
point into a certain solid angle element $\Delta \Omega_k$, 
\begin{equation}
  I_{i,k} \Delta \Omega_k = \frac{1}{4 \pi\Delta V_i \Delta t} \sum_{j \in \Delta V_i, \Delta \Omega_k} l_j \varepsilon_j,
  \label{eq:rec:vbased_est_I}
\end{equation}
the specific intensity can be reconstructed by means of a volume-based
estimator \citep[see e.g.][]{Lucy2005}. Similarly, a further restriction to
segments of packets with a frequency in the interval $[\nu, \nu + \Delta \nu]$,
allows monochromatic radiation field properties to be determined. For
example the monochromatic specific intensity
\begin{equation}
  I_{i,k,p} \Delta \Omega_k \Delta \nu_p = \frac{1}{4 \pi \Delta V_i \Delta t} \sum_{j \in \Delta V_i, \Delta \Omega_k, \Delta \nu_p} l_j \varepsilon_j.
  \label{eq:rec:vbased_est_I_nu}
\end{equation}
Volume-based estimators for moments of the specific intensity ($J$, $H$ and
$K$) can now be easily formulated by including powers of the propagation
direction and summing over all directions. Following this procedure, for
example the total radiation flux can be estimated with \citep[see
e.g.][]{Noebauer2012}
\begin{equation}
  \mathbf{F}_i = \frac{1}{\Delta V_i \Delta t} \sum_{j \in \Delta V_i} \mathbf{n}_j l_j \varepsilon_j.
  \label{eq:rec:vbased_est_F}
\end{equation}

\begin{figure}[htb]
  \centering
  \includegraphics[width=\textwidth]{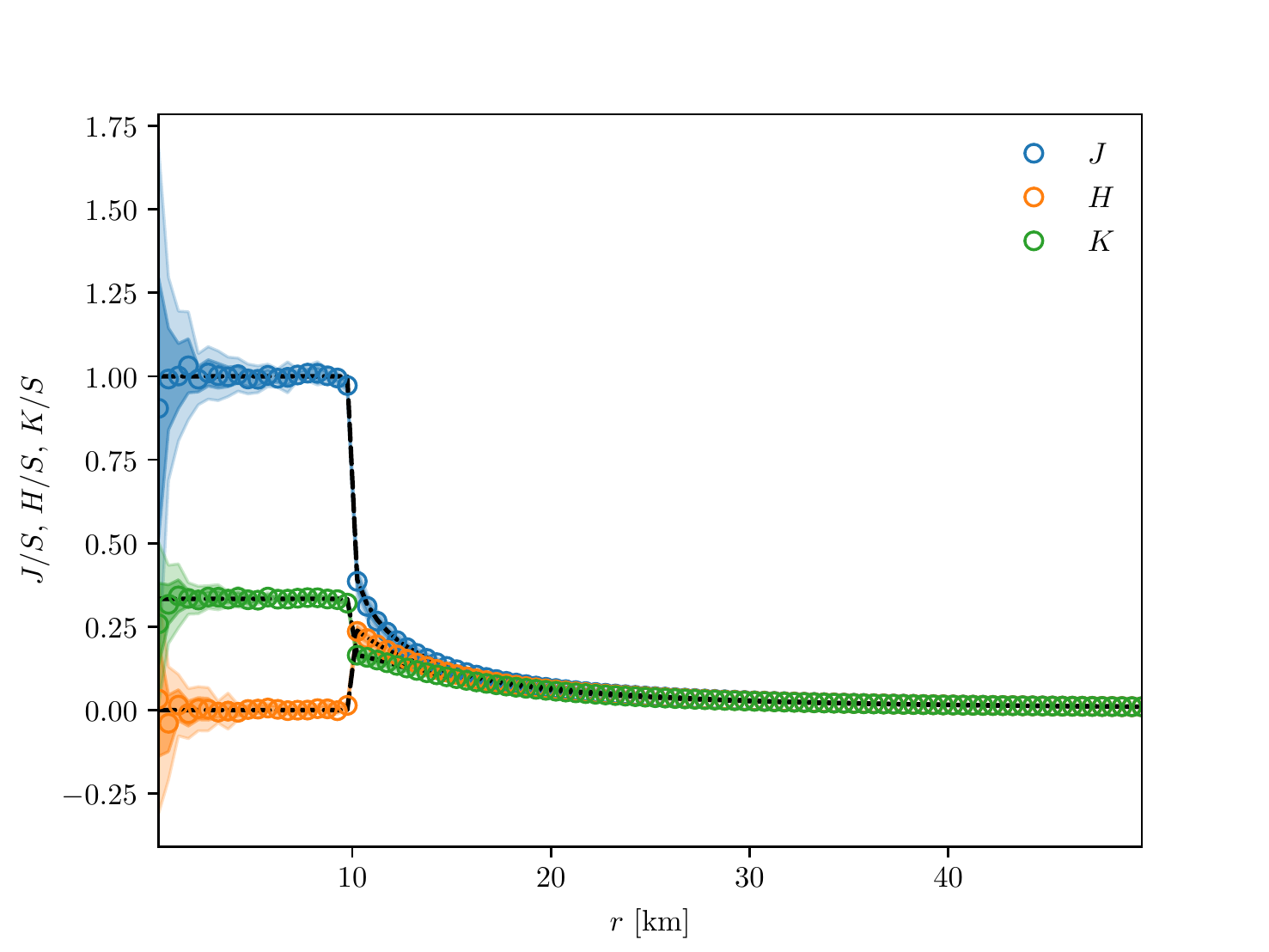}
  \caption{Volume-based estimators for $J$, $H$, and $K$ in the homogeneous
    sphere test (see \Cref{sec:test_hom_sphere}). These are shown
    relative to the adopted source function, $S$, which is assumed to
    be uniform throughout the sphere (see \Cref{sec:test_hom_sphere}). The $1\,\sigma$ and
    $2\,\sigma$ confidence bands (dark and lightly-shaded regions) have been
    constructed by performing ten \gls{MCRT} simulations with $10^5$ packets
    each and different \gls{RNG} seeds. The open symbols represent the mean
    values for the radiation field moments in the different shells.
    Additionally, the analytic solution according to
    \Crefrange{eq:hom_sphere_J_inside}{eq:hom_sphere_K_outside} is included in
  dashed black. The Python tool with which these \gls{MCRT} simulations have
been performed is part of the source code repository distributed with this work
(see \Cref{sec:tool_collection}).}
  \label{fig:homogeneous_sphere_est}
\end{figure}

To demonstrate the capabilities of the volume-based estimator approach to
accurately track the characteristics of the radiation field, we perform a
\gls{MCRT} test simulation of the homogeneous sphere problem presented in
\Cref{sec:test_hom_sphere}. Adopting the parameters suggested by
\citet{Abdikamalov2012} and listed in \Cref{sec:test_hom_sphere}, we perform a
simple time-independent \gls{MCRT} simulation in spherical symmetry, injecting
packets according to the local emissivity and following them until they either
escape from the computational domain or are absorbed. During their propagation
paths, volume-based estimators for $J$, $H$, and $K$ are continuously
incremented. These first three moments of the specific intensity are shown in
\Cref{fig:homogeneous_sphere_est}.
Within the inherent \gls{MC} noise (indicated by uncertainty
  bands\footnote{Note that the increase in uncertainty in the inner regions is
    simply a consequence of the numerical setup. Since the grid has been chosen
    equidistant in $r$ and since the packets carry all the same energy
$\varepsilon$, fewer packets are spawned in inner regions.}), the estimators
agree very well with the analytic solution as outlined in
\Cref{sec:test_hom_sphere}. An example implementation of how the reconstruction
of the moments may be achieved within the volume-based estimator formalism can
be found in the Python program designed for this test problem and distributed
with the tools repository (cf.\ \Cref{sec:tool_collection}). Specifically,
this task is performed by the routine \texttt{update\_estimators} in the
\texttt{mcrt\_hom\_sphere.py} program. Note that due to spherical geometry, not
the instantaneous value of the direction cosine but its mean value along the
trajectory segment (\texttt{mu\_mean}) appears in the estimator increments.

\subsubsection{Constructing volume-based estimators: extracting physical rates}

For many problems, simulation of the radiation field serves not only to predict
synthetic observables but also to determine thermodynamic conditions of the
astrophysical plasma: e.g., often the radiation field is crucial for
determining the ionization/excitation state and heating
\citep[e.g.][]{Mazzali1993, Bjorkman2001, Long2002, Ercolano2003, Ercolano2005,
Ercolano2008}. In such cases, we therefore wish to extract information on the
relevant rates of physical processes in the simulations. Following the
principle outlined in Section~\ref{Sec-directcounting}, this could be achieved
simply by counting the rate at which individual packet events corresponding to
the process in question occur during the simulation. However, such an approach
relies on a sufficient number of such interactions happening to achieve
acceptable statistics and an accurate result. This becomes very challenging in
optically thin regions since very few packets or even none interact. 

Again, the volume-based estimator approach offers a significant improvement
since it takes a broader view and includes the information encoded in the
entire packet propagation paths instead of only considering a series of
isolated snapshots.  In particular, a volume-based estimator can be formulated
for any quantity that depends on the radiation field by applying constructions
similar to those outlined in Section~\ref{Sec:volume_estimators}. The general
principle will be that the rate of energy extracted from the radiation field by
some process can be described in terms of a sum over packet trajectories
weighted by the appropriate absorption coefficient. These energy flow rates can
then be recast in other forms (e.g.\ transitions rates), as required. 

\subsubsection{Example: photoionization rate estimators}

As a concrete example, we illustrate the case of extracting a particular
photoionization rate from a simulation. In particular, the photoionization rate
coefficient\footnote{$\gamma$ gives the number of photoionization events per
second per unit volume per photoionization target atom/ion.} can be written

\begin{equation}
\gamma = 4\pi \int_{\nu_{\mathrm{th}}}^{\infty}\frac{\sigma(\nu)}{h \nu} J_{\nu} \; \rm{d} \nu
\end{equation}
where $\nu_{\mathrm{th}}$ is the threshold frequency and $\sigma_{\nu}$ the cross section
for photoionization at photon frequency $\nu$. Into this expression we
substitute our expression for the MC volume-based estimator for the relevant
moment of the radiation field, in this case \Cref{eq:rec:vbased_est_J}, and
immediately obtain our estimator for $\gamma$
\begin{equation}
\gamma = \frac{1}{\Delta V_i \Delta t} \sum_{j \in \Delta V_i, \nu > \nu_{\mathrm{th}}} \frac{\sigma(\nu)}{h \nu} l_j \varepsilon_j.
\end{equation}
We note that the estimator runs over all packet trajectories on which the
packet frequency is above the photoionization edge ($\nu > \nu_{\mathrm{th}}$) and each
contribution to the sum is multiplied by a factor that depends on the
cross-section of the process for the contributing packet.

\begin{figure}[htb]
  \centering
  \includegraphics[width=\textwidth]{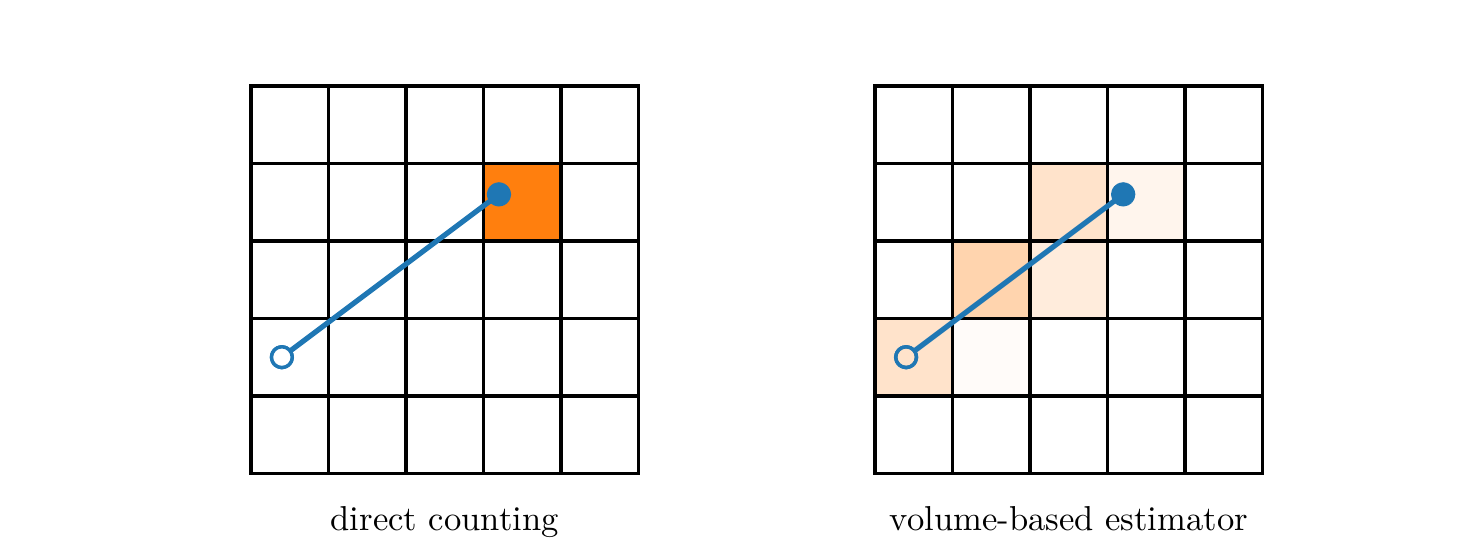}
  \caption{Illustration of the benefits of the volume-based estimator approach
    (right panel) compared to direct counting techniques (left panel). The
    trajectory of a single packet is shown (blue), which is absorbed at the
    location denoted by a filled circle after passing trough a number of cells.
    While in the direct counting approach this packet only contributes to the
    heating rate estimator in the final grid cell, it adds to the local heating
    rate in all cells it passed through in the volume-based estimator scheme.
  The relative contribution to the heating rate in each cell is encoded by the
color transparency in this sketch.}
  \label{fig:rec:vbase_est_illustration}
\end{figure}

With the possibility that all packet propagation segments can in principle
contribute to the estimator, the reconstruction of interaction-based radiation
field properties yields non-zero results as long as at least one packet entered
the grid cell of interest. This significant advantage of the volume-based
estimator approach (compared to estimating rates of processes by direct
counting) is illustrated in \Cref{fig:rec:vbase_est_illustration}.

\subsubsection{Volume-based estimators for energy and momentum flow}

The principle outlined above, and illustrated by the photoionization example is
readily generalised to provide estimators for the rate of any physical process
of interest (see e.g.\ \citealt{Lucy2003}) that can be cast in terms of energy
transfer from the radiation field. The principle is always the same: the
optical depth of each segment is calculated which can be interpreted as the
expected number of interactions a packet would on average experience when
propagating this distance. This information is then used to scale the
contribution of each segment to the estimator and determine the amount of
radiant energy that is absorbed by the process. Of particular relevance to many
problems, including radiation hydrodynamics, is the 
total rate at which energy is transferred from the radiation field into the 
ambient material, i.e.\ the heating rate. If the heating process is for example 
described by a pure absorption coefficient $\chi$, then the amount of radiant 
energy absorbed is \citep[see][]{Lucy1999} 
\begin{equation}
  \Delta \dot E_i = \frac{1}{\Delta V_i c \Delta t} \sum_{j \in \Delta V_i} \chi l_j \varepsilon_j \; .
  \label{eq:rec:vbased_est_dotE}
\end{equation}
We note that the form of this estimator is very similar to that used to
reconstruct $J$ itself, \Cref{eq:rec:vbased_est_J}, which is to be expected
given the close relationship of $J$ to the rate of any radiative heating
processes.

This idea also readily generalises to provide volume-based estimators for
momentum transfer \citep[see e.g.][]{Noebauer2012, Roth2015}, which are
instrumental for \gls{MC}-based \gls{RH} calculations (see
\Cref{Sec:Dynamics}). For continuum driving, the form of these estimators is
very similar to the $F$ estimator, \Cref{eq:rec:vbased_est_F}.

\citet{Lucy1999a} used similar considerations for the formulation of estimators
in applications which are dominated by atomic line interactions that are
treated in the Sobolev limit, such as stellar winds or the ejecta of
thermonuclear \glspl{SN}.  Here, the formulation is slightly more complicated
and the form of the energy/momentum flow rate estimators is different from
those outlined above: they are formed as summations over all packets that have
come into Sobolev resonance within a grid cell \citep[see also][]{Sim2004,
Noebauer2015a}. The principal advantage compared to direct counting still
applies since all resonances contribute, regardless of whether the
packet actually undergoes interaction. Given the potential importance of
forests of weak lines to heating/driving of outflows, as is for example the
case in hot star winds where many weak iron lines drive the outflow
\citep{Vink1999}, this is a critical advantage.

\subsection{Biasing}
\label{Sec:biasing}

In many \gls{MCRT} applications, only a subset of the packet population is of
interest. For example, when creating a synthetic image, only packets that
escape towards the virtual observer are relevant. It is therefore desirable to
selectively invest more effort into propagating packets that are crucial for
the determination of the quantity or process of interest instead of treating
packets that do not contribute.  This selective
increase in statistics can be achieved with the help of so-called
\textit{biasing} techniques. The underlying basic principle is known as
\textit{importance sampling} in the field of \gls{MC} integration. 

The key concept of biasing techniques is to increase the
occurrence of desired packet properties by introducing a new, biasing
\gls{PDF}, $q(x)$, that emphasises the corresponding regions in the parameter
space. We then sample from this \gls{PDF} rather than from the physical one,
$\rho(x)$, in the random processes governing the \gls{MCRT} procedure. In order
to ensure physical consistency, the packet weights have to be adjusted to
counterbalance the artificial over- (and under-) representation of samples from
particular parameter space regions. In particular, the packet weight in the
absence of biasing, $w_{\mathrm{nb}}$, is replaced by 
\begin{equation}
  w(x) = \frac{\rho(x)}{q(x)} w_{\mathrm{nb}}.
  \label{eq:bias:weight}
\end{equation}

Biasing techniques are among the  most popular and widely-used variance
reduction methods \citep[see e.g.][]{Carter1975, Dupree2002}.  In the
following, some of the popular biasing techniques used in astrophysical
applications are briefly described. Most of the presented schemes are designed
to address challenges commonly encountered in dust \gls{RT}, since biasing
techniques are heavily used in this branch of astrophysical research \citep[see
e.g.\ the overview by][]{Steinacker2013}. 

\subsubsection{Biased emission}

\textit{Biased emission} is a simple but powerful illustration of a biasing
scheme. This approach helps in problems where we wish to accurately describe
the emission from sources with very different emissivities. These can be
external sources, such as stars irradiating some environment, or simply the
internal emissivity of the ambient material occupying grid cells of a
computational mesh. Biased emission is frequently used in dust \gls{RT}, for
example by \citet{Yusef-Zadeh1984} and \citet{Juvela2005}. A detailed account of
the technique is given by \citet{Baes2016}. 

For illustrative purposes, we consider a problem that only involves two sources,
with luminosities $L_1$ and $L_2$, and wish to study the case where $L_1 \ll
L_2$.  One approach to simulate the emission, is to spawn $N$ \gls{MC} packets,
each with equal energy (i.e.\ weight)
\begin{equation}
  w = \varepsilon = \frac{L \Delta t}{N}.
  \label{eq:bias:emission_weight}
\end{equation}
Here, the total luminosity ($L = L_1 + L_2$) and the physical duration
corresponding to the \gls{MCRT} simulation ($\Delta t$) appear. Each of these
packets is now associated with one of the two sources according to the discrete
probabilities
\begin{equation}
  p_i = \frac{L_i}{L} \quad \text{for }i\in [1, 2].
  \label{eq:bias:emission_discrete_prob}
\end{equation}
For $L_1 \ll L_2$, this leads to a very uneven distribution of packets, with
the weaker source being represented by very few packets. In the biased emission
approach, an alternative \gls{PDF} is introduced that increases the association
with the weaker source. One possibility would be to choose the uniform
distribution 
\begin{equation}
  p_1 = p_2 = \frac{1}{2}.
  \label{eq:bias:biased_emission_discrete_prob}
\end{equation}
In this case, equal numbers of packets represent the emissions from both
sources. This has to be balanced by adjusting the packet weights according to
\Cref{eq:bias:weight}, leading to packets from the first source representing
less energy
\begin{equation}
  w_1 = 2 \frac{L_1}{L} \varepsilon  = 2 \frac{L_1 \Delta t}{N},
  \label{eq:bias:biased_emission_first_weight}
\end{equation}
and packets from the second source more energy
\begin{equation}
  w_2 = 2 \frac{L_2}{L} \varepsilon  = 2 \frac{L_2 \Delta t}{N}
  \label{eq:bias:biased_emission_second_weight}
\end{equation}
than in the unbiased case. In addition to addressing imbalances in source
luminosities, biased emission often involves preferentially launching packets
into directions of particular interest \citep[cf.][]{Baes2016}.

\subsubsection{Forced scattering}

A well-established biasing technique, already discussed in the context of
neutron transport by \citet{Cashwell1957}, is the \textit{forced scattering}
scheme. This method is often used in dust \gls{RT} applications
\citep[see e.g.][]{Mattila1970, Witt1977, Steinacker2013, Baes2011, Baes2016}
to increase the efficiency in optically thin regions. Here, packets would
otherwise escape without interacting leading to a low dust-scattering
efficiency and challenges in determining heating rates. To circumvent these
difficulties, the interaction probability for packets is biased such that they
are guaranteed to interact before reaching the edge of the computational
domain.  Denoting the optical depth from the current packet position to the
point of escape as $\tau_{\mathrm{edge}}$, the interaction location is drawn
from the biased \gls{PDF} \citep[cf.][]{Steinacker2013}
\begin{align}
  q(\tau)\mathrm{d}\tau = \begin{cases} \frac{\exp(-\tau)}{1 -
      \exp(-\tau_{\mathrm{edge}})} \mathrm{d}\tau&\quad \text{if } \tau \le
    \tau_{\mathrm{edge}}\\0 &\quad \text{if } \tau > \tau_{\mathrm{edge}}\end{cases}
  \label{eq:bias:forced_scattering_pdf}
\end{align}
instead of using \Cref{eq:beer_lambert_sampling}. This ensures that the
interaction location lies between 0 and $\tau_{\mathrm{edge}}$.  Following
\Cref{eq:bias:weight}, the packet weight is modified by  
\begin{equation}
  w(\tau) =  1 - \exp(-\tau_{\mathrm{edge}}).
  \label{eq:bias:forced_scattering_weight}
\end{equation}
If continuously applied in time-independent applications without an absorption
component, this scheme allows packets to propagate indefinitely, with a
continuously decreasing weight. Thus, a termination mechanism (e.g.\ Russian
Roulette, see below) has to be introduced to stop the propagation at a certain
point, typically once the packet weight has dropped below some pre-defined
threshold. Alternatively, forced scattering can also only be applied once for
each packet thus ensuring at least one interaction but leaving the normal
propagation termination mechanism (escape through domain edge) intact. We note,
however, that particular care has to be taken when information is extracted
from \gls{MCRT} simulations employing this technique that also require the
contribution from free-streaming packets.

\subsubsection{Peel-off}

Constructing properties of the emerging radiation field by simply examining the
properties of escaping packets often yields unsatisfactory results,
particularly in multidimensional simulations: typically only a small fraction
of the packet population escapes towards the observer meaning that the
reconstruction will suffer from strong noise. Here, the so-called
\textit{peel-off} technique (sometimes also referred to as \textit{next event
estimate}) helps \citep[e.g.][]{Yusef-Zadeh1984, Wood1999, Baes2011,
Steinacker2013, Lee2017}. In the context of \gls{MCRT} in fast mass outflows,
this method is sometimes referred to as \textit{viewpoint technique} or
\textit{virtual packet} scheme \citep{Woods1991, Knigge1995, Long2002,
Kerzendorf2014, Bulla2015}.  

The peel-off approach introduces ray tracing concepts into the \gls{MC}
simulation. At every interaction point in the simulation, the probability is
calculated that the interaction could have given rise to a packet that
propagated in the direction of the observer and successfully emerged from the
simulation (and so could contribute to the synthetic observables).
Specifically, the weight contributed to the synthetic observables associated
with the interaction of a packet with weight $w$ can be written
\begin{equation}
  w_{\mathrm{obs}} = w \; p(\mathbf{n}_{\mathrm{obs}}) \exp(-\tau_{\mathrm{obs}}).
  \label{eq:bias:peel_off_weights}
\end{equation}
where $p(\mathbf{n}_{\mathrm{obs}})$ is the probability that the interaction
led to reemission of the packet in the direction of the observer
($\mathbf{n}_{\mathrm{obs}}$) and $\exp(-\tau_{\mathrm{obs}})$ describes the
attenuation of the packet as it travels through the total optical depth from
the interaction point to the observer ($\tau_{\mathrm{obs}}$).  The optical
depth is obtained by casting a ray towards the observer and integrating the
opacity along this path. 

Since every interaction any packet performs contributes to the reconstruction,
the improvement in statistics in peel-off methods is substantial. However, the
ray tracing exercise of the peel-off technique adds significantly to the
overall computational effort of the \gls{MCRT} calculation, sometimes even
dominating the computational costs.

We note that variants of methods similar to peel-off have been used in specific
applications.  \citet{Lucy1991, Lucy1999a} introduced a ray tracing technique
for variance reduction specifically designed for applications in which a
photosphere approximation can be adopted and in which the medium is freely
expanding, e.g.\ \gls{SN} ejecta. During the \gls{MCRT} simulation the source
function is reconstructed from the packet interaction histories and then used
in a formal integration step to calculate the emergent radiation field along
cast rays. By relying on this technique, virtually noise-free spectra can be
determined.\footnote{Note, however, that the resulting spectrum will still vary
  once a different \gls{RNG} seed is chosen since the source function used in
the formal integration is determined within a \gls{MCRT} simulation.} Also, as
described by \cite{Bulla2015}, the peel-off technique can be applied
not only to
interaction events but instead to all \gls{MC} packet trajectory elements.
Here, packets can contribute to the synthetic observation even when no
interactions occur: the synthetic observables are obtained by a sum over
contributions from all packet trajectories weighted similar to
\Cref{eq:bias:peel_off_weights} but including an additional multiplicative
term that gives the probability that an interaction event {\it could} have
happened along the trajectory element.

\subsubsection{Further biasing techniques}

In addition to the schemes outlined so far, a variety of other biasing
techniques have been developed and are actively used. Among them are, for
example, techniques called \textit{path length stretching} \citep{Baes2016},
\textit{continuous absorption} \citep[known also as \textit{packet splitting} or
\textit{survival biasing}][]{Carter1975, Steinacker2013, Lee2017} or
\textit{polychromatism} \citep{Jonsson2006, Steinacker2013}. We refer the
reader to the literature for example to the review by \citet{Steinacker2013}
and the book by \citet{Dupree2002} for detailed accounts.  

\subsubsection{Limitations -- Russian Roulette and composite biasing}

Naturally, biasing techniques are not a universal remedy and are also afflicted
by drawbacks.  Here, we highlight some of the more severe limitations and
discuss techniques that have been proposed and developed to alleviate them. 

By design, the increase in statistics in some regions of the parameter space
comes at a cost, namely the decrease of statistics in other regions.
Specifically, the loss of statistics occurs in regions where the biased
\gls{PDF} $q(x)$ is smaller than the original one, $\rho(x)$. This has the
immediate consequence that biasing techniques should be only used if the loss
in statistics happens for packets that are not relevant for the result one is
interested in.  In addition, the packets associated with draws from these
regions experience an increase in their weight, which in principle is unbound.
This poses numerical problems, since a few high-weight packets may then
dominate the \gls{MC} noise. To alleviate this deficit of biasing approaches, a
technique called \textit{composite biasing} has been proposed \citep{Baes2016}.
Here, samples are drawn from a linear combination of the biased and the original
\gls{PDF}:
\begin{equation}
  q^{\star}(x) = (1 - \zeta) \rho(x) + \zeta q(x).
  \label{eq:bias:composite_biasing_pdf}
\end{equation}
As a consequence, the weight adjustment
\begin{equation}
  w^{\star} = \frac{\rho(x)}{q^{\star}(x)} = \frac{1}{(1 - \zeta) + \zeta q(x)/ \rho(x)}
  \label{eq:bias:composite_biasing_weight}
\end{equation}
is limited to 
\begin{equation}
  w^{\star} < \frac{1}{1 - \zeta}.
  \label{eq:bias:composite_biasing_weight_limit}
\end{equation}
If, for example $\zeta = 1/2$, is chosen (as suggested by \citealt{Baes2016}),
the weight increase can at most be a factor of two. Note, however, that in some
biasing techniques (e.g.\ packet splitting) weights are potentially modified
repeatedly. Then, composite biasing limits each consecutive weight change but
as a consequence of the continuous application of biasing, the weights can in
principle still become very large. 

When applying biasing techniques that can act multiple times on the same
packet, also small packet weights can become a hindrance. Packets with very
small weights only contribute insignificantly to the reconstructed property but
roughly the same computational effort has to be invested to follow their
propagation as for important packets. Based on this cost-benefit argument, it
is advisable to terminate the propagation once the weight and thus
importance of a packet has decreased beyond some predefined threshold. In this
context, the so-called \textit{Russian Roulette} method provides a stochastic
framework to remove low-weight packets from the simulation, while still
retaining energy conservation in a statistical sense \citep[see
e.g.][]{Carter1975, Dupree2002}.  In its simplest form, a termination
probability $p_T$ is defined. Whenever a packet enters the roulette, the
termination probability is sampled and the packet propagation is terminated if
the sampling outcome is positive.  Otherwise, the packet survives and its
weight increases to $w / p_T$. This way, the weights of the terminated packets
are distributed probabilistically onto the surviving ones and energy/weight
conservation is ensured statistically.  A detailed description of the Russian
Roulette technique, and more sophisticated realisations, is given by
\citet[][]{Dupree2002}.

\section{Implicit and diffusion Monte Carlo techniques}
\label{Sec:imc_and_ddmc}

Conventional \gls{MCRT} methods, built upon the techniques outlined so far,
inherently rely on explicitly tracking packet flight paths. Although this has a
range of compelling benefits, not least the conceptual ease with which it can
be developed, it has limitations particularly in regard to efficiency for many
applications.  For example, \gls{MCRT} calculations become prohibitively slow
when applied in optically thick media since the number of physical and
numerical events that has to be explicitly tracked increase drastically.
Another challenge is posed in \gls{TRT} applications where successive
absorptions and re-emissions occur frequently. Achieving a stable and accurate
solution of the evolution of the ambient medium and of the radiation field
typically requires a drastic reduction of the size of the physical time step.
In the following, we outline a number of developments and techniques that have
been proposed and are actively used to alleviate these shortcomings.

\subsection{Implicit Monte Carlo}
\label{Sec:imc}

Standard explicit \gls{MC} techniques face challenges when dealing with
\gls{TRT} problems since these involve a rapid succession of absorption and
emission processes. In this situation sufficiently short time steps have to be
used so that the ambient conditions (temperature etc.) can properly react to
absorption--emission imbalances. Otherwise, the radiation source term may
deplete the internal energy reservoir of the ambient material between
successive temperature updates and lead to unphysical conditions (e.g. negative
temperatures). 

These difficulties are addressed by the so-called \textit{Implicit
  Monte Carlo} (\gls{IMC}) method, introduced
in the seminal work by \citet{Fleck1971}. Here, the sequence of absorption and
emission events is replaced by an effective scattering prescription and only
the net imbalance remains as a true absorption and emission contribution.
Despite the name, the \gls{IMC} method does not constitute a truly implicit
solution approach, comparable to techniques encountered in the field of solving
differential equations. Instead, a semi-implicit recasting of the discretized
\gls{RT} equation is performed. This procedure leads to the main advantage of
the \gls{IMC} approach, namely the introduction of unconditional stability. In
the following, we briefly outline the formulation of the \gls{IMC} technique
and discuss some important properties of this approach. For an in-depth
discussion of the method, we refer to the original work by \citet{Fleck1971}
and to the recent detailed review by \citet{Wollaber2016} on the subject.

Following \citet{Fleck1971}, we introduce the \gls{IMC} approach for the
example of the one-dimensional grey \gls{TRT} problem in the absence of
scattering interactions. The derived equations can be easily modified to
account for scatterings and a generalisation to frequency-dependent and
multidimensional problems is possible.\footnote{Indeed, \citet{Fleck1971}
introduce the \gls{IMC} approach both for grey and non-grey applications.} With
these simplifications, the governing equations are
\begin{align}
  \frac{1}{c} \frac{\partial}{\partial t} I + \mu \frac{\partial}{\partial x} I + \chi I &= 
  \frac{1}{2} \chi c a_{\mathrm{R}} T^4, \label{eq:imc:trtI} \\
  \frac{\partial}{\partial t} U &= \chi \int_{-1}^{-1}\mathrm{d}\mu I - \chi c a_{\mathrm{R}} T^4 + S,\label{eq:imc:trtU}
\end{align}
describing the change in specific intensity and internal energy density $U$ in
the presence of a grey, pure-absorption opacity $\chi$ and of an additional
generic source term $S$. Note that due to the assumption of a one-dimensional
slab geometry, the angle-integrated specific intensity $\int_0^{2\pi}
\mathrm{d}\phi I$ is denoted as $I$ for simplicity. In a first step, this
system is slightly modified by introducing the equilibrium radiation energy,
$E$, and its relation to the internal energy of the ambient material by
\begin{align}
  E &= a_{\mathrm{R}} T^4,
  \label{eq:imc:Er}\\
  \beta &= \frac{\partial E}{\partial U} \label{eq:imc:beta},
\end{align}
resulting in
\begin{align}
  \frac{1}{c} \frac{\partial}{\partial t} I + \mu \frac{\partial}{\partial x} I + \chi I &= 
  \frac{1}{2} \chi c E, \label{eq:imc:trtbetaI} \\
  \frac{1}{\beta}\frac{\partial}{\partial t} E &= \chi \int_{-1}^{-1}\mathrm{d}\mu I - \chi c E + S. \label{eq:imc:trtbetaE}
\end{align}
At the heart of the \gls{IMC} method lies a semi-implicit recasting of
\Cref{eq:imc:trtbetaE} to approximate $E$ and thus the emission term in
\Cref{eq:imc:trtbetaI}. For this purpose, a discrete version of
\Cref{eq:imc:trtbetaE} is considered, in which all time-continuous quantities
are replaced by appropriate time averages, which we denote by
bars\footnote{\citet{Fleck1971} point out that different time-averaging
prescriptions can in principle be chosen for the various quantities.}:
\begin{equation}
  \frac{1}{\bar \beta} \frac{E^{n+1} - E^n}{ \Delta t^n} + \bar \chi c \bar E = \bar \chi \int_{-1}^1 \mathrm{d} \mu \bar I + \bar S.
\end{equation}
Here, the time discretization $\Delta t^n = t^{n+1} - t^n$ has been introduced.
As proposed by \citet{Fleck1971}, this equation is solved for $E^{n+1}$ by
invoking the specific time-centering scheme 
\begin{equation} 
  \bar E = \alpha E^{n+1} + (1 - \alpha) E^n 
\end{equation} 
for the radiation energy with the time-centering parameter $0 \le \alpha \le
1$. Here, $\alpha = 0$ would revert back to the traditional, fully explicit
\gls{MC} scheme. Finally, all remaining time-averaged values are
re-interpreted as their time-continuous counterparts and re-inserted into
\Cref{eq:imc:trtbetaI}, yielding the modified \gls{RT} equation
\begin{equation}
  \frac{1}{c} \frac{\partial}{\partial t} I + \mu \frac{\partial}{\partial x} I
  + f \chi I  + (1 - f) \chi I = \frac{1}{2} (1 - f) \chi \int_{-1}^{1}\mathrm{d}\mu I + \frac{1}{2} f \chi c E + \frac{1}{2} (1 - f) S,
  \label{eq:imc:mod_radtrans_eq}
\end{equation}
after introducing the so-called \textit{Fleck factor}
\begin{equation}
  f = \frac{1}{1 + \alpha \beta c \Delta t \chi}.
\end{equation}

\begin{figure}[htb]
  \centering
  \includegraphics[width=\textwidth]{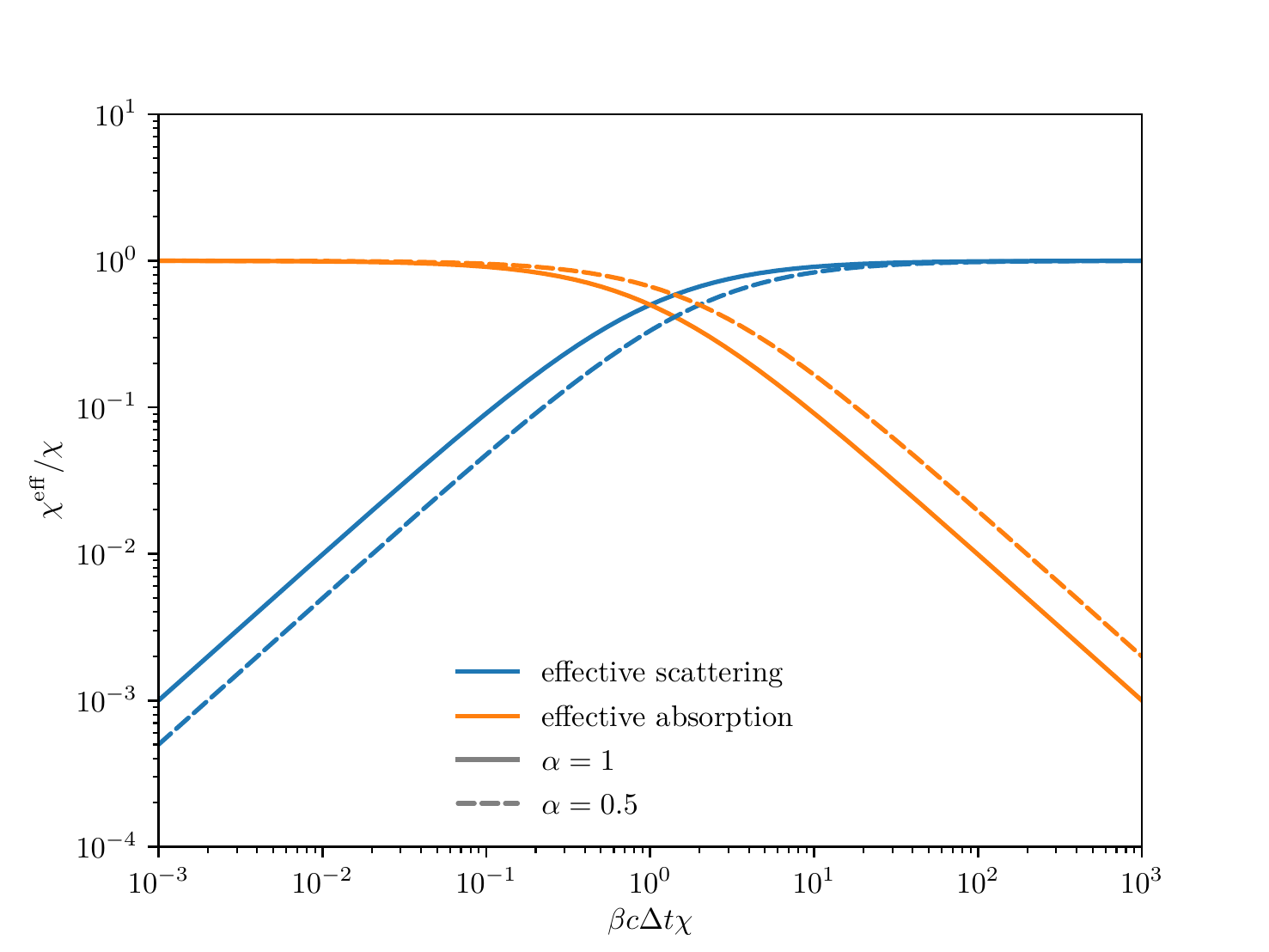}
  \caption{Illustration of the main consequence of the \gls{IMC} scheme, namely
    the introduction of effective scatterings at the cost of true absorptions
    as a function of $\beta c \Delta t \chi$.  For large values of $\beta c
    \Delta t \chi$ (i.e. long time steps (large $\Delta t$) or strong
    radiation-matter coupling (large $\beta$)) , the effective scattering
    opacity $\chi^{\mathrm{scat}}_{\mathrm{eff}} = (1 - f) \chi$ can become
  orders of magnitudes larger than the remaining ``net'' absorption term,
$\chi^{\mathrm{abs}}_{\mathrm{eff}} = f \chi$. The illustration shows the
behaviour for two choices for the time-centering parameter $\alpha$.} 
  \label{fig:fleck_factor}
\end{figure}

Compared to the original \gls{RT} equation, \Cref{eq:imc:trtI}, the
\gls{IMC} semi-implicit recasting reduces the importance of physical absorption
interactions by the factor $f$. At the same time, this reduction is compensated by the
introduction of terms that formally behave as a scattering contribution whose
strength is governed by $(1 - f) \chi$. From the definition of the Fleck factor
it is apparent that, as the time steps become large or as the coupling between
the internal and radiation energy pool becomes strong (i.e.\ $\beta$ and/or
$\chi$ become large), the contribution of the true absorption--emission
interactions to the transfer process are reduced and replaced by effective
scatterings. This behaviour of the \gls{IMC} scheme is illustrated in 
\Cref{fig:fleck_factor} and leads to unconditional stability (i.e.\ also holds
for $\Delta t \rightarrow \infty$) as long as a time-centering parameter  $0.5
\le \alpha \le 1$ is chosen.

This unconditional stability constitutes the main advantage of \gls{IMC} and a
substantial improvement over conventional \gls{MCRT} approaches.  This
beneficial property has led to widespread adoption of the \gls{IMC} scheme. In
the astrophysics community, \gls{IMC} schemes are predominantly applied in the
field of \gls{RT} in \gls{SN} ejecta.  \citet{Abdikamalov2012} have
incorporated the method in a \gls{MCRT} scheme for neutrino transport,
\citet{Wollaeger2013, Wollaeger2014} have developed a \gls{MC} tool for
\gls{RT} in \glspl{SN} based on \gls{IMC} and recently \citet{Roth2015} have
included \gls{IMC} into the \gls{MCRT} code \textsc{Sedona} \citep{Kasen2006}
and demonstrated its utility in one-dimensional radiation hydrodynamical
calculations.

The stability benefit of \gls{IMC} does, however, come at a cost and some of
the less desirable features of this technique should not go unmentioned. In
general, the construction of the governing \gls{IMC} equations introduces a
time discretization error which is formally of $\mathcal{O}(\Delta t)$. As a
consequence, the scheme becomes increasingly inaccurate as the time step
becomes larger. Moreover, \citet{Wollaber2016} cite the so-called
\textit{maximum principle violation} which can occur within \gls{IMC}
calculations as its main weakness. Here, temperatures within a computational
domain can non-physically exceed the imposed boundary temperatures in the
absence of internal sources. \citet{Larsen1987} formulate a time step
constraint under which these violations may be avoided. However, these
conditions are very restrictive and limit the applicability of \gls{IMC}. More
information about the maximum principle violation, and about efforts to
alleviate it within the \gls{IMC} framework as well as other drawbacks, such as
accurately reproducing the diffusion limit, the introduction of damped
oscillations or teleportation errors, are summarized by \citet{Wollaber2016}.

Finally, we note that the linearisation, semi-implicit recasting and
discretisation proposed by \citet{Fleck1971} and reviewed here constitutes only
one possibility to improve numerical stability. The recent review by
\citet{Wollaber2016} provides a comprehensive overview of a number of
alternative approaches. In particular, we draw attention to the family of
techniques, mainly shaped by Brooks and collaborators
\citep[e.g.][]{Brooks1989, Brooks2005}, denoted \gls{SIMC}, which leave the
thermal emission term formally unknown by introducing unknown \textit{symbolic}
packet weights. This technique may be denoted as a truly implicit \gls{MC}
method in the same sense as applied in the field of solving differential
equations \citep[see][]{Wollaber2016}.

\subsection{Efficient Monte Carlo techniques in optically thick media}
\label{sec:optically_thick}

While conventional \gls{MC} techniques are well suited for problems with a
moderate or low optical depth, their efficiency decreases dramatically in
optically thick applications. In a pure scattering environment, packets are
frequently deflected by collisions and their propagation effectively becomes a
random walk. Explicitly following and treating the multitude of interactions as
is required in conventional \gls{MC} approaches becomes very inefficient and
computationally expensive. The situation is similar in problems with high
absorption opacities. At first glance the short packet trajectories due to
rapid truncation by frequent absorption events seem to argue for a efficient
application of \gls{MC} techniques in this regime.  However, in
equilibrium/steady-state problems this would need to be countered by very large
numbers of quanta to describe the propagation while in explicit time-dependent
\gls{MC} treatments, small time steps are required to ensure numerical
stability (see discussion in \Cref{Sec:imc}). As detailed above, the \gls{IMC}
approach offers a solution to the time-step problem since it ensures
unconditional stability. However, the \gls{IMC} approach suffers equally in
efficiency in the optically thick regime since the Fleck factor is very small
in such situations and the vast majority of interactions proceed as effective
scatterings. 

A number of authors have developed techniques that improve the efficiency of
\gls{MC} calculations in optically thick regimes. These acceleration techniques
replace the conventional \gls{MC} transport process by a diffusion treatment
that efficiently transports \gls{MC} quanta through regions of high optical
depth. The appropriate probabilities for these transport processes are found by
a stochastic interpretation of the diffusion equation that constitutes the
correct physical limit for \gls{RT} processes in the presence of high
opacities. Typically, these \gls{MC} diffusion techniques are interfaced with a
conventional, often \gls{IMC} transport approach to yield a hybrid scheme that
efficiently solves \gls{RT} in problems with varying optical thickness. In the
following, we briefly outline two popular flavours of these diffusion
techniques, which predominantly differ in how the diffusion regions, in which
the normal transport simulation is switched off, are treated. These are the
so-called \textit{random walk} or \gls{MRW} techniques originally developed by
\citet{Fleck1984} and the \textit{Discrete Diffusion MC} (\gls{DDMC}) methods \citep[see e.g.][and references
therein]{Densmore2007}.

\subsubsection{Modified Random Walk}
\label{sec:mrw}

The \gls{RW}, or \gls{MRW} as coined by \citealt{Min2009}) was developed by
\citet{Fleck1984} as an extension to their \gls{IMC} method (see 
\Cref{Sec:imc}) to improve the computational efficiency in applications with
regions of high optical depth. The main idea underlying this approach is the
introduction of spherical diffusion regions whenever the optical depth is high.
Instead of following the multitude of effective scatterings in these regions
with \gls{IMC}, the conventional packet transport process is switched off and
replaced by a diffusion procedure. Here, packets are able to traverse the
diffusion regions in one-step processes. The probabilities governing this
propagation mode are derived by \citet{Fleck1984} by examining the statistical
properties of the random walk process and the solution to the
diffusion equation. While the original \gls{MRW} scheme has been derived for
\gls{IMC} applications, it naturally applies to explicit \gls{MC} approaches as
well after setting the Fleck factor to 1. 

\begin{figure}[htb]
  \centering
  \includegraphics[width=\textwidth]{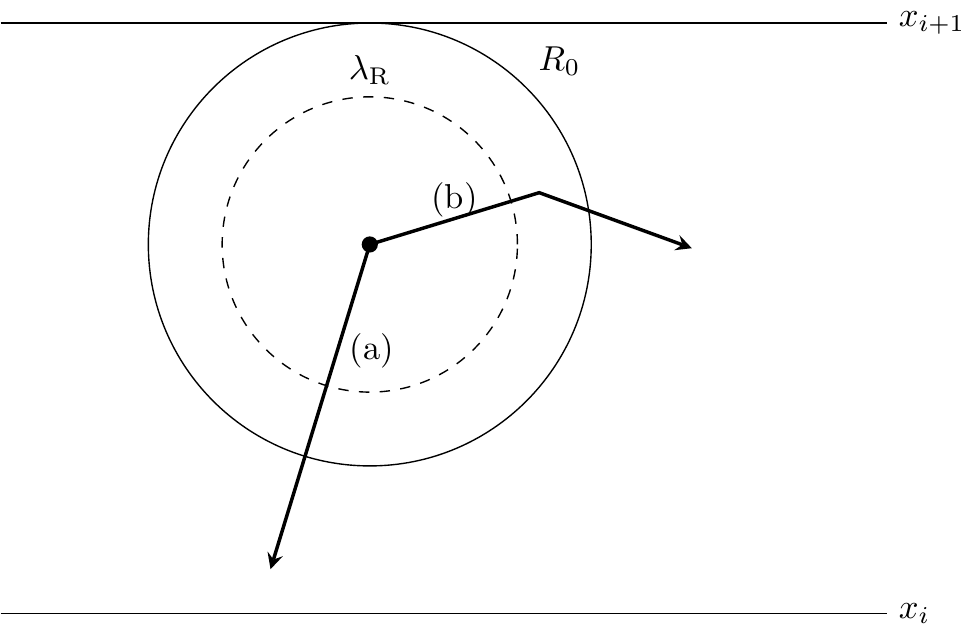}
  \caption{Illustration of the construction of diffusion spheres in the
    \gls{MRW} approach. The sketch illustrates the current packet location as a
    thick black dot. With this location as the origin, a sphere with radius
    $R_0$ that fully fits into the current grid cell (defined in the
      plane-parallel case by the boundaries $x_i$ and $x_{i+1}$) and has
      maximum volume is constructed. Two possible packet propagation paths in
      the normal transport scheme are shown. In the first case (a), the packet
      leaves the sphere without interacting while in the second case (b) the
      packet scatters inside the sphere after covering the distance
      $l_{\mathrm{col}}$. Since $\lambda_{\mathrm{R}}$ is smaller than $R_0$ in
      the illustrated situation, the \gls{MRW} diffusion scheme would be
      switched on in case (b) while the packet would be transported normally in
    case (a).  This illustration is adapted from \citet[][Fig.\ 3]{Fleck1984}.}
  \label{fig:mrw:diffusion_sphere}
\end{figure}

We briefly outline the \gls{MRW} procedure and refer to the original work by
\citet{Fleck1984} for a detailed derivation. Diffusion spheres, originating
from the current location of all packets are constructed. The radii
$R_0$ of the spheres are chosen such that they entirely lie within their host
grid cells but occupy the largest possible volume (see illustration in 
\Cref{fig:mrw:diffusion_sphere}).
In these homogeneous and isothermal spheres, the explicit \gls{MC} packet
transport may be replaced by a diffusion solution. However, this replacement is
only accurate if the packets are expected to perform a multitude of
interactions as they propagate in the sphere so that the diffusion
approximation becomes valid. \citet{Fleck1984} translate this requirement into
the following criteria for the activation of the diffusion process:
\begin{align}
  R_0 & > \lambda_{\mathrm{R}},\label{eq:mrw:applicability_cond1}\\
  l_{\mathrm{col}} & < R_0.
  \label{eq:mrw:applicability_cond2}
\end{align}
Here, $\lambda_{\mathrm{R}}$ denotes the Rosseland mean free
path\footnote{See e.g. \citet{Hubeny2014} for definition and discussion of the
  Rosseland mean opacity.} and
$l_{\mathrm{col}}$ the physical distance the packet has to cover to the next
interaction as determined in the standard \gls{MC} transport propagation
procedure (see \Cref{Sec:Propagation}). These conditions ensure that packets
are expected to perform multiple interactions\footnote{\citet{Fleck1984} argue
that the Rosseland mean free path tends to be much smaller than the Planck mean
free path, which describes the typical distance between collisions.}  within
the sphere and are guaranteed to interact at least once. As soon as these
conditions apply, the normal \gls{IMC} transport of packets is stopped in the
sphere and a diffusion procedure is started (see
\Cref{fig:mrw:diffusion_sphere}).  
This process is govern by transport rules obtained from considering how far packets could propagate under diffusion conditions as a function of time.
Specifically, the probability of finding a packet at distance $r$ from its
initial position (which is by construction at $r=0$) after time $t$ is given by 
\begin{equation}
  \psi(r, t) = \frac{1}{R_{0}^2} \sum_{n=1}^{\infty} \left(
  \frac{n}{r} \right) \exp \left[ - \left( \frac{\pi n}{R_0} \right)^2
  D c t
  \right] \sin \left( \frac{n \pi r}{R_0} \right).
  \label{eq:mrw:psi}
\end{equation}
Here, $D$ is the diffusion constant for the process
\citep[see][Eq.~21]{Fleck1984}.  This result can be used to determine the
probability that the packet still resides within the sphere after time $t$,
which is 
\begin{equation}
  P_{\mathrm{R}}(t) = 4 \pi \int_{0}^{R_{0}} \psi(r, t) r^2 dr.
  \label{eq:mrw:residence_prob}
\end{equation}
Consequently, at any given time (e.g.\ the end of a time step), the fate of the
packet can be decided by the random number experiment 
\begin{align}
  0 < \xi \leq P_{\mathrm{T}}(t) && \text{packet escapes sphere},
  \label{eq:mrw:packet_fate_decision_esc}\\
  P_{\mathrm{T}}(t) < \xi \leq 1 && \text{packet remains in sphere},
  \label{eq:mrw:packet_fate_decision_rem}
\end{align}
with the escape probability
\begin{equation}
  P_{\mathrm{T}}(t) = 1 - P_{\mathrm{R}}(t).
  \label{eq:mrw:escape_prob}
\end{equation}
If the packet escapes, its position is updated to a location uniformly drawn
from the surface of the diffusion sphere. A new direction is assigned by
sampling the cosine distribution about the normal to the surface of the sphere
and finally the propagation time is advanced to the time of escape which is
found by
\begin{equation}
  P_{\mathrm{T}}(t) = \xi.
  \label{eq:mrw:escape_time}
\end{equation}
This identity is solved using the same random number that determined the
decision about the escape from the diffusion sphere, i.e.\ in the experiment
described by \Cref{eq:mrw:packet_fate_decision_esc}. If the packet remains in
the sphere after time $t$ has elapsed, it is moved within the diffusion sphere.
A new direction is drawn uniformly and its location is updated by determining a
new sphere with radius $R_1$
\begin{equation}
  4 \pi \int_0^{R_1} \psi(r, t) r^2 \mathrm{d}r = \xi P_{\mathrm{R}}(t)
  \label{eq:mrw:new_sphere}
\end{equation}
and placing the packet randomly onto its surface.

\begin{figure}[htb]
  \centering
  \includegraphics[width=\textwidth]{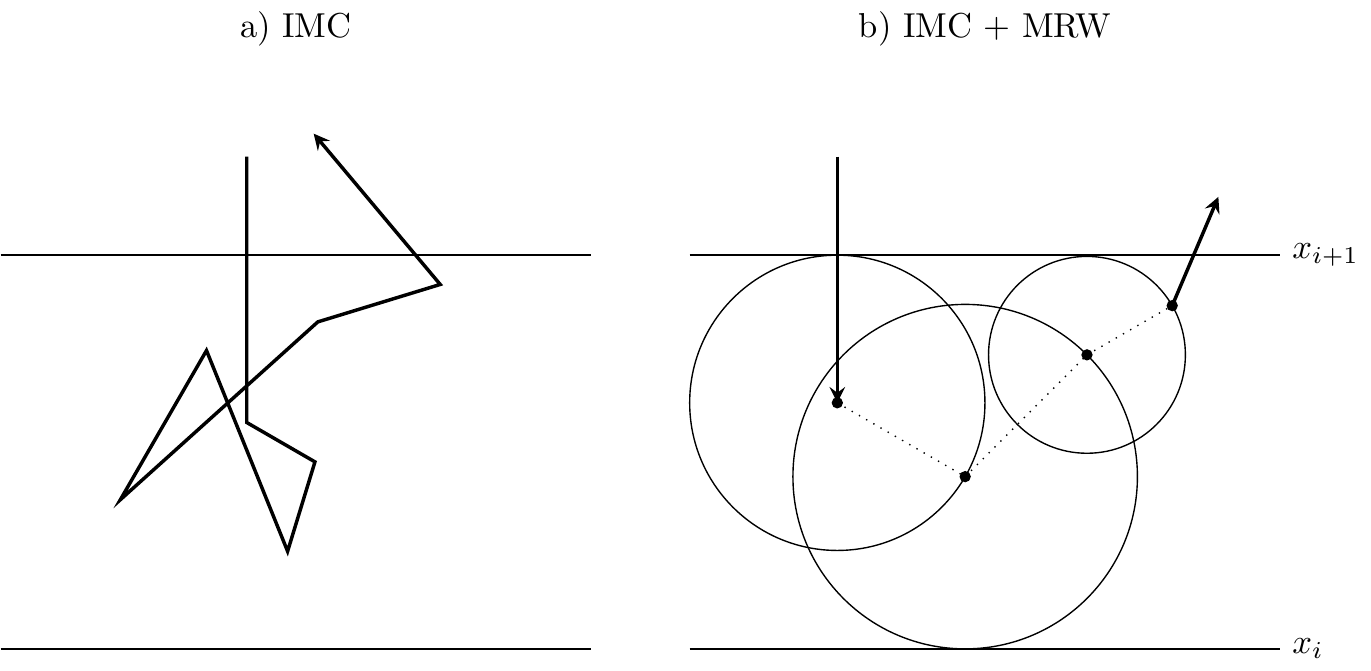}
  \caption{Illustration of the \gls{MRW} propagation process. On the left hand
    side, the situation in an optically thick cell is shown if only a
    conventional \gls{MC} transport scheme (in this case \gls{IMC}) is shown.
    The packet will scatter multiple times which is time consuming to simulate
    directly.  Instead, when the \gls{IMC} approach is coupled with a \gls{MRW}
    scheme, packets can much more efficiently step through optically thick
    regions, as illustrated in the right panel. As long as the applicability
    criteria for \gls{MRW},
    \Crefrange{eq:mrw:applicability_cond1}{eq:mrw:applicability_cond2}, are met
    the explicit packet transport procedure (denoted by thick solid lines) is
    switched off and the packet can traverse the diffusion spheres in simple
    one-step processes (denoted by the thin dotted lines). In the illustration
    shown here, it is assumed that the packet escapes the diffusion spheres by
    virtue of the \gls{MC} experiment in \Cref{eq:mrw:packet_fate_decision_esc}
    during the time step. At this point, it is placed randomly (according to a
    uniform distribution) onto the surface of the sphere (black dots). Only
    close to the cell boundaries ($x_i$ and $x_{i+1}$), the diffusion spheres
  are often too small to activate \gls{MRW} and the normal \gls{MC} transport
scheme has to be used (cf.\ last segment of the depicted packet
trajectory).}
  \label{fig:mrw:propagation}
\end{figure}

Once the packet resumes its propagation (either in a new time step or after
escaping from a diffusion sphere), the criteria in
\Cref{eq:mrw:applicability_cond1} and \Cref{eq:mrw:applicability_cond2} are
again used to determine whether the packet is transported via normal (I)\gls{MC}
procedures or by defining a new diffusion sphere. This technique is
schematically illustrated in \Cref{fig:mrw:propagation}. 
Following such a \gls{MRW} scheme can significantly increase the efficiency of
\gls{MCRT} calculations in optically thick regimes \citep{Fleck1984}. However,
it still faces efficiency problems if packets are close to grid cell
boundaries. In this region, the diffusion spheres can become too small to allow
the criteria in \Cref{eq:mrw:applicability_cond1} and
\Cref{eq:mrw:applicability_cond2} to activate the diffusion procedure. Thus,
the inefficient transport scheme has to be used to propagate packets in these
situations \citep[see comments by][]{Densmore2012}.

Recently, the \gls{MRW} approach has been applied in astrophysical \gls{RT}
problems by \citet{Min2009} and \citet{Robitaille2010}. There, the scheme is
incorporated into \gls{MC} approaches to dust \gls{RT} and specifically helps
to transport packets through optically thick parts of dusty discs. However, it
seems very challenging to adapt this scheme to applications in which complex
opacities, particularly Sobolev-type line opacities, have to be taken
into accounted.

\subsubsection{Discrete Diffusion Monte Carlo}
\label{sec:ddmc}

In the \gls{MRW} scheme, only spherical subregions of grid cells are designated
diffusion zones. As outlined above, constraints imposed on the size of the
sphere lead to efficiency problems when packets are located close to grid cell
boundaries. This drawback is eliminated in other \gls{MC} diffusion approaches.
In so-called Discrete Diffusion Monte Carlo (\gls{DDMC}) techniques, entire grid cells are treated as diffusion
regions. Within, \gls{DDMC} packets are generated that can traverse these cells
efficiently in one-step processes. The propagation rules for this procedure are
again extracted from a probabilistic interpretation of the discretized
diffusion equation. In analogy to the \gls{MRW} method, \gls{DDMC} schemes are
commonly used in hybrid approaches in combination with \gls{IMC} transport
techniques to ensure an efficient applicability to problems with regions of
different optical thickness \citep[see e.g.][]{Gentile2001, Densmore2007}.

\gls{DDMC} techniques have their origin in neutron transport problems (see
overview by \citealt{Densmore2007}) but a popular variant designed for photon
\gls{RT} has been presented by \citet{Densmore2007}. Another flavour of the
diffusion technique has been developed by \citet{Gentile2001} and is often
referred to as \gls{IMD}\footnote{\citet{Densmore2007} still classifies the
\gls{IMD} approach as a member of the class of \gls{DDMC} techniques.}. The
main difference with respect to the \gls{DDMC} approach by \citet{Densmore2007}
lies in the treatment of how time is tracked by the \gls{DDMC} packets. While
both \gls{DDMC} and \gls{IMD} have originally been presented for grey problems,
multi-group extensions appropriate for frequency-dependent applications have
already been devised, in particular by \citet{Densmore2012} and
\citet{Cleveland2010} respectively.

Of the \gls{DDMC} schemes, the variant of \citet{Densmore2007, Densmore2012}
seems to currently have experienced the most attention in the astrophysical
community.  \citet{Abdikamalov2012} have developed a hybrid
\gls{DDMC}-\gls{IMC} approach for neutrino transport in core-collapse
\glspl{SN} and \citet{Wollaeger2013} and \citet{Wollaeger2014} have introduced
a \gls{MC} method for \gls{RT} in \gls{SN} ejecta, constructed around a
\gls{DDMC}-\gls{IMC} core. Consequently, we only focus on the \gls{DDMC} scheme
of \citet{Densmore2007} in the following, where we briefly highlight the
guiding principles of discrete diffusion techniques. We refer the reader to
\citet{Gentile2001}, \citet{Cleveland2010} and \citet{Cleveland2015} for details about the closely
related \gls{IMD} approach.

The \gls{DDMC} scheme as presented by \citet{Densmore2007} is formulated for
grey and static diffusion problems. Extensions for frequency dependent problems
and moving media are introduced by \citet{Densmore2012} and by
\citet{Abdikamalov2012} respectively. The \gls{DDMC} approach begins with the
designation of a subset of grid cells in the computational domain, which are
considered sufficiently optically thick, as \gls{DDMC} diffusion zones. For
simplicity in presenting the governing principles, we assume that there is one
continuous subregion in the domain in which \gls{DDMC} is active and which
consists of $N$ cells. The interfaces of these cells lie at $x_{j-1/2}$ and
$x_{j+1/2}$ for $j=1\cdots N$. The diffusion cells are logically separated into
interior cells (i.e.\ cells with neighbouring \gls{DDMC} cells at both sides)
and interface cells (i.e.\ cells which lie at the interface between \gls{DDMC}
  and transport regions or lie at the domain boundary). In the diffusion cells,
  \gls{DDMC} particles are generated at the beginning of each simulation (time)
  step based on the active radiation field from the previous time step,
  internal emission due to source terms or, in the interface cells, due to
  influx of normal \gls{IMC} packets or due to inflow imposed by the domain
  boundary conditions. The magnitude of all these processes and rules for how
  the \gls{DDMC} particles propagate are obtained from a discretized diffusion
  equation. For this purpose, the basic \gls{IMC} transport equation,
  \Cref{eq:imc:mod_radtrans_eq}, is considered as the starting point
  in the \gls{DDMC}, and its zeroth moment is discretized in space, yielding
\begin{equation}
  \frac{1}{c}\frac{\mathrm{d}}{\mathrm{d} t} \phi_j + \frac{1}{\Delta x_j} (F_{j+1/2} - F_{j-1/2}) - f_{n,j} \sigma_{n,j} \phi_j = f_{n,j} \sigma_{n,j} a c T^4_{n,j}.
  \label{eq:ddmc:diff_eq}
\end{equation}
Here, cell-averaged scalar intensities
\begin{equation}
  \phi_j(t) = \frac{1}{\Delta x_j} \int_{x_{j-1/2}}^{x_{j+1/2}} \mathrm{d}x \int_{-1}^{1}\mathrm{d}\mu I(x, t; \mu) 
  \label{eq:ddmc:phi}
\end{equation}
and cell-edge fluxes
\begin{equation}
  F_{j+1} = F(x_{j+1/2}, t) = \int_{-1}^{1}\mathrm{d}\mu \mu I(x_{j+1/2}, t; \mu)
  \label{eq:ddmc:flux}
\end{equation}
are used. \Cref{eq:ddmc:diff_eq} is closed by using Fick's diffusion law \citep{Fick1855}
\begin{equation}
  F_{j+1/2} = - \frac{1}{3 \sigma_n} \left. \frac{\partial \phi}{\partial x} \right |_{x=x_{j+1/2}}.
  \label{eq:ddmc:ficklaw}
\end{equation}
An appropriate representation of the spatial derivative is found by
finite-differencing. This leads to a time-continuous diffusion
equation, discretized in space, which takes the form
\begin{align}
  \frac{1}{c}\frac{\mathrm{d}}{\mathrm{d} t} \phi_j + (\sigma_{\mathrm{L},j} + \sigma_{\mathrm{R}, j} + f_{n,j} \sigma_{n,j})\phi_j = \nonumber\\
  f_{n,j} \sigma_{n, j} a c T^4_{n,j} + \frac{1}{\Delta x_j}(\sigma_{\mathrm{L}, j+1} \phi_{j+1} \Delta x_{j+1} + \sigma_{\mathrm{R},j-1} \phi_{j-1} \Delta x_{j-1})
  \label{eq:ddmc:disc_diff_interior}
\end{align}
for interior cells. Here, left and right leakage opacities
\begin{align}
  \sigma_{\mathrm{L},j} &= \frac{2}{3 \Delta x_{j}} \frac{1}{\sigma^{+}_{n,j-1/2} \Delta x_j + \sigma^{-}_{n, j-1/2} \Delta x_{j-1}}\label{eq:ddmc:left_leakage},\\
  \sigma_{\mathrm{R},j} &= \frac{2}{3 \Delta x_{j}} \frac{1}{\sigma^{-}_{n,j+1/2} \Delta x_j + \sigma^{+}_{n, j+1/2} \Delta x_{j+1}},
  \label{eq:ddmc:right_leakage}
\end{align}
are defined in terms of the face-averaged opacities $\sigma^{+}_{j-1/2}$ and
$\sigma^{-}_{j+1/2}$. In this nomenclature, the superscripts denote which
neighbouring cell is used for the opacity calculation. In particular,
$\sigma^{-}_{j+1/2}$ is based on the material properties of cell $j$ and
conversely, $\sigma^{+}_{j+1/2}$ uses the information from cell $j+1$.
According to \citet{Densmore2007}, this procedure is necessary to avoid
propagation problems in cases where one of the opacities becomes very large. As
the authors point out, in addition to relying on the material properties of the
appropriate neighbouring cell, the use of a common cell-edge temperature is
vital for overcoming these problems \citep[see discussion by][]{Densmore2007}.

\Cref{eq:ddmc:disc_diff_interior} builds the foundation of the \gls{DDMC}
scheme and sets the propagation behaviour of \gls{DDMC} particles after a
probabilistic interpretation has been performed. It describes a
time-dependent\footnote{A crucial difference in the \gls{IMD} scheme is that
the time-derivative in the diffusion equation is discretized by finite
differences as well \citep[cf.][]{Gentile2001}.} evolution equation for the
scalar intensity and thus for the population of \gls{DDMC} particles. In this
view, the terms on the \gls{RHS} of \Cref{eq:ddmc:disc_diff_interior} describe
processes that increase $\phi_j$, and thus the population of \gls{DDMC}
particles in cell $j$. This can either be by emission (first term on \gls{RHS})
or by leakage of \gls{DDMC} particles from neighbouring cells (second and third
terms on \gls{RHS}). Conversely, the term on the \gls{LHS} captures all
processes which reduce $\phi_j$ and thus remove \gls{DDMC} particles from cell
$j$. Again, this can occur via leakage into one of the neighbouring cells or
through an absorption event. In this interpretation, \gls{DDMC} packets are not
associated with explicit location or direction information but only with their
current host grid cell. They are propagated by considering the time to the end
of the time step and the time to the next collision, which can be determined
analogously to the corresponding procedure in conventional \gls{MC} transport
by
\begin{equation}
  t_{\mathrm{col}} = -\frac{1}{c} \frac{1}{\sigma_{\mathrm{L}, j} + \sigma_{\mathrm{R}, j} + f_{n,j} \sigma_{n,j}} \log \xi.
  \label{eq:ddmc:collision_time}
\end{equation}
Here, collisions can either refer to an absorption or leakage event. The exact
nature of the collision can be established in a random number experiment
similar to the decision between scattering and absorption in conventional
\gls{MC} transport (see \Cref{Sec:Propagation_abs_scat}) based on the relative
magnitudes of the terms appearing in the denominator of
\Cref{eq:ddmc:collision_time}. In the case of absorption, the propagation of
the \gls{DDMC} particle is terminated, otherwise its internal clock is advanced
by $t_{\mathrm{col}}$ and it continues its propagation in the new cell until
the end of the time step is reached or it is absorbed.

An equation similar to \Cref{eq:ddmc:disc_diff_interior} is found for
\gls{DDMC} interface cells, which are at the edge of the diffusion regions,
after imposing appropriate boundary conditions. Instead of relying on the
Marshak boundary condition, \citet{Densmore2007} propose a condition inspired
by the asymptotic diffusion-limit. This ensures an accurate behaviour of the
\gls{DDMC} scheme in situations in which the incoming transport packet
population has a very anisotropic angular distribution too
\citep[see][]{Densmore2007}. The resulting space-discretized diffusion equation
has the same structure as for interior cells apart from an additional source
term that describes the influx of radiation from the transport region (or from
  outside of the computational domain if the interface is at the domain edge).
  This source term can be converted into a probability which is sampled every
  time a \gls{MC} packet from the transport region or from the domain boundary
  condition impinges onto the diffusion region to decide whether the packet is
  converted into a \gls{DDMC} particle or reflected back. The complementary
  process of \gls{DDMC} packets leaking out of the diffusion region is handled
  by placing them isotropically onto the interface. Such packets then continue
  propagating according to the conventional \gls{MC} transport scheme.

\section{MCRT and dynamics}
\label{Sec:Dynamics}

In \Cref{Sec:Estimators} we reviewed how estimators can be constructed to
determine the rate of transfer of energy and momentum from the radiation field
to the ambient medium.  This transfer can become dynamically important and
drastically affect the evolution of a system. In the astrophysical realm,
prominent examples for such circumstances include radiatively driven mass
outflows from hot stars \citep[see review by][]{Puls2008} or accretion discs
\citep[e.g.][]{Proga1998,Proga2000,Proga2004}, the star formation process
\citep[see review by][and references therein]{McKee2007} or the shock outbreak
phase in \glspl{SN} \citep[see e.g.\ overview in][]{Mihalas1984}. In situations
such as these, a decoupled treatment of hydrodynamics and \gls{RT} is no longer
accurate but a coupled \gls{RH} solution approach has to be followed. 

Historically, \gls{RH} studies have been typically performed with deterministic
solution techniques. But particularly in the field of line-driven winds from
hot stars, there is a substantial literature based on \gls{MC} studies by
\citet{Abbott1985} and, among others, \citet{Lucy1993},
\citet{Vink1999, Vink2000} and \citet{Mueller2008}. The main motivation for relying on \gls{MC} schemes certainly
lies in their ease of treating the Sobolev-type line opacities encountered in
these winds. Specifically, a \gls{MC} calculation is used to determine the
momentum deposition in the outflow material according to which a steady-state
wind structure is calculated. In addition, fully dynamic \gls{RH} approaches
which rely on \gls{MC} methods have been developed and applied. For example,
\citet{Nayakshin2009} and \citet{Acreman2010} coupled \gls{SPH} approaches with
\gls{MCRT} calculations. \citet{Haworth2012} investigated triggered star
formation with a \gls{RH} approach in which the gas temperature is adjusted by
a \gls{MC}-based photo-ionization calculation. \citet{Harries2015} and
\citet{Harries2017}
continued the development of \gls{MC}-based \gls{RH} methods for star formation
problems. \citet{Noebauer2012} and \citet{Roth2015} introduced \gls{MC}-based
\gls{RH} techniques with a general-purpose scope, with a particular focus on
\gls{IMC} techniques in the latter. Implicit \gls{MC} diffusion methods were
coupled with hydrodynamics calculations by \citet{Cleveland2015}.  This limited
list of examples illustrates that the possibility of using \gls{MCRT}
techniques in fully dynamic applications is actively researched and developed.
In the following, we briefly sketch how energy and momentum transfer terms may
be reconstructed from \gls{MCRT} calculations and included in fluid dynamics
calculations.

\subsection{Reconstructing energy and momentum transfer terms}

The full \gls{RH} problem can be formulated in terms of the
normal fluid dynamical equations describing mass, momentum and energy
conservation but modified by terms capturing the energy and momentum exchange
mediated by radiation--matter interactions. Following \citet{Mihalas2001}, we
present the equations in the \gls{LF} and refer the reader to the standard
literature, in particular to \citet{Mihalas1984}, for more details and full
derivations:
\begin{align}
  & \frac{\partial \rho}{\partial t} + \sum_i \frac{(\rho v^i)}{\partial x^i} = 0,\label{eq:dynamics:mass}\\
  & \frac{\partial (\rho v^i)}{\partial t} + \sum_j \frac{\partial}{\partial x^j} (\rho v^i v^j + p \delta^{ij}) = G^i - \frac{v^i}{c} G^0,\label{eq:dynamics:momentum}\\
  & \frac{\partial }{\partial t} \left[\rho \left(\frac{1}{2} v^2 + e \right)\right] + \sum_i \frac{\partial}{\partial x^i} \left[ \rho v^i \left(\frac{1}{2} v^2 + e \right) + p v^i\right] = c G^0.\label{eq:dynamics:energy}
\end{align}
Here, the material density $\rho$, pressure $p$ and specific internal energy
$e$ appear. Also, the Kronecker delta $\delta^{ij}$ is used.  On the right-hand
side of the energy and momentum equation, the components of the so-called
radiation force appear, $G^0$ and $G^i$, encoding energy and momentum exchange
between the radiation field and the ambient material. These are defined by
\begin{align}
  -c G^0 & = \int_0^{\infty}\mathrm{d}\nu \int \mathrm{d}\Omega (\eta_{\nu} - \chi_{\nu} I_{\nu}),\\
  -c G^i & = \int_0^{\infty}\mathrm{d}\nu \int \mathrm{d}\Omega
           (\eta_{\nu} - \chi_{\nu} I_{\nu})n^i \; .
\end{align}

Relying on similar techniques as outlined in \Cref{Sec:volume_estimators}, the
radiation force components can be reconstructed from the ensemble of \gls{MC}
packet histories. This procedure is particularly simple if we adopt a thermal
equilibrium emission coefficient\footnote{I.e., assume $\eta_{\nu}  = \chi_{\mathrm{a}} B_{\nu}$.}
and treat 
the emission and
scattering processes as isotropic in the \gls{CMF}. In this case, the
radiation force in the \gls{CMF} is given by
\begin{align}
  c G^0_0 &= \int_0^{\infty} \mathrm{d}\nu_0 \; \chi_{\mathrm{a},0}(\nu_{0}) (c E_{0\nu} - 4 \pi B_{\nu_0}),\\
  c G^i_0 &= \int_0^{\infty} \mathrm{d}\nu_0 \; 
            \chi_{\mathrm{tot},0}(\nu_{0}) F^i_{0\nu} \; ,
\end{align}
where $\chi_{\mathrm{tot},0}(\nu_{0})$ and
$\chi_{\mathrm{a},0}(\nu_{0})$ are the total and
absorption parts of the opacity (see \Cref{Sec:Propagation_abs_scat})
in the \gls{CMF}. $E_{0\nu}$ and
$F^i_{0\nu}$ are, respectively the \gls{CMF} specific radiation energy
density and flux, and $B_{\nu_0}$ is the Planck function. 

For grey opacities, this simplifies further to 
\begin{align}
  c G^{0}_{0} & = c \; \chi_{\mathrm{a},0} \left( E_0 - a_{\mathrm{R}} T_{0}^{4} \right),\\
  c G^{i}_{0} & = \chi_{\mathrm{tot},0}
                F^{i}_{0}\label{eq:radforcecmf} \;
\end{align}
where $a_{\mathrm{R}}$ is the radiation constant and $T_{0}$ the temperature.
In this case, the radiation force may be reconstructed using the efficient volume-based
estimators for the radiation energy density and flux $E_0$ and $\mathbf{F}_0$
which already have been presented in \Cref{Sec:volume_estimators}. For
frequency-dependent material functions, one may introduce appropriate frequency
averages of the opacities, as for example done by \citet{Roth2015} and retain
the analogous equations as above. Alternatively, the opacities may be included
in the volume-based averaging process.
\begin{align}
  G^0_0 &= \frac{1}{\Delta V c \Delta t} \sum \chi_{\mathrm{a},0}(\nu_0) l_0
          \varepsilon_0 - 4\pi \int_0^{\infty} \mathrm{d}\nu_0 \;
          \chi_{\mathrm{a},0}(\nu_0) B_{\nu_0} \; ,\\
  G^i_0 &= \frac{1}{\Delta V c \Delta t} \sum \chi_{\mathrm{tot},0}(\nu_0)  l_0 \mu_0
          \varepsilon_0 \; .
\end{align}
This may be interpreted as summing packet energies and momenta, and weighting
the individual contributions by the probability that a packet interacts along
the trajectory element $l_i$ (see also the discussion about reconstructing
heating rates with volume-based estimators in \Cref{Sec:volume_estimators}). In
the above formulation, we evaluate the frequency-dependent opacity at the
beginning of each individual packet trajectory element according to the
instantaneous packet frequency. This naturally assumes that the opacity varies
only mildly along the trajectory segment. In situations, in which this is not
fulfilled, alternative formulations have to be devised. In astrophysical
applications, this occurs for example whenever bound-bound processes, i.e.\
interactions with atomic line transitions, are important, as in line-driven
mass outflows from hot stars or in \gls{SNIa} ejecta (cf.\
\Cref{Sec:line_ints_sobolev}). Here, the opacity varies strongly whenever
photons resonate with a line transition. For such applications, the energy and
momentum transfer terms may be reconstructed as proposed by
\citet[][]{Lucy1999a} and \citet{Noebauer2015a}.\footnote{Note, however, that the
  applicability of the Sobolev approximation \citep{Sobolev1960} to line
  opacity is assumed in these radiation force estimators.} 

Reconstructing the momentum deposition based on \cref{eq:radforcecmf} as
sketched above relies on the radiation flux in the \gls{CMF}. As pointed out by
\citet{Roth2015}, volume-based estimators as derived previously involve the
cancellation of contributions from packets propagating in opposite directions.
In particular in the diffusion regime, in which the net flux is expected to be
very small, such estimators suffer from high \gls{MC} shot noise. Thus,
\citet{Roth2015} proposed an alternative reconstruction scheme for this regime,
based on the first moment of the transfer equation, which reduces to
\begin{equation}
  G^{i} = -\sum_j \frac{\partial P^{ij}}{\partial x^j}
\end{equation}
under diffusive conditions \citep{Mihalas2001}.  Now, the momentum deposition
depends on the radiation pressure tensor which can be easily reconstructed
without relying on cancellation effects.

As an alternative to the \gls{CMF}-based reconstruction approaches detailed
above, the radiation force components can also be determined in the \gls{LF}.
A corresponding reconstruction procedure within the volume-based estimator
approach was outlined by \citet{Noebauer2012}. 

\subsection{Coupling to fluid dynamics}

\begin{figure}[htb]
  \centering
  \includegraphics[width=\textwidth]{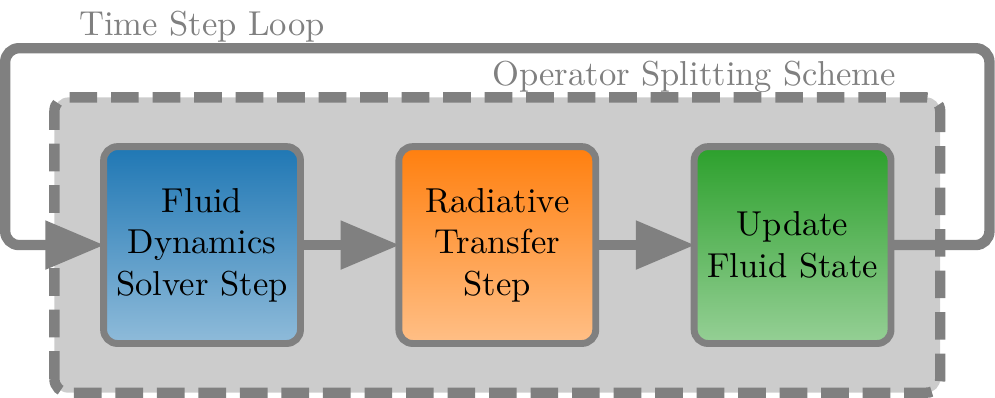}
  \caption{Illustration of a simple Godunov-splitting approach to \gls{MC}-based
  \gls{RH}.}
  \label{fig:dynamics:operator_splitting}
\end{figure}

Once the energy and momentum transfer terms are available via the radiation
force components they can be coupled to a fluid dynamical calculation.
Typically, an operator-splitting approach \citep[see e.g.][for a detailed
explanation of the operator splitting principle]{LeVeque2002} is used to tackle
the \gls{RH} problem. This is a widely used technique to deal
with source terms in hydrodynamical equations (e.g.\ gravity, nuclear energy
release, etc.) and is part of many \gls{MC}-based \gls{RH}
approaches \citep[e.g.][]{Noebauer2012, Roth2015}. Implementing the simplest
incarnation of this technique, the so-called \textit{Godunov splitting}
\citep[cf.][]{LeVeque2002}, a \gls{RH} time step would then
proceed as outlined in \Cref{fig:dynamics:operator_splitting}.
It begins with a pure hydrodynamical solver call, assuming the absence of any
source terms on the right hand side of
\Crefrange{eq:dynamics:momentum}{eq:dynamics:energy} due to \gls{RT}.
The new fluid state thus determined is then used to solve the \gls{RT}
problem using \gls{MC} techniques. From the ensemble of packet
trajectories, energy and momentum transfer between the ambient material and the
radiation field can be reconstructed using the concepts detailed above.
According to these transfer terms, the fluid momentum and energy are updated
and the time step is complete.

\subsection{Example application}

As originally suggested by
\citet{Ensman1994}, solving the structure of radiative shocks has become a
standard test problem for \gls{RH} solution techniques. In these shocks, a
radiative precursor emerging from the shocked domain penetrates the upstream
material pre-heating and compressing it (for a detailed overview of these
phenomena, we refer the reader to \citealt{Zeldovich1969}). Depending on the
strength of the pre-heating, \textit{sub-} and \textit{super-critical} shocks
are distinguished. The temperature in the precursor region remains below that
of the shocked material in the sub-critical case but reaches it in
super-critical shocks. Thanks to the seminal works by \citet{Lowrie2007} and
\citet{Lowrie2008}, analytic steady-state solutions are available for these
shocks. 

As a test of the methods, \citet{Noebauer2012} and \citet{Roth2015} have used
operator-splitting techniques to successfully calculate the structure of
radiative shocks with \gls{MC}-based \gls{RH} approaches.  Here, we discuss the
success of these tests -- further details about the physical and numerical
setup of these simulations are given in \Cref{sec:radiative_shocks}.

\begin{figure}[htb]
  \centering
  \includegraphics[width=\textwidth]{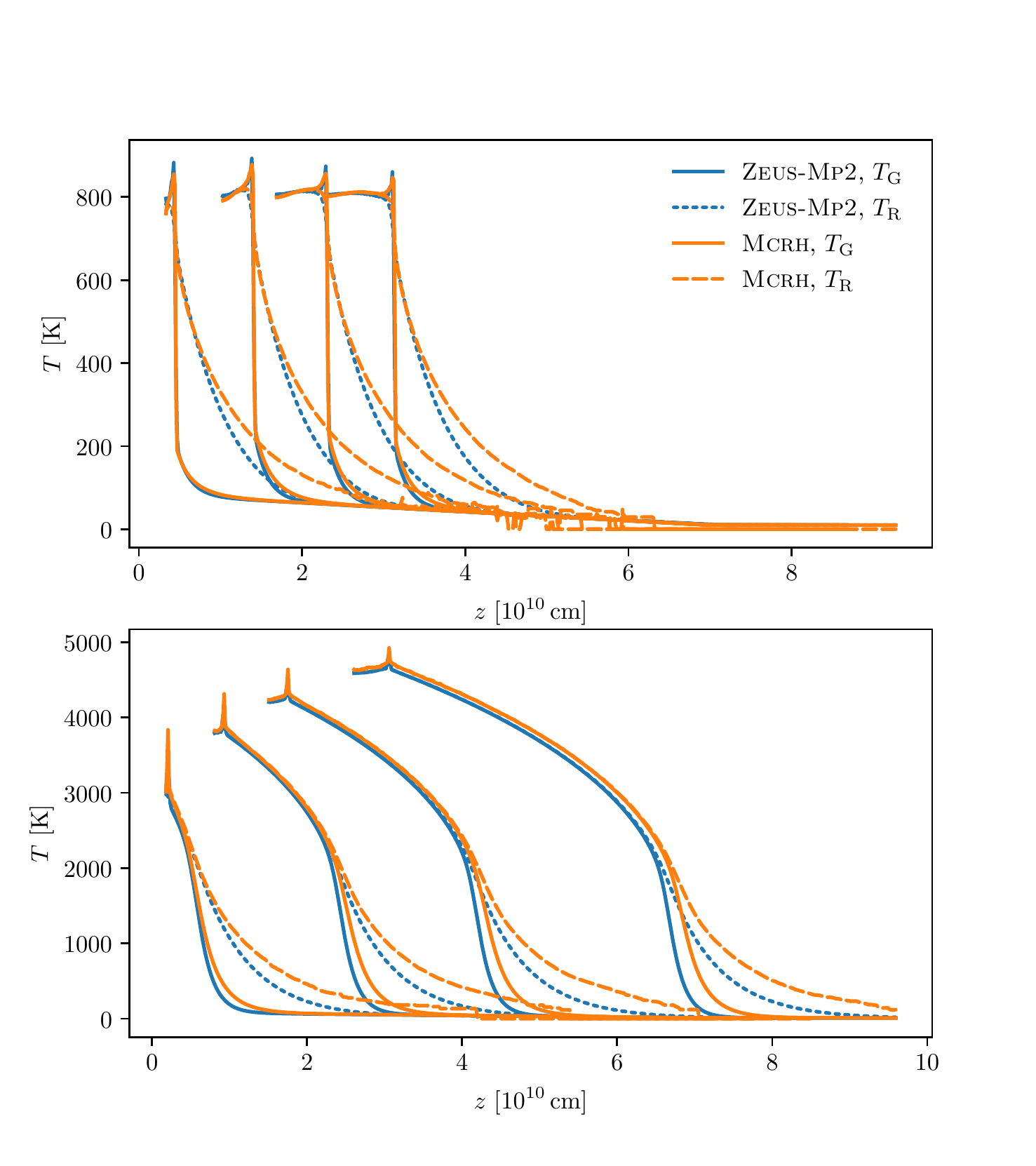}
  \caption{The temperature structure of sub-critical (top panel) and
    supercritical radiative shocks (bottom panel), calculated with
    \textsc{Mcrh} (orange) and \textsc{Zeus-Mp2} (blue). The gas (solid lines)
    and radiation temperature (dashed and dotted lines) are shown for 4
    different snapshots. These are $5.5 \times 10^3, 1.7 \times 10^4, 2.8
    \times 10^4$ and $\SI{3.8e4}{s}$ for the subcritical and $8.6 \times 10^2,
    4.0 \times 10^3, 7.5 \times 10^3$ and $\SI{1.3e4}{s}$ for the supercritical
    case. This illustration is adapted from \citet[][figs.\ 5 and
    6]{Noebauer2012}. More details about the setup are provided in
    \Cref{sec:radiative_shocks}.}
  \label{fig:dynamics:radiative_shocks}
\end{figure}

\begin{figure}[htb]
  \centering
  \includegraphics[width=\textwidth]{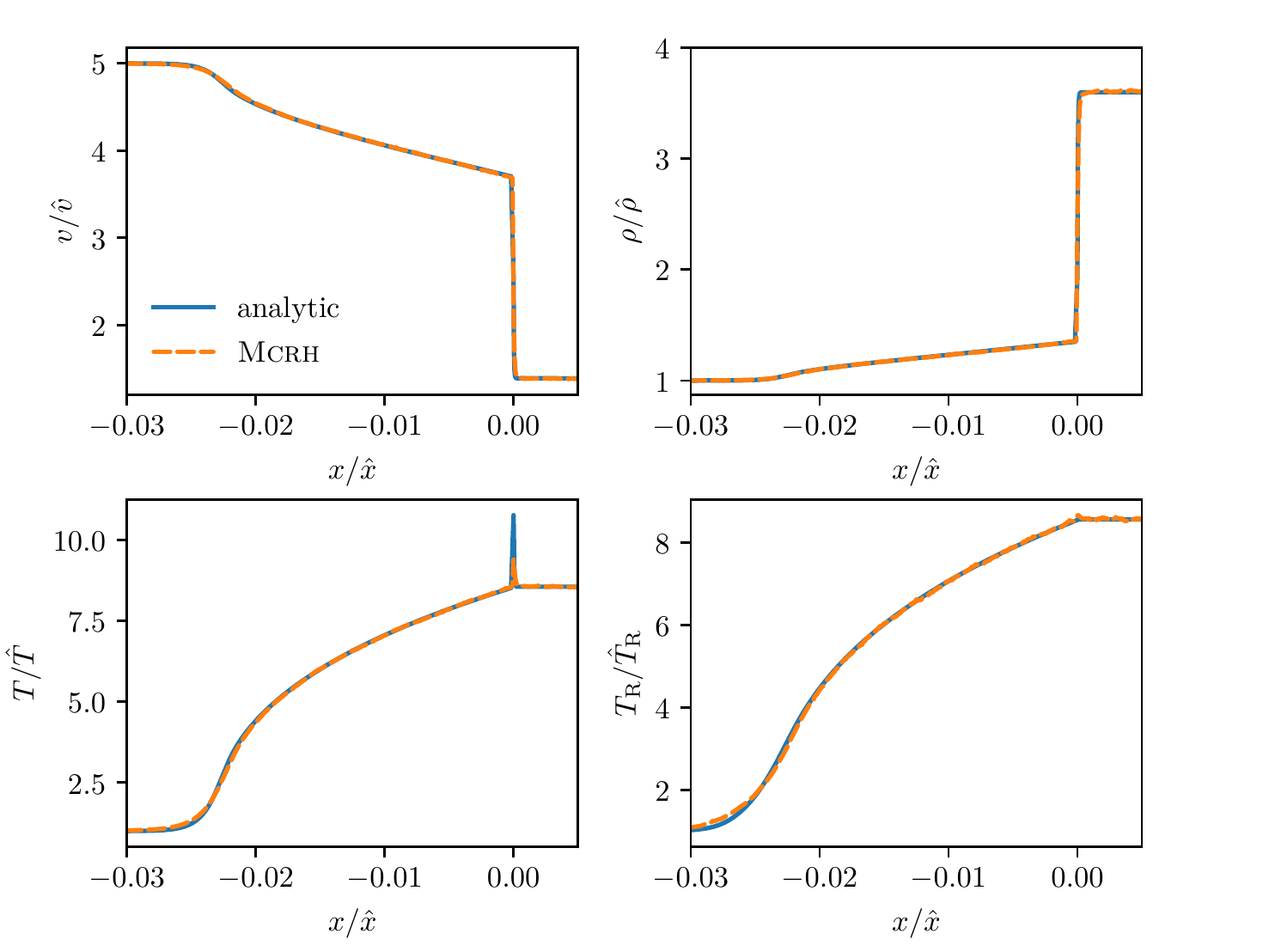}
  \caption{Comparison between the semi-analytic solution (blue solid)
  for the steady radiative shock with $M=5$ according to \citet{Lowrie2008} and
  the corresponding \textsc{Mcrh} results (orange dashed). From top left in
  clockwise direction, velocity, density, radiation temperature and gas
  temperature are shown. All quantities are displayed in their non-dimensional
  form. For details on this process and the numerical setup, consult 
  \Cref{sec:radiative_shocks}.}
  \label{fig:dynamics:steady_shocks}
\end{figure}

\Cref{fig:dynamics:radiative_shocks} shows the time evolution of the structure
of sub- and supercritical non-steady radiative shocks solved with the
\gls{MC}-based approach \textsc{Mcrh} \citep{Noebauer2012}, compared with the
results of calculations performed with the finite-difference approach
\textsc{Zeus-Mp2} \citep{Hayes2003, Hayes2006}. In addition,
\Cref{fig:dynamics:steady_shocks} shows the structure of a steady radiative
shock with Mach number $M = 5$ obtained with \textsc{Mcrh} in comparison with
the analytic predictions following the solution strategy developed by
\citet{Lowrie2008}\footnote{A Python implementation for this task can be found
at \url{https://github.com/unoebauer/public-astro-tools}}.
In both cases, the results of the \gls{MC} simulation agree very well
with the
reference calculation and the semi-analytic predictions respectively.

\subsection{Challenges and limitations}

Despite being conceptually simple and easily implemented, \gls{MC}-based
radiation hydrodynamical approaches relying on the operator splitting
techniques suffer from limitations. For a successful application of
operator-splitting, strict limits have to be set on the duration of the time
step. These restrictions are imposed by the characteristic time scales of the
source terms, most notably the heating and cooling terms in the energy
equations. The difficulties arising from these time-scale limits are best
illustrated at the example of \gls{TRT} (see also 
\Cref{Sec:imc}). If thermal emission is stronger than the corresponding
absorption of radiation, a characteristic cooling time can be
formulated\footnote{To
  $\mathcal{O}(v/c)$.} \citep[see][for an analogous definition]{Harries2011}
\begin{equation}
  t_{\mathrm{cool}} = \frac{\rho e}{cG^0}.
  \label{eq:dynamics:cooling_time}
\end{equation}
If a global time step larger than this value is chosen, thermal emission during
the \gls{RT} sub-step will completely deplete the internal energy content of
the ambient material and unphysical states with negative internal energy are induced
in the final step. This restriction renders the simple time-explicit operator
splitting \gls{MC} \gls{RH} approach inefficient in the stiff source term
regime, i.e.\ in situations in which the characteristic radiative time scales
are much shorter than the typical fluid-flow time scales, which in explicit schemes
are given by the Courant criterion \citep{Courant1928}.\footnote{The
  Courant condition essentially limits the duration of a simulation
  time step relative to the grid-cell crossing time for the
  characteristic fluid waves.}

The stiff source term problem is not unique to the \gls{MC} \gls{RH} problem
but a general challenge when dealing with source terms in hydrodynamical
calculations  \citep[cf.][]{LeVeque2002}. A common approach to address this
problem is to rely on implicit solution techniques. In this context, the
\gls{IMC} techniques outlined in \Cref{Sec:imc} seem very promising. In fact,
\citet{Roth2015} coupled an \gls{IMC} \gls{RT} scheme with a fluid dynamical
calculation and successfully applied it to test problems in which the radiative
time scales are smaller than the fluid-flow time scales. Nevertheless, as stressed
in \Cref{Sec:imc}, \gls{IMC} methods are not truly implicit in the traditional
sense and also suffer from other potential downsides, e.g.\ maximum principle
violation \citep[cf.][]{Wollaber2016}.

A completely different approach to the stiff source term problem was suggested
by \citet{Miniati2007}. An unsplit Godunov scheme was developed, consisting of
a modified predictor and a semi-implicit corrector step which incorporates the
effects of the source term. This method was adapted to \gls{RH} by
\citet{Sekora2010} and \citet{Jiang2012}. In principle, the hybrid Godunov
approach could also be utilised in \gls{MC}-based \gls{RH} calculations, but a
successful application of this scheme in conjunction with \gls{MCRT} methods
has yet to be demonstrated. 

Notwithstanding the challenges, \gls{MC}-based techniques constitute a valuable
alternative approach to \gls{RH}. Such methods offer the possibility to benefit
from the same advantages that \gls{MC} techniques already bring to pure
\gls{RT} calculations, namely a straightforward generalization to
multidimensional geometries and the ease with which complex interaction
processes are incorporated.

\section{Example astrophysical application}
\label{Sec:example}

We conclude this article by presenting a concrete example from our own
experience of how \gls{MCRT} methods can be used to solve \gls{RT} problems in
astrophysics. In this first version of our {\it Living Review}, we
will focus on a discussion of calculating synthetic
spectra for \glspl{SNIa}. This example, makes use of many techniques outlined in
this review, particularly, the indivisible energy packet
scheme (cf. \Cref{Sec:epackets}), a variant of the macro-atom scheme (cf.\
\Cref{Sec:macroatom}), volume-based estimators (cf.\
\Cref{Sec:volume_estimators}) and the peeling-off technique for variance
reduction (cf.\ \Cref{Sec:biasing}). Throughout the discussion we make use of the open source code
\textsc{Tardis} \citep{Kerzendorf2014, Kerzendorf2018}, which is readily
available\footnote{The code can be obtained from
\url{https://github.com/tardis-sn/tardis}} for inspection (or use) by the
interested reader. 

In future verstions of this {\it Review} we will plan to gradually
extend our discussion of examples. In particular, we aim to summarise closely
related work on the modelling of fast outflows for other classes of
astrophysical sources such as hot stars and accretion disk winds (see
references in
\Cref{Sec:History}). Such applications also make use of many of the
techniques outlined in this review and are generally quite closely
related to the methods used in the \gls{SNIa} example discussed here. 
The most important difference, arguably, is that the
SN problem often requires only an homologous velocity law, which leads to a
number of simplifications (see
\Cref{Sec:line_ints_sobolev}). In contrast, more general
stellar/disk wind applications require that more complicated velocity
fields are considered.

\subsection{Type Ia supernovae}

\glspl{SNIa} are transient events that have been instrumental in establishing
our currently accepted cosmological standard model and are still widely used in
precision cosmology \citep[see e.g.][]{Goobar2011}. In particular,
\citet{Riess1998} and \citet{Perlmutter1999} pioneered the use of \glspl{SNIa}
as standardisable distance indicators to map out the recent expansion history
of our Universe, finding an accelerated expansion. Apart from their relevance
in cosmological studies, \glspl{SNIa} play an important role in many other
branches of astrophysics as well, for example in galactic chemical evolution
\citep[e.g.][]{Kobayashi1998, Seitenzahl2017}.  Notwithstanding the importance
of \glspl{SNIa}, a full understanding of the exact nature of these transients
still remains elusive and a range of proposed models remain under study (see
e.g.\ \citealt{Hillebrandt2000, Hillebrandt2013, Roepke2017, Roepke2018}).  One
important strategy to study \glspl{SNIa} is to model their observed spectra with
the aim of inferring the ejecta composition and structure as a means to
understand the explosion itself.  \textsc{Tardis}, which we use for this
demonstration, is a tool aimed at this problem in which highly parameterized
and flexible \gls{RT} simulations are used to interpret observations.

\gls{MCRT} methods are well-suited for calculating synthetic observables in
\glspl{SNIa}. Due to the absence of hydrogen and helium and the dominance of heavy 
elements
in the ejecta of \glspl{SNIa}, \gls{RT} is mainly driven by bound-bound
interactions. As a consequence, \gls{SNIa} spectra show no true continuum but
rather a pseudo-continuum, generated by the flux redistribution achieved in a
multitude of non-resonant line interactions.  This property in combination with
the fact that many models predict anisotropies in the overall morphology and
chemical structure of \gls{SNIa} ejecta make \gls{MCRT} an attractive choice
for treating \gls{RT}.  Popular numerical approaches relying on \gls{MCRT} for
\gls{SNIa} studies include \textsc{Artis} \citep{Kromer2009}, \textsc{Sedona}
\citep{Kasen2006}, \textsc{SuperNu} \citep{Wollaeger2013, Wollaeger2014},
\textsc{Tardis} \citep{Kerzendorf2014, Kerzendorf2018, Vogl2018}, the scheme
developed by \citet{Mazzali1993} and \citet{Mazzali2000} and \textsc{Sumo}
\citep{Jerkstrand2011, Jerkstrand2012}.

\subsection{Model type Ia supernova}

Since \textsc{Tardis} was specifically designed as a highly parameterized
\gls{MCRT} approach for spectral synthesis in \glspl{SNIa}, it adopts a number
of simplifications. For a detailed overview we refer to the original
publication by \citet{Kerzendorf2014} and the publicly available
documentation\footnote{\url{http://tardis.readthedocs.io/en/latest/}}. Here, we
only highlight some of the key aspects of the \gls{MCRT} machinery of
\textsc{Tardis}.

Similar to the approach by \citet{Mazzali1993}, \textsc{Tardis} adopts the
elementary \gls{SN} model of \citet{Jeffery1990}. Here, the \gls{SN} ejecta are
approximated as spherically symmetric and divided into two domains, the
continuum-forming region and the atmosphere. A \textit{photosphere} separates
both regions. It is assumed that thermalization processes are only relevant below
the photosphere and that interactions in the atmosphere are either electron
scatterings in the Thomson limit or line interactions. \textsc{Tardis} follows
the spectral synthesis process in the atmosphere with a time-independent,
frequency-dependent \gls{MCRT} approach. Packets are launched from the
photosphere at the inner computational boundary from a thermal distribution
according to the photospheric temperature and followed as they propagate
through the envelope until escaping through either boundary.  An important
aspect of the \textsc{Tardis} approach is the determination of a
self-consistent plasma state and photospheric temperature, which is achieved
using volume-based estimator techniques akin to those outlined in
\Cref{Sec:volume_estimators} in an iterative process. Only after a converged
plasma state has been found, the final synthetic spectrum is calculated.
\textsc{Tardis} includes electron scattering and bound-bound interactions
relying on the Sobolev-approximation (see \Cref{Sec:line_ints_sobolev}).
Fluorescence can be treated either using the downbranching scheme by
\citet{Lucy1999a} or a simplified version of the macro atom scheme by
\citet[][see \Cref{Sec:macroatom}]{Lucy2002, Lucy2003}. To reduce the \gls{MC}
noise in the synthetic spectra, a variant of the peel-off technique can be
used, referred to as \textit{virtual packet scheme} (see \Cref{Sec:biasing}).
Different assumptions about the ioniziation and excitation state can be adopted
but for the \textsc{Tardis} simulations presented below, a modified nebular
approximation \citep[see][]{Mazzali1993} was used together with a
dilute-Boltzmann excitation treatment.  Finally, \textsc{Tardis} relies on a
discrete representation of the ejecta state in terms of density and velocity on
a spherical grid. For each grid cell, the mass density, the velocity at the
cell interfaces and the chemical composition have to be specified. Internally,
perfect homology is assumed, for example when progressing through the Sobolev
line interaction scheme (see \Cref{Sec:line_ints_sobolev}).

\subsection{Spectral synthesis with MCRT}

\begin{figure}[htb]
  \centering
  \includegraphics[width=\textwidth]{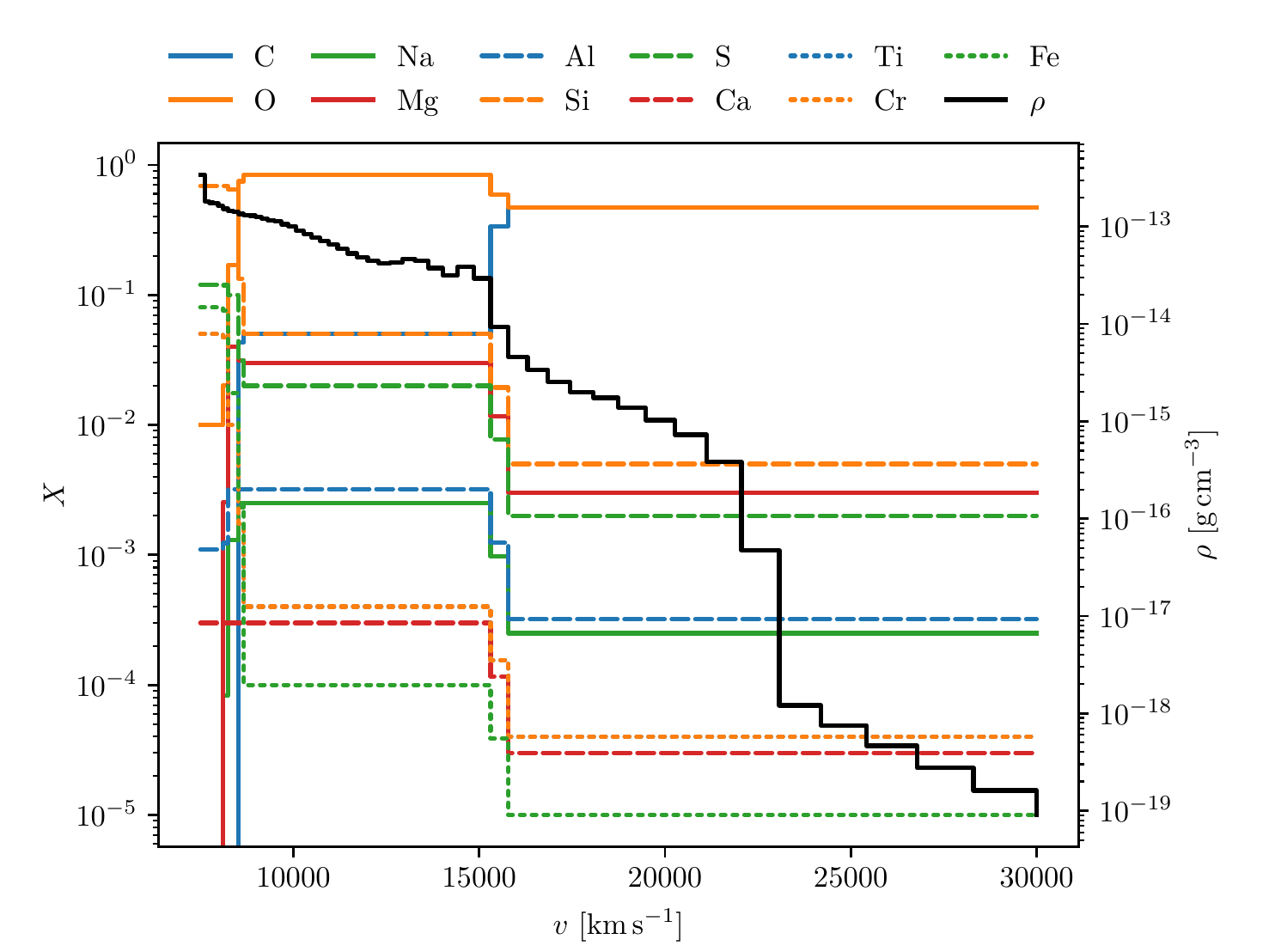}
  \caption{Density and composition of the input model used in the
    \textsc{Tardis} calculation for SN~2005bl. The mass fractions ($X$) of all
    elements that are present in the model are shown. This setup has been
    adapted by \citet{Barbosa2016} from
    \citet{Hachinger2009}.}
  \label{fig:tardis_input_model}
\end{figure}

As an illustration, we use \textsc{Tardis} to calculate a synthetic spectrum
for the \gls{SNIa} SN~2005bl.\footnote{This sub-luminous \gls{SNIa} belongs to
  a peculiar sub-class of these transients, which is named after the
  prototypical event, SN~1991bg. SN~2005bl is well-studied and a spherically
symmetric approximation to its ejecta structure has been previously estimated
by \citet{Hachinger2009} using the abundance tomography method developed by
\citet{Stehle2005}.} We consider the epoch three days before maximum light in
the $B$-band which corresponds to \SI{14}{d} after explosion. We adopt the
stratified chemical composition derived by \citet{Hachinger2009} and use a
density profile similar to the famous W7 explosion model by \citet{Nomoto1984}.
This setup, which is shown in \Cref{fig:tardis_input_model}, has been
previously used by \citet{Barbosa2016} to establish the suitability of
\textsc{Tardis} for abundance tomography studies. All necessary configuration
and data files to repeat the \textsc{Tardis} calculations are included in the
repository published as part of this review (see \Cref{sec:tool_collection}).

\begin{figure}[htb]
  \centering
  \includegraphics[width=\textwidth]{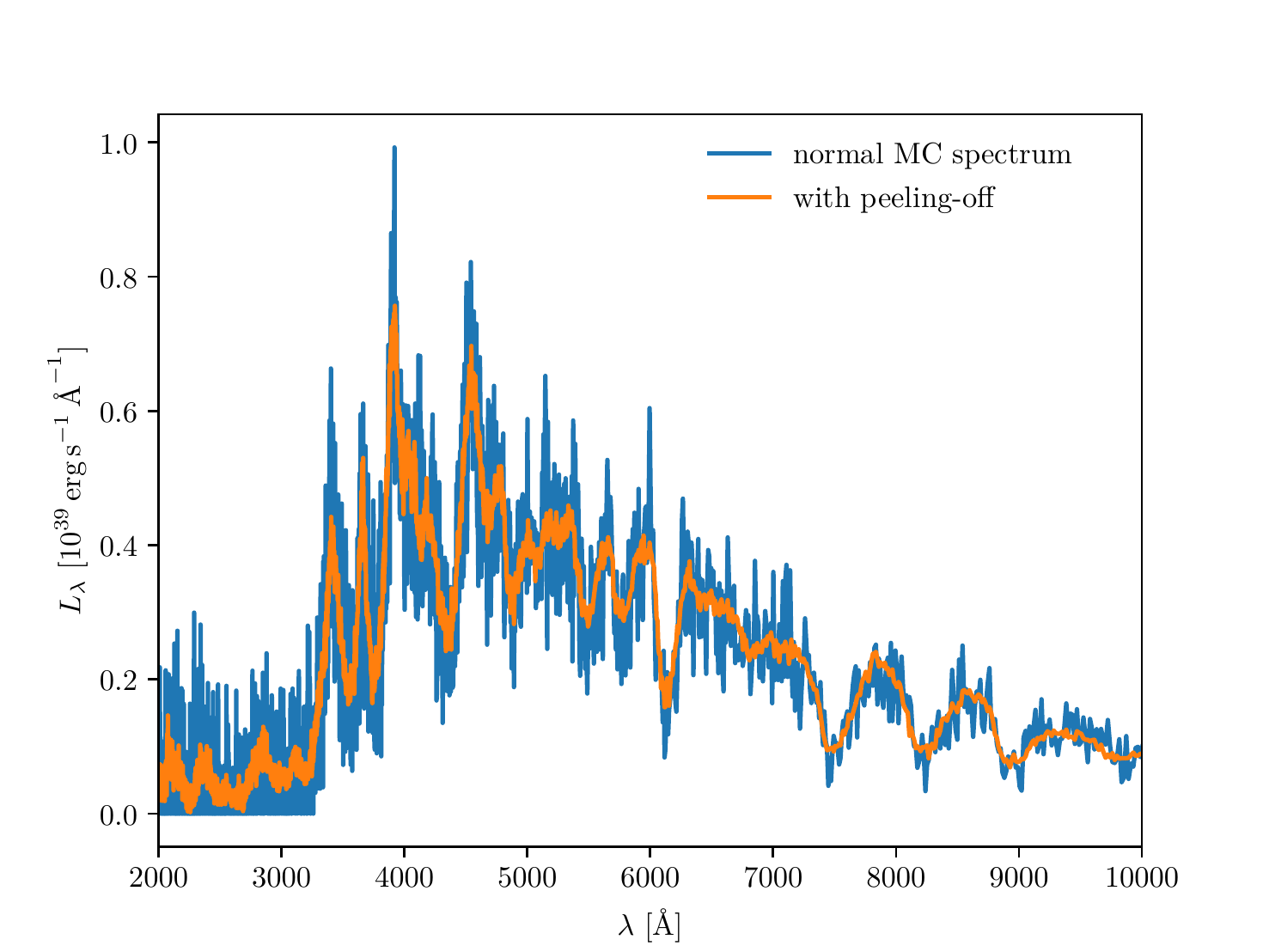}
  \caption{\textsc{Tardis} synthetic spectra for the \gls{SN} model constructed
    for SN~2005bl at \SI{3}{d} before $B$-band maximum. The spectrum
    constructed from the escaping \gls{MC} packets is shown in blue and
    exhibits high \gls{MC} noise. In addition, the synthetic spectrum which is
    generated from the virtual packets and which suffers from much less
  \gls{MC} noise, is shown in orange.}
  \label{fig:tardis_spectrum}
\end{figure}

\Cref{fig:tardis_spectrum} shows the main product of a \textsc{Tardis}
calculation, namely the synthetic spectrum for the model setup.
Since the optical depth of the constructed \gls{SN} atmosphere is rather high,
many \gls{MC} packets injected at the lower boundary are back-scattered onto
the photosphere and lost for the spectral synthesis process. Only a small
fraction of the launched packets reach the ejecta surface and contribute to the
emergent spectrum, leading to a substantial amount of \gls{MC} noise. This
situation can be significantly improved by using the implemented virtual
packet scheme.  Whenever a \gls{MC} packet is launched or interacts, a
pre-defined number of virtual packets (ten in the current \textsc{Tardis}
simulation) are spawned and propagated towards the ejecta surface along rays
that are cast in directions drawn from the emission profile of the
corresponding process. The optical depth to the surface is calculated along
these rays and the energy of the virtual packet decreased by a corresponding
attenuation factor (see \Cref{Sec:biasing} or \citealt{Kerzendorf2014} for more
details). \Cref{fig:tardis_spectrum} also includes the synthetic spectrum
generated from the virtual packets which has a much lower noise level than the
spectrum that is based on the real packet population.

\begin{figure}[htb]
  \centering
  \includegraphics[width=\textwidth]{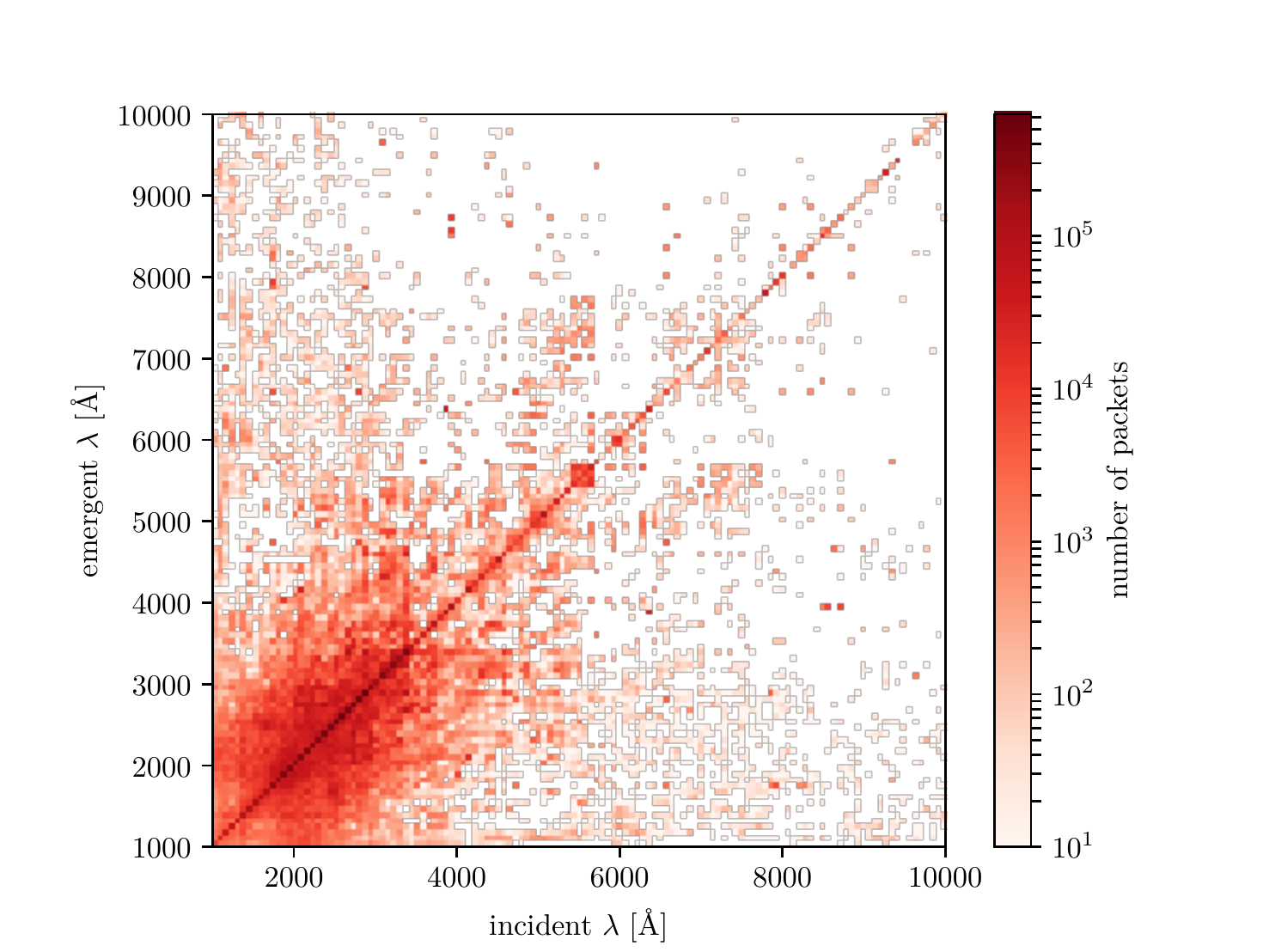}
  \caption{Histogram of the incident and emergent wavelength of all escaping
    \gls{MC} packets in the last line interaction from the \textsc{Tardis}
    simulation of SN~2005bl. Resonance interactions can be found on the
    diagonal, fluorescence above it and inverse-fluorescence process below it.
    The macro atom scheme of \citet{Lucy2002,Lucy2003} was used in this
    simulation (see also \Cref{Sec:macroatom}).}
  \label{fig:tardis_redistribution}
\end{figure}

One advantage of \gls{MCRT} lies in the diagnostic possibilities this approach
offers. Details about the interactions packets experienced can be easily
recorded and used to examine the radiation--matter coupling or to investigate
the origin of particular features in the SED of the emergent radiation field.
In the following, we highlight only some possible applications of these
capabilities. For simplicity, we will only focus on the last interaction
\gls{MC} packets performed before escaping through the outer
boundary\footnote{This limitation has only book-keeping reasons. There are no
conceptual obstacles to record and diagnose the entire interaction histories of
all packets.}. \Cref{fig:tardis_redistribution} illustrates the importance of
non-resonant line interactions when calculating synthetic spectra for
\glspl{SNIa}. All emergent packets have been binned according to their incident
and emergent wavelengths in their last line interaction. In the \textsc{Tardis}
simulations shown here, the macro atom scheme is used to treat non-resonant
interactions within the indivisible energy packet paradigm.
While \Cref{fig:tardis_redistribution} shows that many packets have interacted
resonantly (cf.\ diagonal where $\lambda_{\mathrm{in}} =
\lambda_{\mathrm{out}}$), the fluorescence and inverse-fluorescence regions
above and below the diagonal are also densely populated.

\begin{figure}[htb]
  \centering
  \includegraphics[width=\textwidth]{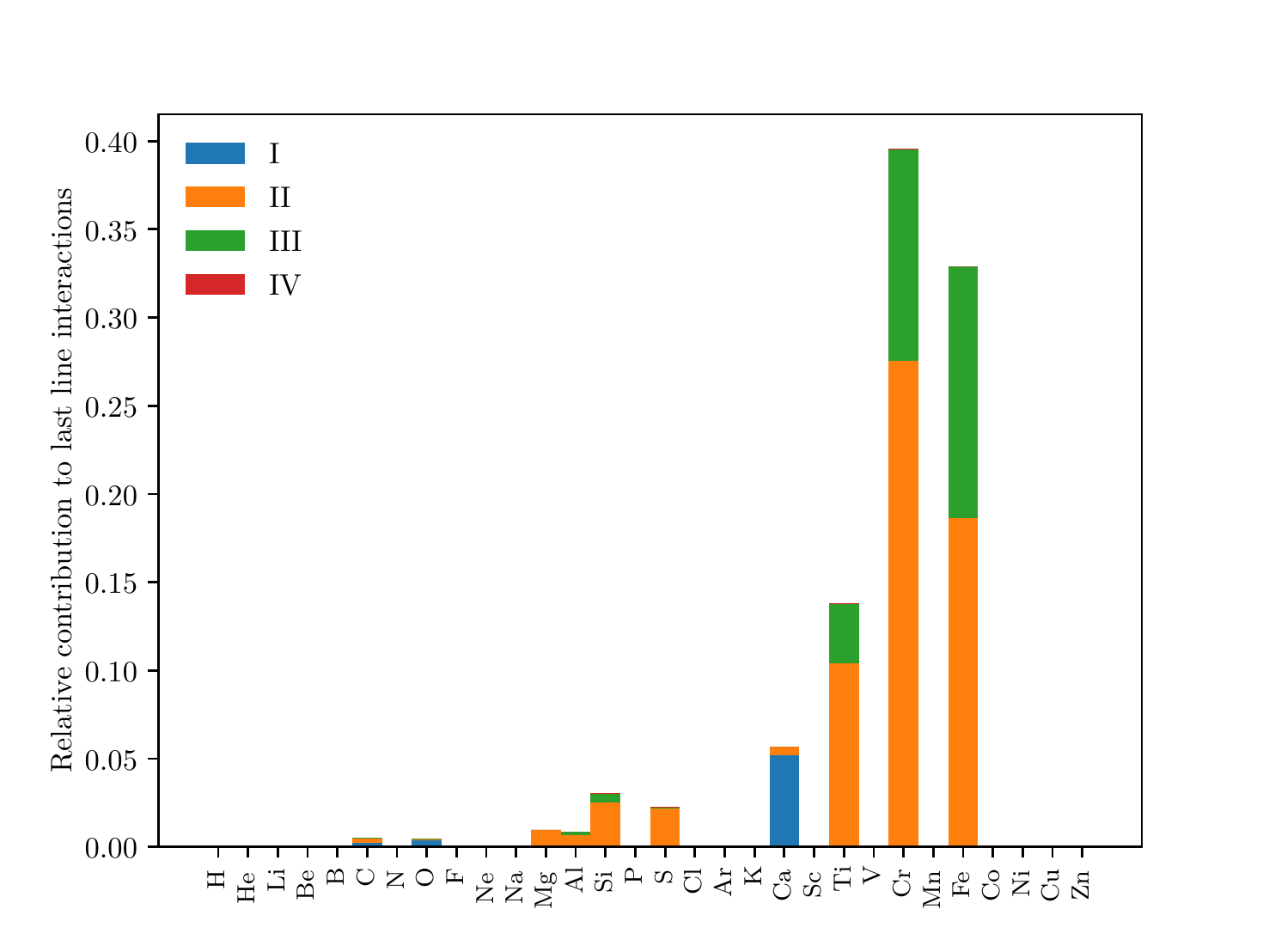}
  \caption{Contribution of the different ions (note that ionization stage I
  corresponds to neutral atoms) to the last line interactions escaping \gls{MC}
packets experienced in the \textsc{Tardis} simulation of SN~2005bl.}
  \label{fig:tardis_ion_contrib}
\end{figure}

In analogy to extracting information about the wavelength redistribution,
details about the interaction process can be recorded just as easily.
\Cref{fig:tardis_ion_contrib} shows which ions predominantly contribute to the
last line interactions \gls{MC} packets experience in the \textsc{Tardis}
simulation of SN~2005bl.
It clearly illustrates, that singly- and doubly- ionized iron-group elements
are the dominant interaction partners, followed by the intermediate-mass
elements in the same ionization state.

\begin{figure}[htb]
  \centering
  \includegraphics[width=\textwidth]{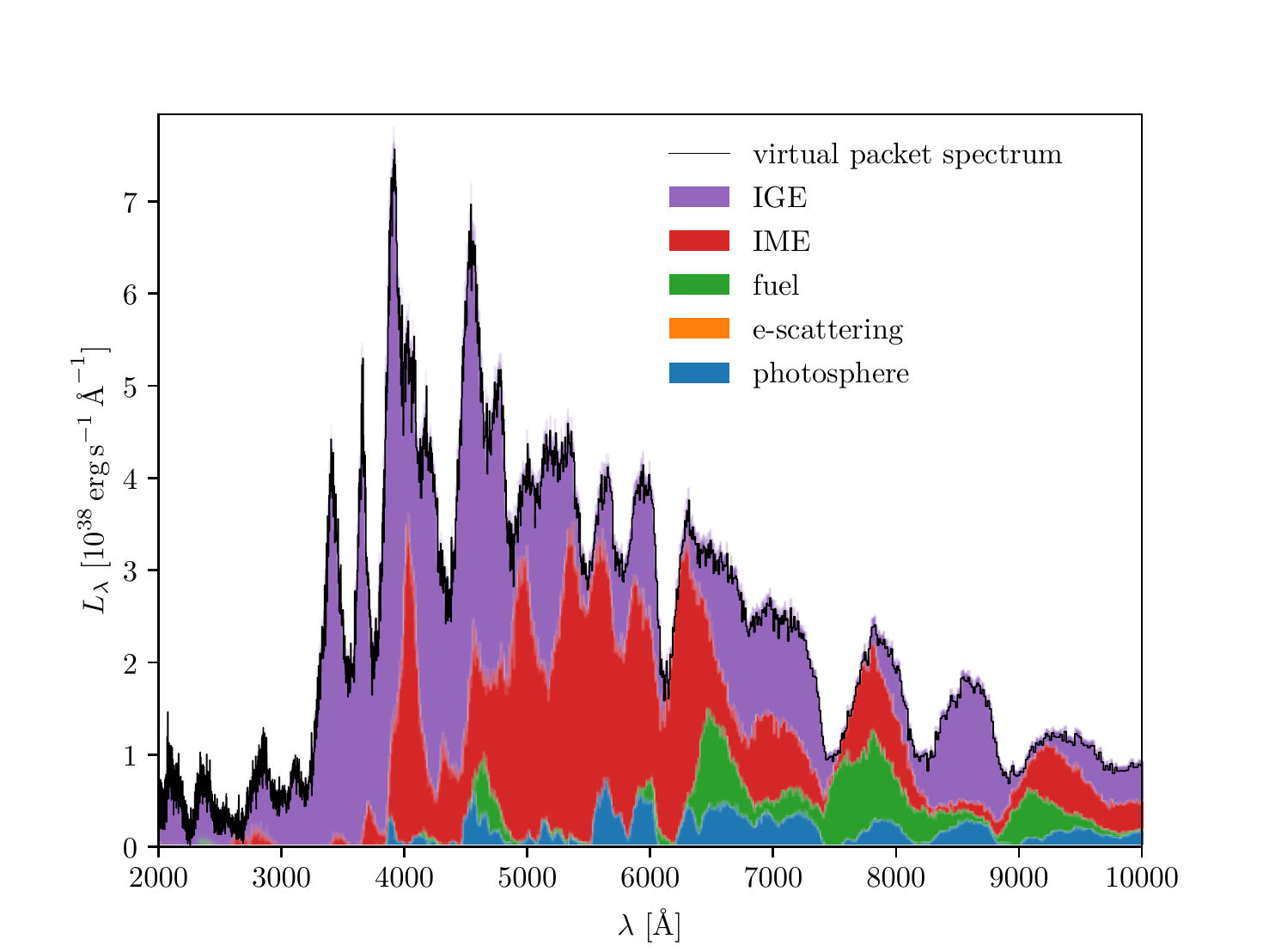}
  \caption{Illustration of the contributions of the various elemental groups to
    the final emergent spectrum in the \textsc{Tardis} simulation of SN~2005bl.
    In particular, the contribution of each escaping (virtual) \gls{MC} packet
    to the final spectrum is colour-coded according to the last interaction
    partner.  In addition to the contributions of the different elemental
    groups, packets that escaped without interacting straight from the inner
    boundary are shown as well (``photosphere''), together with packets that
    performed electron scatterings as their last interactions.}
  \label{fig:tardis_kromer_plot}
\end{figure}

Finally, we combine the information about the interaction partner and the
wavelength change into a visualization proposed by M.~Kromer \citep[see
e.g.][]{Kromer2009}. This provides detailed information about the spectrum
formation process. The contribution of each escaping packet to the emergent
spectrum is colour-coded according to the atomic number of the last interaction
partner and plotted at the location of the emergent wavelength. This procedure
can be performed on the level of individual elements, or as we chose to do here
for simplicity, by elemental groups. \Cref{fig:tardis_kromer_plot} shows the
synthetic spectrum calculated with \textsc{Tardis} and how the different
elemental groups, fuel (C, N, O, Ne), intermediate-mass elements (Na
through Sc) and iron-peak elements, contribute.

\section{Summary and conclusions}
\label{Sec:Phew!}

In this work, we provide an overview of some of the \gls{MCRT} techniques used
in astrophysics. We have presented a variety of evidence that this approach has
evolved into a competitive and very successful method to solve radiative
transfer problems. With its probabilistic approach, \gls{MCRT} offers a number
of compelling advantages that make this technique ideal for a variety of
astrophysical applications. Whenever irregular multidimensional geometries are
encountered or complex interaction processes, particularly scatterings, have to
be accounted for, \gls{MCRT} methods are typically a good choice for addressing
radiative transfer problems. For this reason, the \gls{MCRT} framework finds
wide-spread application in astrophysics, from modelling mass-outflows from
stars and accretion discs, to simulating radiative transfer through dusty
environments or studying ionization on cosmological scales. Recently,
\gls{MCRT} schemes have even been included in fully dynamic radiation
hydrodynamics calculations.

Relying on \gls{MCRT} approaches, however, always comes at the cost of
introducing statistical fluctuations into the solution process. Nevertheless, a
variety of variance reduction techniques have been developed over the years to
keep this noise component under control -- many of these methods have been
reviewed in this work. Also, conventional \gls{MCRT} approaches are ill-suited
for the application to optically thick environments and to problems with short
cooling time scales. Extensions and modifications, particularly \gls{MC}
diffusion schemes and the \gls{IMC} approach, have been developed to alleviate
these deficiencies and have already found their application in astrophysical
\gls{MCRT} calculations.

Finally, we want to emphasize an important aspect of \gls{MCRT} methods, the
value of which should not be under-rated: the \gls{MCRT} approach of performing
a simulation of radiative transfer by following the propagation of packets is
very intuitive since it closely resembles the microphysical processes realised
in nature. Furthermore, the fundamental \gls{MCRT} concepts are quite simple
and basic computer programs can be developed quickly with only a handful of
instructions. The directness of the physics and simplicity of the algorithms
also mean that it is typically fairly easy to develop codes by gradually
upgrading the physics: incorporating new physical processes rarely requires any
fundamental overhaul. All this, together with the fact that many
state-of-the-art \gls{MCRT} simulation codes for astrophysical applications are
open source and freely available, makes the entrance barrier quite low for the
adoption of \gls{MCRT}. As the continuous increase in the availability of
computational resources seems to hold and since \gls{MC} calculations can
easily be distributed over multiple computation units, it seems more than
likely that the success \gls{MCRT} will continue. 

\begin{acknowledgements}
\label{acknowledgments}

We dedicate this work to the memory of Leon~B.\ Lucy who, with his many
invaluable contributions, had a profound influence on the development of Monte
Carlo radiative transfer techniques in astrophysics. One of us (SAS) had the
privilege of working with LBL during his time at Imperial College London and
wishes to acknowledge the large number of insightful discussions
with LBL.

We both (UMN/SAS) wish to thank Knox~S.\ Long who has, for over a decade, been a
collaborator and sounding board for many projects. UMN first came into contact with
Monte Carlo radiative transfer when working with KSL, an experience that
instilled a fascination for the subject that ultimately lead to the development
of this work.  

Nick Higginbottom, Wolfgang Kerzendorf, Christian Knigge, James Matthews, Jorick Vink and Christian Vogl are
thanked for many fruitful discussions on topics included in this review. We are
also very grateful to R\'emi Kazeroni and J\'er\^ome Guilet for their help with
the original publications in French that were used for the historical sketch of
the Monte Carlo approach.

Finally, we would like to express our sincere thanks to Markus Kromer and
Wolfgang Hillebrandt for their thoughtful comments and suggestions in
the prepration of this review, and for many interesting and productive
collaborations over the years.

\end{acknowledgements}

\glsaddall

\printnoidxglossary[type=abbrev]


\appendix
\normalsize 
\section{Test problems}

For this review, we use a number of simple test problems to illustrate
various techniques relevant for \gls{MCRT} calculations. In the
following, we introduce these test problems.

\subsection{Homogeneous sphere}
\label{sec:test_hom_sphere}

A frequently used test setup to verify and validate numerical approaches to
\gls{RT} is that of radiation emerging from a homogeneous sphere \citep[see
e.g.][]{Smit1997, Abdikamalov2012}. In particular, we consider a homogeneous
sphere of radius $R$ composed of a material with constant opacity $\chi$ and
emissivity $\eta$ (and thus with a constant source function $S$).  The sphere
is assumed to be surrounded by vacuum. The structure of the steady-state
radiation field inside and outside of the sphere can be obtained from the
formal solution to the transfer equation \citep[cf.][]{Smit1997} and follows
\begin{equation}
  I(r, \mu) = S \left( 1 - \exp[-\chi s(r,\mu)] \right)
  \label{eq:hom_sphere_I}
\end{equation}
with
\begin{align}
  s(r, \mu) = \begin{dcases*}
    r\mu + R g(r, \mu) & if $r < R$ \\
    2 R g(r, \mu) & if $r \ge R$, $\mu_{\star} \le \mu \le 1$\\
    0 & else
  \end{dcases*}
  \label{eq:hom_sphere_s}
\end{align}
and 
\begin{align}
  g(r, \mu) & = \sqrt{1 - \left( \frac{r}{R} \right)^2 (1 - \mu^2)}\label{eq:hom_sphere_g},\\
  \mu_{\star} & = \sqrt{1 - \left(\frac{R}{r} \right)^2}.
\end{align}
After performing the appropriate angle averaging (cf.\ \Cref{sec:RT}), the
moments of the specific intensity are obtained. Inside the sphere ($r < R$),
they follow \citep[cf.][]{Smit1997}
\begin{align}
  J(r) & = S \left[ 1 - \int_0^1\mathrm{d} \mu \cosh(\chi r \mu) \exp(-\chi R g(r, \mu)) \right], \label{eq:hom_sphere_J_inside}\\
  H(r) & = S \int_0^1 \mathrm{d}\mu \mu \sinh(\chi r \mu) \exp(-\chi R g(r, \mu)), \label{eq:hom_sphere_H_inside}\\
  K(r) & = S \left[ \frac{1}{3} - \int_0^{1}\mathrm{d} \mu \mu^2 \cosh(\chi r \mu) \exp(-\chi R g(r, \mu)) \right],
\end{align}
while outside ($r \ge R$), they are given by 
\begin{align}
  J(r) & = \frac{1}{2} S \left[ (1 - \mu_{\star}) - \int_{\mu_{\star}}^1 \mathrm{d}\mu \exp(-2 \chi R g(r, \mu)) \right] ,\\
  H(r) & = \frac{1}{2} S \left[ \frac{1}{2} (1 - \mu_{\star}^2) - \int_{\mu_{\star}}^1 \mathrm{d}\mu \mu \exp(-2 \chi R g(r, \mu)) \right], \label{eq:hom_sphere_H_outside}\\
  K(r) & = \frac{1}{2} S \left[ \frac{1}{3} (1 - \mu_{\star}^3) - \int_{\mu_{\star}}^1 \mathrm{d}\mu \mu^2 \exp(-2 \chi R g(r, \mu)) \right]. \label{eq:hom_sphere_K_outside}
\end{align}

\begin{figure}[htb]
  \centering
  \includegraphics[width=\textwidth]{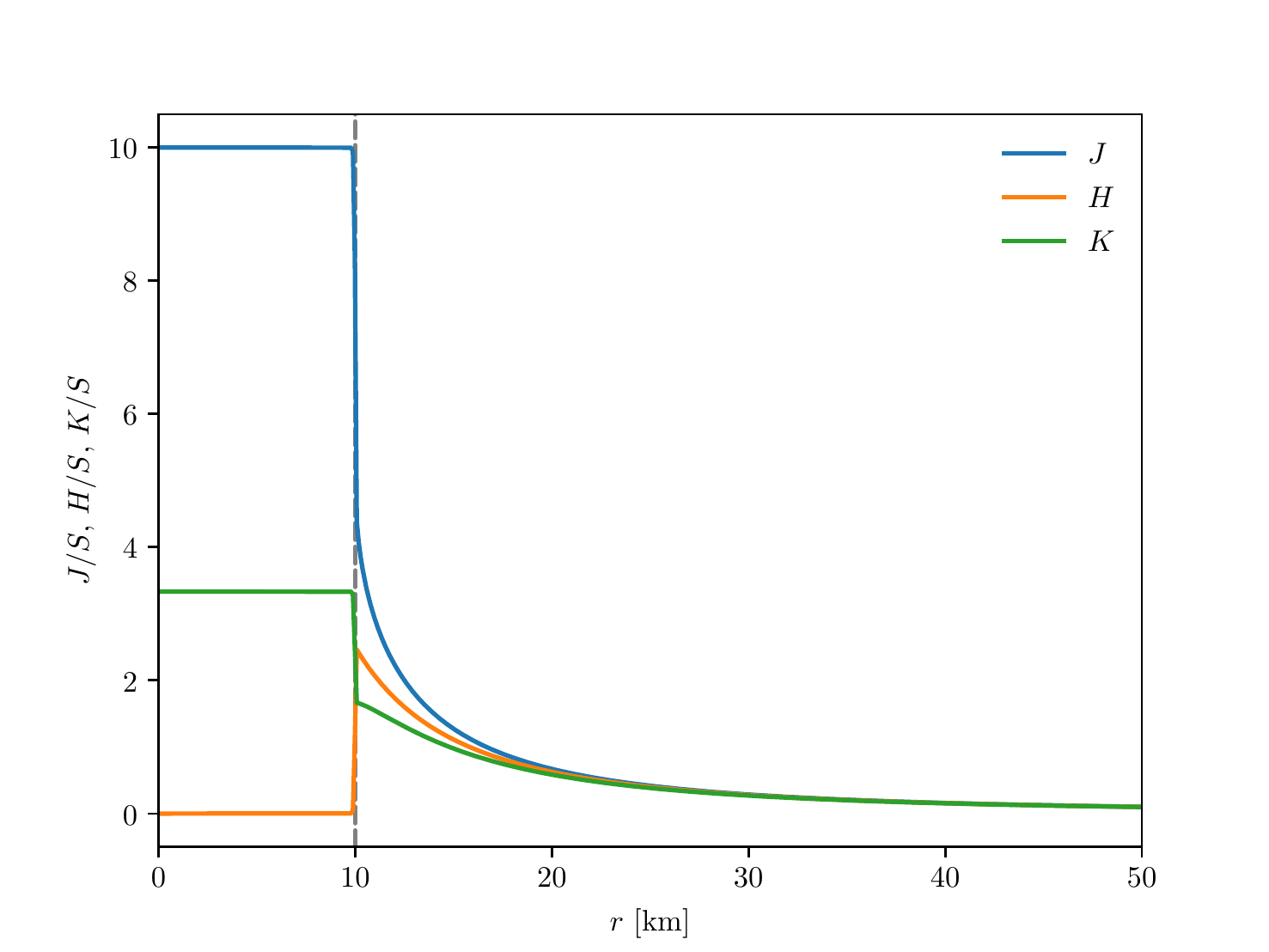}
  \caption{The first three moments of the specific intensity for 
the steady-state radiation field in the homogeneous sphere problem
    as given by
    \Crefrange{eq:hom_sphere_J_inside}{eq:hom_sphere_K_outside}. The extent of
  the homogeneous sphere is indicated by a dashed grey line and the
  moments are expressed in units of the source function, $S$.}
  \label{fig:homogeneous_sphere_J_H}
\end{figure}

For the test calculations presented in this work, specifically in
\Cref{Sec:Estimators}, we adopt the parameters suggested by
\citet{Abdikamalov2012}. In particular, a homogeneous sphere with radius $R =
\SI{10}{km}$ and the constant absorption opacity $\chi = \SI{2.5e-4}{cm^{-1}}$
and source function $S = \SI{10}{erg.cm^{-2}.s^{-1}}$ on the inside is
considered.  In the \gls{MCRT} test simulations the sphere is embedded in a
computational domain that extends out to $r = \SI{50}{km}$ and is divided into
100 equidistant shells.  \Cref{fig:homogeneous_sphere_J_H} shows the analytic
solution for this homogeneous sphere test problem in terms of $J$, $H$, and $K$
according to \Crefrange{eq:hom_sphere_J_inside}{eq:hom_sphere_K_outside}.

\begin{figure}[htb]
  \centering
  \includegraphics[width=\textwidth]{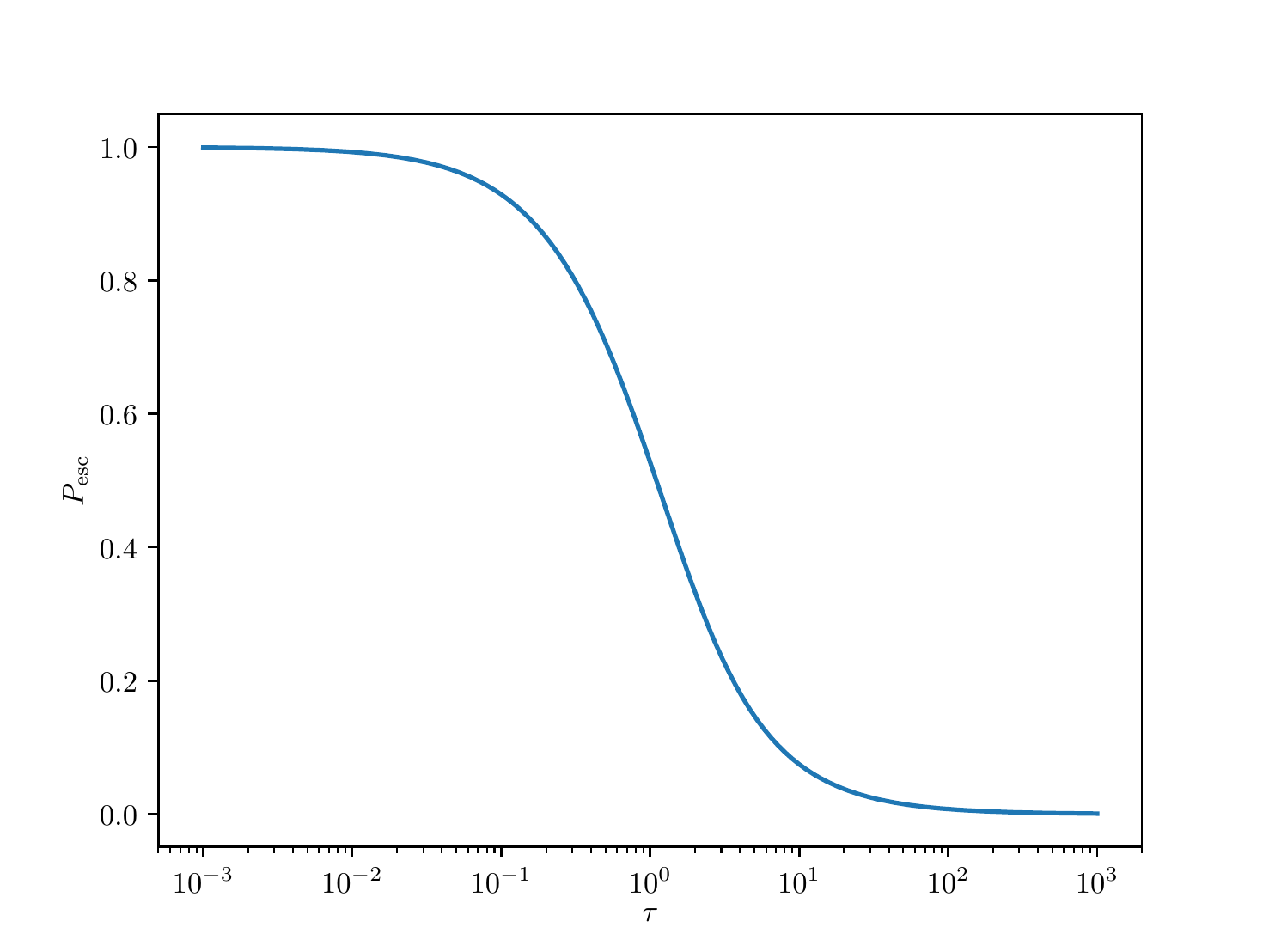}
  \caption{Escape probability from a homogeneous sphere according to
    \Cref{eq:hom_sphere_escape_probability} as a function of its optical
  depth, $\tau = \chi R$.}
  \label{fig:hom_sphere_escape_probability}
\end{figure}

The homogeneous sphere problem is often discussed from a slightly different
viewpoint, namely when the probability of photons escaping from such a sphere
is of interest. This question has been discussed in detail by \citet[][Appendix
2]{Osterbrock1974} and \citet[][Section 4.5]{Osterbrock2006} in the context of
gaseous nebulae and an analytic expression is derived and presented there. By
considering the emergent flux at the surface of the sphere (i.e.\
\Cref{eq:hom_sphere_H_outside}, evaluated at $r=R$) and relating it to the
expected flux in the absence of absorption (i.e.\ $\chi=0$), the escape
probability as a function of optical depth ($\tau$) 
\begin{equation}
  p(\tau) = \frac{3}{4 \tau} \left[ 1 - \frac{1}{2 \tau^2} + \left( \frac{1}{\tau} + \frac{1}{2 \tau^2} \right) \exp(-2 \tau) \right]
  \label{eq:hom_sphere_escape_probability}
\end{equation}
is obtained (see also \Cref{fig:hom_sphere_escape_probability}). Throughout
\Cref{Sec:Propagation} and \Cref{Sec:Estimators}, we determine the escape
probability from the homogeneous sphere with \gls{MCRT} simulations and compare
the results to the predictions according to
\Cref{eq:hom_sphere_escape_probability}. 

\subsection{Line profile for a sphere in homologous expansion}
\label{sec:line_test}

To test the line formation process implemented in Sobolev-based \gls{MCRT}
schemes in \Cref{Sec:line_ints_sobolev} we use a simple setup, which aims at
predicting the H Lyman $\alpha$ line profile emanating from a homologous flow composed
of neutral hydrogen.  Specifically, we consider a spherical domain with an
inner and outer boundary at $R_{\mathrm{inner}}$ and $R_{\mathrm{outer}}$. The
material is assumed to be in perfect homologous expansion, i.e.\ $v = r/t$, and
we set the extent of the domain by choosing the time $t = \SI{13.5}{d}$ and the
minimum and maximum material velocities $v_{\mathrm{min}} = 10^{-4}\,c$ and
$v_{\mathrm{max}} = 10^{-2}\,c$. We neglect time-dependence and follow the
propagation of $N = 10^5$ packets that are emitted from the lower boundary at
$R_{\mathrm{inner}}$. Their initial frequency is uniformly drawn from the
interval corresponding to the wavelength range $\lambda_{\mathrm{min}} =
\SI{1185}{\angstrom}$ to $\lambda_{\mathrm{max}} = \SI{1255}{\angstrom}$. We
assume that line interactions proceed resonantly at the natural wavelength
$\lambda_{\mathrm{line}} = \SI{1215}{\angstrom}$ and that their strength
throughout the flow is given by a constant Sobolev optical depth
$\tau_{\mathrm{s}} = 1$.  All packets are followed until they either escape
through the outer domain edge at $R_{\mathrm{outer}}$ and contribute to the
line profile or are back-scattered onto the inner boundary and are discarded.
For this illustration, we only include the Doppler effect and further simplify
the transformations by working in the weakly-relativistic limit, thus setting
$\gamma = 1$. 

The \gls{MCRT} simulation starts by launching the packets at the inner
boundary. These packets are initialized with $r_{\mathrm{ini}} = R_{\mathrm{inner}}$, a
frequency sampled from
\begin{equation}
  \nu = \nu_\mathrm{min} + \xi (\nu_\mathrm{max} - \nu_\mathrm{min}),
\end{equation}
a distance to the next interaction, $\tau$, given by
\Cref{eq:beer_lambert_sampling} and an initial propagation direction drawn from
\Cref{eq;photosphere_init_mu}. At the beginning of the packet propagation, the
distance to the outer domain edge is calculated
\begin{equation}
  l_{\mathrm{edge}} = -\mu_{\mathrm{ini}} r_{\mathrm{ini}} + \sqrt{(\mu_{\mathrm{ini}} r_{\mathrm{ini}})^2 - r_{\mathrm{ini}}^2 + R_{\mathrm{outer}}},
  \label{eq:line_test_distance_to_edge}
\end{equation}
together with the distance to the Sobolev point
\begin{equation}
  l_{\mathrm{s}} = c t \left( 1 - \frac{\nu_{\mathrm{line}}}{\nu} \right) - r_{\mathrm{ini}} \mu_{\mathrm{ini}}.
  \label{eq:line_test_distance_to_sobolev}
\end{equation}
If $l_{\mathrm{edge}} < l_{\mathrm{s}}$, the packet escapes without interacting
and contributes to the emergent spectrum with $\nu_{\mathrm{esc}} = \nu$.
Otherwise, it reaches the Sobolev point where its properties are updated to
\begin{align}
  r_{\mathrm{s}} &= \sqrt{r_{\mathrm{ini}}^2 + l_{\mathrm{s}} + 2 l_{\mathrm{s}} r_{\mathrm{ini}} \mu_{\mathrm{ini}}}, \label{eq:line_test_sobolev_r}\\
  \mu &= l_{\mathrm{s}} + \mu_{\mathrm{ini}} \frac{r_{\mathrm{ini}}}{r},\label{eq:line_test_sobolev_mu}
\end{align}
and it comes into resonance with the line transition. If 
\begin{equation}
  \tau < \tau_{\mathrm{s}},
  \label{eq:line_test_sobolev_interaction}
\end{equation}
the packet interacts at the Sobolev point. In this case, the \gls{LF} frequency
is updated according to the energy conservation principles outlined in
\Cref{Sec:mixed_frame}:
\begin{equation}
  \nu = \nu \frac{1 - \beta \mu_{\mathrm{i}}}{1 - \beta \mu_{\mathrm{e}}}.
\end{equation}
Here, $\mu_{\mathrm{i}}$ is the incident propagation direction determined by
\Cref{eq:line_test_sobolev_mu} and $\mu_{\mathrm{e}}$ denotes the direction
into which the packet emerges after scattering. In homologous flows, the Sobolev escape
probability is (approximately) direction independent. Thus, a new propagation
direction can be drawn isotropically and the packet propagation then continues.
In this simple illustration, the ensuing propagation process is trivial. Since
the packet only redshifts in a homologous flow, it cannot come into resonance
and interact with the line transition again. What remains is to determine
whether the packet has been backscattered or if it propagates towards the outer
domain boundary. If
\begin{equation}
  \mu < - \sqrt{1 - \left(\frac{R_{\mathrm{inner}}}{r_{\mathrm{s}}}\right)^2},
\end{equation}
the packet intersects the inner boundary and is discarded. Otherwise, it
escapes and contributes with $\nu_{\mathrm{esc}} = \nu$ to the line profile.
The final line profile is determined after all packets have been processed by
binning the frequencies of the escaping ones.

\subsection{Model supernova}
\label{sec:lucy2005_test}

\citet{Lucy2005} presented a simple but powerful test problem to verify the
performance of \gls{RT} schemes for the calculation of \gls{SNIa} light curves.
In this spherically symmetric setup, radiative energy is non-uniformly
generated throughout the calculation. This simulates the energy liberated in
the decay of non-uniformly distributed $^{56}$Ni.  In particular, spherically
symmetric ejecta in homologous expansion with a total mass of $M_{\mathrm{tot}}
= 1.39\,M_{\odot}$ and uniform density are considered. The maximum material velocity is assumed to
be $v_{\mathrm{max}} = \SI{e4}{km.s^{-1}}$ and the composition is chosen as
follows: the inner ejecta regions up to $M_r = 0.5\,M_{\odot}$ are entirely
made up from $^{56}$Ni. Its abundance then drops linearly until it reaches zero
at $M_r = 0.75\,M_{\odot}$. The remaining mass is composed of carbon and oxygen
in equal parts. The energy released in the decay of radioactive material is
first distributed into \gls{MC} packets representing $\gamma$-rays, which are propagated
through the model \gls{SN}.  For these $\gamma$-packets, a purely absorptive
specific cross section, $\kappa = \SI{0.03}{cm^2.g^{-1}}$ is assumed. When
$\gamma$-packets are absorbed, they are instantly converted into \gls{MC}
packets representing the ultraviolet-optical-infrared radiation
field\footnote{Note that this test is performed without a specific frequency
association.}. Interactions for these packets are treated as isotropic
resonant scatterings and their strength is given by a uniform specific cross
section $\sigma = \SI{0.1}{cm^2.g^{-1}}$. A time-dependent \gls{MCRT}
simulation is performed by following the packets which represent the energy
release from the decay until they leave through the ejecta surface. We note
that we do not start the \gls{MCRT} simulation promptly after explosion but
at $t = \SI{3}{d}$. This is advised since the ejecta are initially very
optically thick and virtually opaque, rendering an application of a
conventional \gls{MCRT} approach (without the techniques outlined in
\Cref{Sec:Dynamics}) to determine the emergent radiation field inefficient and
unnecessary. We simply expand the ejecta in accordance with the homologous
velocity law to the start time of the simulation and keep track of the
radioactive energy release up to this point. After accounting for adiabatic
cooling, this energy constitutes the seed radiation field at the start of the
\gls{MCRT} simulations. We refer the reader to \citet{Lucy2005} and \citet{Noebauer2012}
for more details on the physical and numerical setup of this test problem.

\subsection{Radiative shocks} \label{sec:radiative_shocks}

Determining the structure of radiation-dominated shocks has become a standard
test problem to verify and validate the performance of numerical schemes for
\gls{RH}. The theoretical foundations for an understanding and
description of these phenomena have been laid out by \citet{Raizer1957a} and
\citet{Zeldovich1957a}\footnote{The original publications in Russian are
  \citet{Raizer1957} and \citet{Zeldovich1957}.}. Their findings are 
summarized in the text book by \citet{Zeldovich1967}. In
contrast to their hydrodynamical counterparts, radiation flows from the hot
shocked domain into the cold unshocked region and pre-heats and pre-compresses
the flow there. As a consequence, the sharp shock front becomes washed out.
Depending on the amount of pre-heating one distinguishes between sub- and
supercritical radiative shocks. In the latter, the material right in front of
the hydrodynamic shock is heated to the temperature of the shocked material
beyond the relaxation region.  For more details, we refer to the standard
literature on this subject, for example \citet{Zeldovich1967},
\citet{Sincell1999}, and \citet{Lowrie2008}.

\paragraph{Non-Steady Radiative Shocks: }

\citet{Ensman1994} examined a number of different test problems to verify and
validate numerical approaches to \gls{RT} and \gls{RH}. In particular, it was
proposed to solve the time-dependent structure of non-steady radiative shocks.
Over the years, this original suggestion has become a standard test to validate
new numerical \gls{RH} approaches
\citep[e.g.][]{Turner2001, Hayes2003, Hayes2006, Commerccon2011, Noebauer2012,
Kolb2013, Roth2015, Sijoy2015}.

The non-steady radiative shock calculations presented in this work follow
closely the suggestion by \citet{Ensman1994}. Specific details about the setup
are given by \citet{Noebauer2012}. The shock calculations are carried out in
a plane-parallel computational domain of size $L = \SI{7e10}{cm}$. The shock is
generated by directing the flow, which has a uniform grey absorption opacity of
$\chi = \SI{3.1e-10}{cm^{-1}}$ and initial uniform density of $\rho =
\SI{7.78e-8}{g.cm^{-3}}$, towards the left, reflecting boundary. Typically,
two realisations of the test problem are considered that differ in the bulk
velocity of the flow. With $v = \SI{-6e5}{cm.s^{-1}}$, a subcritical shock
emerges, while a super-critical shock is generated with $v =
\SI{-2e6}{cm.s^{-1}}$. Finally, an initial temperature structure with a slight
gradient is chosen\footnote{\citet{Ensman1994} found it necessary to include
this slight gradient to avoid numerical problems in their calculations.}:
\begin{equation}
  T(x) = \SI{10}{K} + \frac{L - x}{L} \SI{75}{K}.
  \label{eq:rhshocks_Tini}
\end{equation}
Following the suggestion by \citet{Hayes2003}, the shock structure calculated
is displayed
as a function of a pseudo-Lagrangian coordinate:
\begin{equation}
  z = x - v t.
  \label{eq:rhshocks_coordinate}
\end{equation}
This allows us to easily compare with the results of \citet{Ensman1994} who used
a Lagrangian code.

\paragraph{Steady Radiative Shocks: }

While the calculation of the evolving structure of non-steady radiative shocks
has become a standard test problem for radiation hydrodynamics, its value is
somewhat limited by the lack of an analytic solution to the test
problem.
Thanks to the developments by \citet{Lowrie2007} and \citet{Lowrie2008}, this
is not the case for steady radiative shocks: an analytic solution technique,
first based on the equilibrium diffusion description of radiative
transfer \citep{Lowrie2007} and later generalized by relying on non-equilibrium
diffusion \citep{Lowrie2008}, has been presented.  Following the original work,
solving the structure of steady radiative shocks and comparing it to the
predictions obtained with the solution technique of \citet{Lowrie2008} has
quickly become an integral part of the standard test suite for \gls{RH} approaches \citep[e.g.][]{Sekora2010, Holst2011, Zhang2011,
Davis2012, Jiang2012, Ramsey2014, Gonzalez2015, Roth2015}.

\citet{Lowrie2007} and \citet{Lowrie2008} based their solution strategy on the
non-dimensional form of the radiation hydrodynamical equations. In particular,
reference length, mass density, material temperature and material sound speeds
are used to convert the relevant physical quantities into their non-dimensional
counterparts, which we denote with the tilde symbol:
\begin{align}
  \tilde{x} &= \frac{x}{\hat{x}} & & \text{spatial coordinate}\\
  \tilde{\rho} &= \frac{\rho}{\hat{\rho}} & & \text{material density}\\
  \tilde{v} &= \frac{v}{\hat{a}} & & \text{material velocity}\\
  \tilde{T} &= \frac{T}{\hat{T}} & & \text{material temperature}\\
  \tilde{\theta} &= \frac{\theta}{\hat{T}} & &\text{radiation temperature}\\
  \tilde{\sigma} &= \frac{\sigma \hat{x} c}{\hat{a}} & &\text{cross section}\\
  \tilde{\kappa} &= \frac{c}{3 \sigma \hat{a} \hat{x}} &
                                   &\text{radiation diffusivity} \; ,
\end{align} 
where reference values for each quantity are denoted by a hat symbol. 
While not required for the applicability of the solution approach, the steady
radiation shock test problem is typically set up assuming a uniform absorption
cross section. In this case, a particular shock realisation can be
specified by choosing values for the dimensionless quantity
\begin{equation}
  \tilde{\mathcal{P}} = \frac{a_{\mathrm{R}} \hat{T}^4}{\hat{\rho} \hat{a}^2}
  \label{dq:non_dim_quant_P}
\end{equation}
the radiation diffusivity $\tilde{\kappa}$, an adiabatic index $\gamma$, an
absorption cross section $\tilde{\sigma}$, a value for the shock Mach number
$M$ and a reference length. By convention, the pre-shock state is given by $\tilde{\theta} =
1$, $\tilde{T} = 1$, $\tilde{\rho} = 1$ and $\tilde{v} = M$.

In \Cref{Sec:Dynamics}, we present the results of a $M=5$ test calculation,
with $ \tilde{\mathcal{P}}  = 10^{-4}$, $\tilde{\kappa} = 1$,
$\tilde{\sigma} = 10^6$ and $\tilde{x} = 1$, obtained with the
\gls{MC}-based \gls{RH} approach \textsc{Mcrh} \citep{Noebauer2012}. For the
numerical setup in \textsc{Mcrh}, which relies on the dimensional form of
physical quantities, the reference values $\hat{x} =
\SI{1}{cm}$, $\hat{a} = \SI{1.7310e7}{cm.s^{-1}}$, $\hat{\rho} =
\SI{0.3449}{g.cm^{-3}}$ and $\hat{T} = \SI{1.0810e6}{K}$ were used. A
plane-parallel computational domain was chosen in which a central discontinuity
separates two regions with constant fluid states. The jump fulfils the jump
conditions of the radiation-plus-matter system as given by \citet{Lowrie2007}.
Simple outflow boundary conditions are used for the hydrodynamic
sub-system and
\gls{MC} packets reaching the boundaries freely escape. To counteract this
outflow, inflow boundary conditions are set up for the radiative subsystem.
Specifically, inflowing radiation is imposed with a rate corresponding to that
of a thermal radiation field at the temperature given by the jump conditions
and integrated over one hemisphere. The system is then followed until a
steady-state has established which is then compared to the analytic predictions
according to \citet{Lowrie2008}\footnote{We use a Python implementation of the
solution strategy which is available at
\url{https://github.com/unoebauer/public-astro-tools}}.

\section{Software collection}
\label{sec:tool_collection}

As part of this review, we freely distribute a number of \gls{MCRT} Python
programs used in the test calculations together with the configuration files
for the \textsc{Tardis} simulation presented in \Cref{Sec:example}. The
interested reader can find all these in the GitHub repository at
\url{https://github.com/unoebauer/mcrtreview-tools.git}. It contains the
following tools and data files:
\begin{itemize}
 \item Python tool \texttt{mcrt\_hom\_sphere.py} to perform simple,
    time-independent \gls{MCRT} simulations for the homogeneous sphere problem (cf.\
    \Cref{sec:test_hom_sphere})
  \item Python script \texttt{mcrt\_escape\_prop.py} with which a simple \gls{MCRT}
    simulation to determine the escape probability from a homogeneous sphere
    can be performed (cf. \Cref{sec:test_hom_sphere} and
    \Cref{Sec:Propagation_abs_scat}) 
  \item Python script \texttt{mcrt\_pcyngi.py} to determine the P-Cygni line
    profile formed in a homologously expanding spherical flow (cf.\
    \Cref{sec:line_test} and \Cref{Sec:line_ints_sobolev})
   \item \textsc{Tardis} setup files for the spectral synthesis calculations
    presented in \Cref{Sec:example}. The setup consists of 
    \begin{itemize}
      \item the configuration file \texttt{tardis\_sn2005bl\_m3\_config.yml}
      \item the density structure file \texttt{tardis\_sn2005bl\_m3\_density.dat}
      \item the chemical composition file \texttt{tardis\_sn2005bl\_m3\_abundances.dat}
      \item text file containing information about the \textsc{Tardis} simulation facilitating reproduction; \texttt{info.rst}
    \end{itemize}
\end{itemize}


\bibliographystyle{spbasic}      
\bibliography{references}   

\end{document}